\documentstyle[preprint,aps,epsf,amssymb,amsbsy,amstext,subeqnarray,amsfonts]{revtex}
\draft
\tighten

\begin{document}

\def\barr{\begin{array}}
\def\earr{\end{array}}

\title{Deeply virtual electroproduction of photons and mesons on the
  nucleon : leading order amplitudes and power corrections}
\author{M. Vanderhaeghen$^a$, P.A.M. Guichon$^b$ and M. Guidal$^c$ }
\address{$^a$ Institut f\"ur Kernphysik, Johannes Gutenberg Universit\"at, D-55099 Mainz, Germany}
\address{$^b$ CEA-Saclay, DAPNIA/SPhN, F-91191 Gif-sur-Yvette, France}
\address{$^c$ IPN Orsay, F-91406 Orsay, France}
\date{\today}
\maketitle

\begin{abstract}
We estimate the leading order amplitudes for exclusive photon and meson electroproduction
reactions at large \( Q^{2} \) in the valence region in terms of skewed quark
distributions. As experimental investigations can currently only be envisaged
at moderate values of \( Q^{2} \) , we estimate power corrections due to the
intrinsic transverse momentum of the partons in the meson wavefunction and in
the nucleon. To this aim the skewed parton distribution formalism is generalized
so as to include the parton intrinsic transverse momentum dependence. Furthermore,
for the meson electroproduction reactions, we calculate the soft overlap type
contributions and compare with the leading order amplitudes. We give first estimates
for these different power corrections in kinematics which are relevant for experiments
in the near future.

\end{abstract}

\section{Introduction}

Much of the internal structure of the nucleon has been revealed during the last
two decades through the \textit{inclusive} scattering of high energy leptons
on the nucleon in the Bjorken -or ``Deep Inelastic Scattering\char`\"{} (DIS)-
regime defined by \( Q^{2},\nu \rightarrow \infty  \) and \( x_{B}=Q^{2}/2M\nu  \)
finite. Simple theoretical interpretations of the experimental results can be
reached in the framework of QCD, when one sums over all the possible hadronic
final states. \\
 \indent With the advent of the new generation of high-energy, high-luminosity
lepton accelerators combined with large acceptance spectrometers, a wide variety
of \textit{exclusive} processes in the Bjorken regime are considered as experimentally
accessible. In recent years, a unified theoretical description of such processes
has emerged through the formalism introducing new generalized parton distributions,
the so-called `Off-Forward Parton Distributions' (OFPD's), commonly also denoted
as skewed parton distributions. It has been shown that these distributions,
which parametrize the structure of the nucleon, allow to describe, in leading
order perturbative QCD (PQCD), various exclusive processes in the near forward
direction. The most promising ones are deeply Virtual Compton Scattering (DVCS)
and longitudinal electroproduction of vector or pseudoscalar mesons (see Refs.\cite{Ji97}-\cite{Collins97}
and references therein to other existing literature). Maybe most of the recent
interest and activity in this field has been triggered by the observation of
Ji \cite{Ji97}, who showed that the second moment of these OFPD's measures
the contribution of the spin and orbital momentum of the 
quarks to the nucleon spin. 
Clearly this may shed a new light on the ``spin-puzzle''. \\
 \indent In section \ref{sec:sec2}, we introduce the definitions and conventions
of the skewed parton distributions. We also present the modelizations for these
distributions that will be used in our cross section estimates. \\
 \indent In section \ref{sec:sec3}, the leading order PQCD amplitudes for DVCS
and longitudinal electroproduction of mesons are presented in some detail. This
is an extension of our previous works \cite{ourprl,vcsrev}. \\
 \indent In section \ref{sec:sec4}, we investigate the power corrections to
the leading order amplitudes when the virtuality of the photon is in the region
\( Q^{2}\approx 1-20 \) GeV\( ^{2} \). The correction due to the intrinsic
transverse momentum dependence is known to be important to get a successful
description of the \( \pi ^{0}\gamma ^{*}\gamma  \) transition form factor,
for which data exist in the range \( Q^{2}\approx 1-10 \) GeV\( ^{2} \). For
the pion electromagnetic form factor in the transition region before asymptotia
is reached, the power corrections due to both the transverse momentum dependence
and the soft overlap mechanism are quantitatively important. We will therefore
take these form factors as a guide to calculate the corrections due to the parton's
intrinsic transverse momentum dependence in the DVCS and hard meson electroproduction
amplitudes. In addition, for the meson electroproduction amplitude, an estimate
is given for the competing soft overlap mechanism, which - in contrast to the
leading order perturbative mechanism - does not proceed through one-gluon exchange.
\\
 \indent In section \ref{sec:sec5}, we give numerical estimates for different
observables for DVCS and for the electroproduction of vector mesons \( \rho ^{(\pm ,0)} \),
\( \omega  \) and pseudoscalar mesons \( \pi ^{(\pm ,0)} \), \( \eta  \).
We give several examples of experimental opportunities to access the OFPD's
at the current high-energy lepton facilities~: JLab (\( E_{e}\geq  \) 6 GeV)
\cite{propo98107,bertin}, 
HERMES (\( E_{e} \)=27 GeV) and COMPASS (\( E_{\mu } \)=200 GeV). \\
 \indent Finally, we give our conclusions in section \ref{sec:sec6}.

\section{Review of Skewed Parton Distributions}

\label{sec:sec2}

\subsection{Preliminary : bilocal operators and gauge invariance}

\label{sec:bilocal}

\begin{figure}[ht]
\epsfxsize = 12cm
\centerline{\epsffile{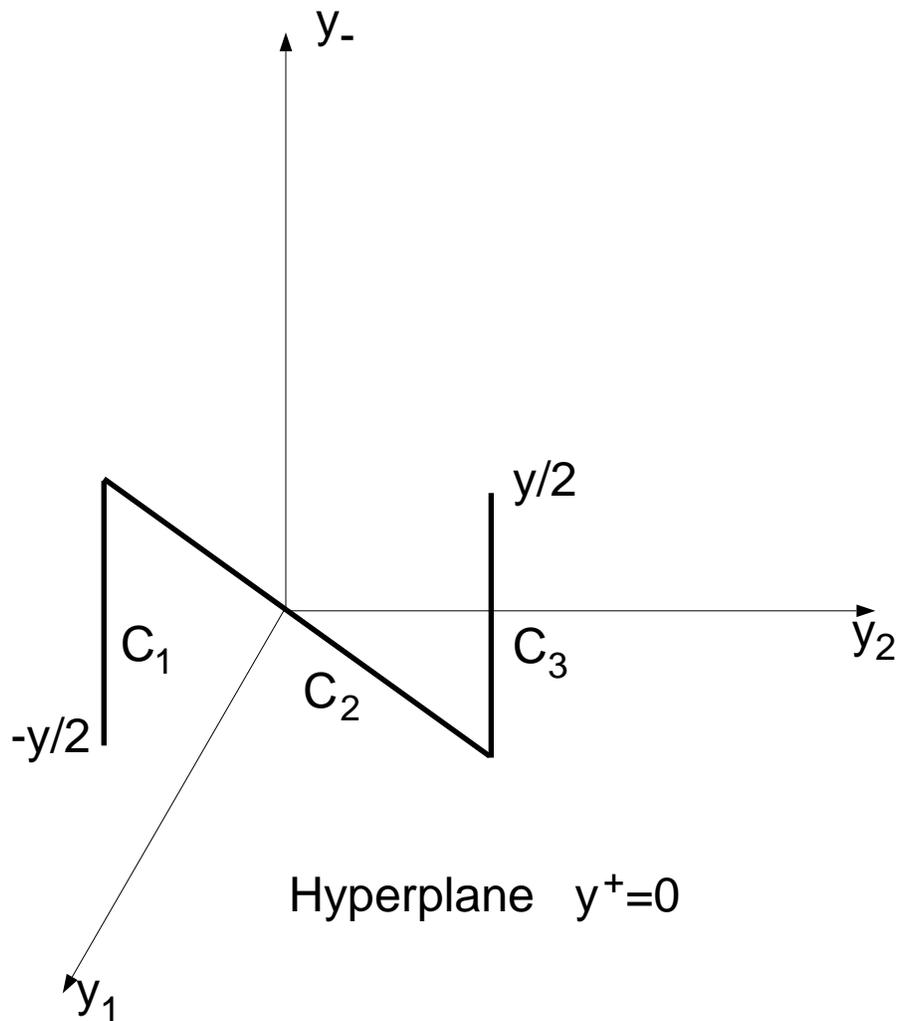} }
\caption{\small Path chosen for the gauge invariant definition of 
a bilocal product of quark fields. Remark that the segment $C_2$ lies
in the plane $y^- = 0$.}
\label{fig:path}
\end{figure}

In the following we often encounter bilocal products of quark fields \( \psi  \)
at equal light-cone time of the form 
\begin{equation}
\label{eq:biloc1}
\overline{\psi }(-y/2)\psi (y/2)\mid _{y^{+}=0}\; ,
\end{equation}
 where \( y^{+} \) is the light-cone time, with \( y^{\pm }=(y^{0}\pm y^{3})/\sqrt{2} \).
At leading twist one has the further restriction \( \vec{y}_{\perp }=(y^{1},y^{2},0)=0 \),
that is the bilocal product needs only be evaluated along a light-cone segment
of length \( y^{-} \). For our purposes we need to consider the more general
case where only \( y^{+} \) is set to zero 
( \(y^{+}\) = 0 is understood in the remaining of this section). \\
 \indent The product of fields at different points is not invariant under a
local color gauge transformation. To enforce gauge invariance, one replaces
Eq.~(\ref{eq:biloc1}) by 
\begin{equation}
\label{eq:biloc2}
\overline{\psi }(-y/2)L(-y/2\rightarrow y/2)\psi (y/2),
\end{equation}
 where the link operator \( L \) is defined by 
\begin{equation}
\label{eq:biloc3}
L(-y/2\rightarrow y/2)=P(e^{-i\int _{C}A.dl}).
\end{equation}
 In Eq.~(\ref{eq:biloc3}), the path ordered product is evaluated along a curve
\( C \) joining the points \( -y/2 \) and \( y/2 \) and \( A \) is the matrix
valued color gauge field. So in a general gauge, bilocal products of quark fields
are not defined independently of the gauge field. To simplify the analysis it
is therefore convenient to choose a curve \( C \) and a gauge such that the
link operator reduces to the identity. \\
 \indent Among the many possibilities we choose the curve shown on Fig.~\ref{fig:path},
where it is understood that \( y^{+}=0 \). We then split \( L \) into 3 factors
\( L_{1},L_{2},L_{3} \) corresponding to the segments \( C_{1},C_{2},C_{3} \)
indicated on Fig.~\ref{fig:path}. We first go to the gauge where \( A^{+}=0 \)
everywhere. This is always possible if one ignores the problems associated with
the boundary conditions at infinity. Since we only deal with localised systems
this is not a true restriction. In this gauge we have \( L_{1}=L_{3}=1. \)
While staying in the gauge \( A^{+}=0 \) we still have the freedom to make
a gauge transformation which depends only on \( (y^{1},y^{2}). \) This allows
us to go to a gauge where \( y^{1}A^{1}+y^{2}A^{2}=0 \) \emph{in the plane}
\( y^{-}=0. \) This is the 2-dimensional version of the radial, or Schwinger-Fock,
gauge \cite{Leib87}. In this gauge we have \( L_{2}=1 \) which completes our
argument, that is there exists a curve and a gauge such that the link operator
is the identity operator. In the following it will always be understood that
the bilocal quark field products (at \( y^{+}=0 \)) are in this gauge 
\footnote{ or any other one which allows to replace the link operator 
by the identity }, which is thus a convenient choice for model evaluations 
of the matrix elements of these operators.

\subsection{Definitions of skewed parton distributions}

\begin{figure}[ht]
\epsfxsize=11 cm
\epsfysize=6. cm
\centerline{\epsffile{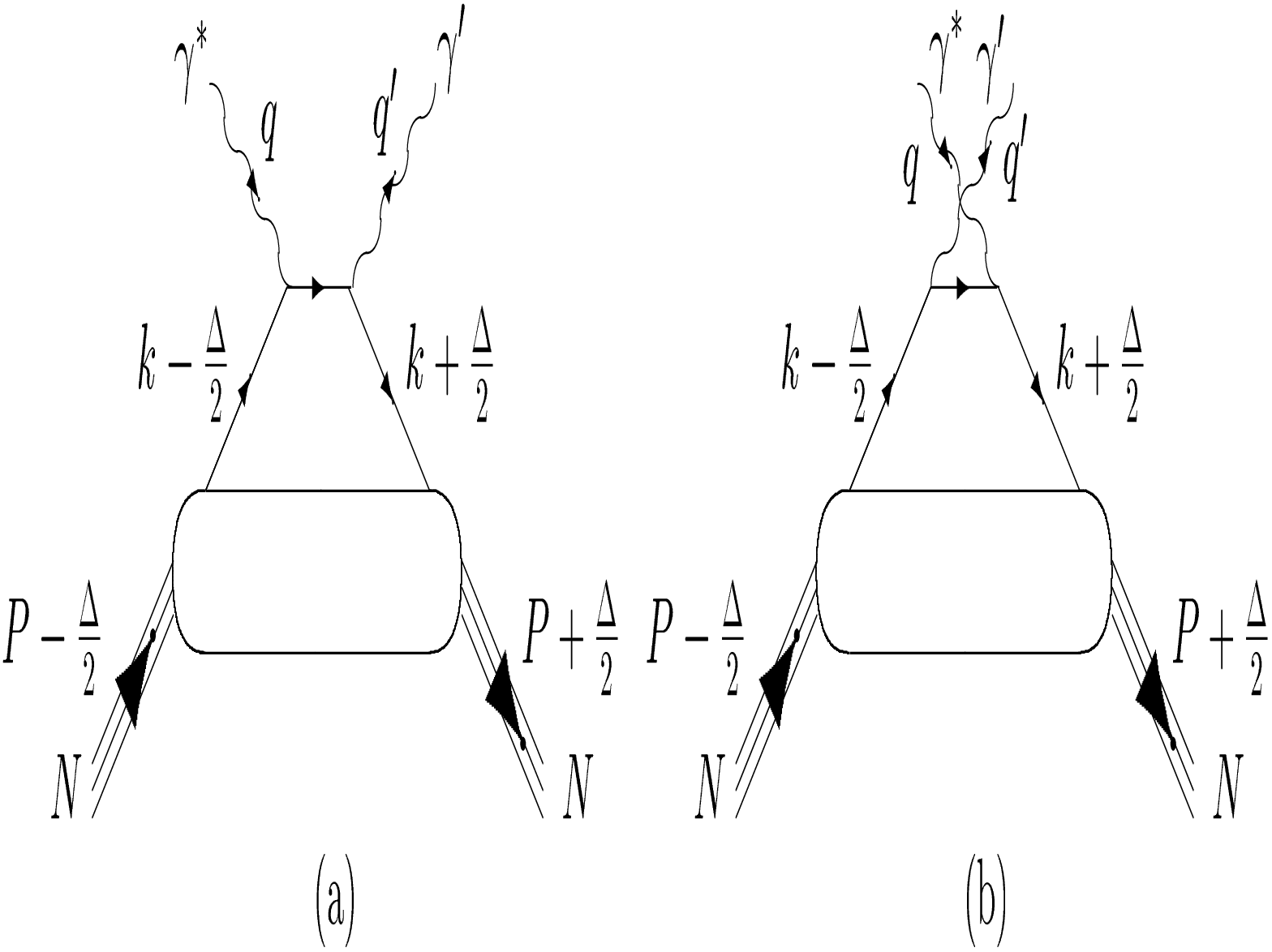}}
\vspace{-1.5cm}
\caption[]{\small ``Handbag'' diagrams for DVCS.}
\label{fig:handbags}
\end{figure}

To set the framework of this work, we briefly review how the nonperturbative
nucleon structure information enters the leading order PQCD amplitude for DVCS
and hard meson electroproduction. \\
 \indent For DVCS, Ji \cite{Ji97} and Radyushkin \cite{Rady96} have shown
that the leading order amplitude in the forward direction can be factorized
in a hard scattering part (which is exactly calculable in PQCD) and a soft,
nonperturbative nucleon structure part as is illustrated in Fig.~\ref{fig:handbags}.
In these so-called ``handbag\char`\"{} diagrams, the nucleon structure information
can be parametrized, at leading order PQCD, in terms of four generalized structure
functions, which conserve quark helicity. It was shown in Refs.~\cite{Hood98,Ji98},
that at leading twist, two more functions appear which involve a quark helicity
flip. However, the authors of Ref.~\cite{Diehl99} have shown that in reactions
which would involve these helicity flip distributions (such as exclusive electroproduction
of transversely polarized vector mesons), this contribution vanishes due to
angular momentum and chirality conservation in the hard scattering. This was
argued \cite{Diehl99} to hold at leading order in \( 1/Q \) and to all orders
in perturbation theory. In this work, we will restrict ourselves to the four
distributions which conserve quark helicity. Then, the matrix element of the
bilocal quark operator, representing the lower blob in Figs.~\ref{fig:handbags}a
and \ref{fig:handbags}b, can be expressed at leading twist, in the notation
of Ji, in terms of the OFPD's \( H,\tilde{H},E,\tilde{E} \) as : 
\begin{eqnarray}
 &  & {{P^{+}}\over {2\pi }}\, \int dy^{-}e^{ixP^{+}y^{-}}\langle p^{'}|\bar{\psi }_{\beta }(-{y\over 2})\psi _{\alpha }({y\over 2})|p\rangle {\Bigg |}_{y^{+}=\vec{y}_{\perp }=0}\nonumber \\
 &  & ={1\over 4}\left\{ ({\gamma ^{-}})_{\alpha \beta }\left[ {H^{q}(x,\xi ,t)\; \bar{N}(p^{'})\gamma ^{+}N(p)\; +\; E^{q}(x,\xi ,t)\; \bar{N}(p^{'})i\sigma ^{+\kappa }{{\Delta _{\kappa }}\over {2m_{N}}}N(p)}\right] \right. \nonumber \\
 &  & \hspace {.7cm}\left. +({\gamma _{5}\gamma ^{-}})_{\alpha \beta }\left[ {\tilde{H}^{q}(x,\xi ,t)\; \bar{N}(p^{'})\gamma ^{+}\gamma _{5}N(p)\; +\; \tilde{E}^{q}(x,\xi ,t)\; \bar{N}(p^{'})\gamma _{5}{{\Delta ^{+}}\over {2m_{N}}}N(p)}\right] \right\} \; ,\label{eq:qsplitting} 
\end{eqnarray}
 where \( \psi  \) is the quark field, \( N \) the nucleon spinor and \( m_{N} \)
the nucleon mass. In writing down Eq.~(\ref{eq:qsplitting}) one uses a frame
where the virtual photon momentum \( q^{\mu } \) and the average nucleon momentum
\( P^{\mu } \) (see Fig.~\ref{fig:handbags} for the kinematics) are collinear
along the \( z \)-axis and in opposite direction. We denote the lightlike vectors
along the positive and negative \( z \)-directions as \( \tilde{p}^{\mu }=P^{+}/\sqrt{2}(1,0,0,1) \)
and \( n^{\mu }=1/P^{+}\cdot 1/\sqrt{2}(1,0,0,-1) \) respectively, and define
light-cone components \( a^{\pm } \) by \( a^{\pm }\equiv (a^{0}\pm a^{3})/\sqrt{2} \).
In this frame, the physical momenta have the following decomposition : 
\begin{eqnarray}
 &  & P^{\mu }={1\over 2}\left( p^{\mu }+p'^{\mu }\right) \; =\tilde{p}^{\mu }+{{\bar{m}^{2}}\over 2}\, n^{\mu }\; ,\\
 &  & q^{\mu }=-\left( 2\xi ^{'}\right) \, \tilde{p}^{\mu }+\left( {{Q^{2}}\over {4\xi ^{'}}}\right) \, n^{\mu }\; ,\\
 &  & \Delta ^{\mu }\equiv p^{'\mu }-p^{\mu }=-\left( 2\xi \right) \, \tilde{p}^{\mu }+\left( \xi \, \bar{m}^{2}\right) \, n^{\mu }+\Delta ^{\mu }_{\perp }\; ,\label{eq:dvcsxi} \\
 &  & q^{'\mu }\equiv q^{\mu }-\Delta ^{\mu }=-2\left( \xi ^{'}-\xi \right) \, \tilde{p}^{\mu }+\left( {{Q^{2}}\over {4\xi ^{'}}}-\xi \, \bar{m}^{2}\right) \, n^{\mu }-\Delta ^{\mu }_{\perp }\; ,\label{eq:dvcsqp} 
\end{eqnarray}
 where the variables \( \bar{m}^{2} \), \( \xi ^{'} \) and \( \xi  \) are
given by 
\begin{eqnarray}
 &  & \bar{m}^{2}={m_{N}}^{2}-{{\Delta ^{2}}\over 4}\; ,\label{eq:dvcskin} \\
 &  & 2\xi ^{'}={{P\cdot q}\over {\bar{m}^{2}}}\, \left[ -1+\sqrt{1+{{Q^{2}\, \bar{m}^{2}}\over {(P\cdot q)^{2}}}}\right] \; \stackrel{Bj}{\longrightarrow }\; \frac{x_{B}}{1-\frac{x_{B}}{2}}\; ,\\
 &  & 2\xi =2\xi ^{'}\, {{Q^{2}-\Delta ^{2}}\over {Q^{2}+\bar{m}^{2}(2\xi ^{'})^{2}}}\; \stackrel{Bj}{\longrightarrow }\; \frac{x_{B}}{1-\frac{x_{B}}{2}}\; .\label{eq:dvcskin3} 
\end{eqnarray}
 In Eq.~(\ref{eq:qsplitting}), the OFPD's \( H^{q},E^{q},\tilde{H}^{q},\tilde{E}^{q} \)
are defined for one quark flavor (\( q=u,d \) and \( s \)). The functions
\( H \) and \( E \) are helicity averaged whereas \( \tilde{H} \) and \( \tilde{E} \)
are helicity dependent functions. They depend upon three variables : \( x \),
\( \xi  \), \( t \). The light-cone momentum fraction \( x \) is defined
by \( k^{+}=xP^{+} \). The variable \( \xi  \) is defined by \( \Delta ^{+}=-2\xi \, P^{+} \),
where \( \Delta \equiv p^{'}-p \) and \( t=\Delta ^{2} \). Note that \( 2\xi \rightarrow x_{B}/(1-x_{B}/2) \)
in the Bjorken limit. The support in \( x \) of the OFPD's is \( \left[ -1,1\right]  \)
and a negative momentum fraction corresponds with the antiquark contribution.
\\
 \indent To simplify the presentation, we will take \( \xi ^{'}=\xi  \) in
the following. As shown in Eqs.~(\ref{eq:dvcskin}-\ref{eq:dvcskin3}), these
two variables have the same value in the Bjorken limit. The kinematic variable
\( \xi  \), which represents the longitudinal momentum fraction of the transfer
\( \Delta  \), is bounded by 
\begin{equation}
0<\xi <{{\sqrt{-\Delta ^{2}}/2}\over {\bar{m}}}<1\; .
\end{equation}
\\
 \indent A glance at Figs.~\ref{fig:handbags} a,b shows that the active quark
with momentum \( k-\Delta /2 \) has longitudinal (+ component) momentum fraction
\( x+\xi  \), whereas the one with momentum \( k+\Delta /2 \) has longitudinal
momentum fraction \( x-\xi  \). As noted by Radyushkin \cite{Rady96}, since
negative momentum fractions correspond to antiquarks, one can identify two regions
according to whether \( |x|>\xi  \) or \( |x|<\xi  \) .

\begin{figure}[ht]
\epsfxsize=16 cm
\centerline{\epsffile{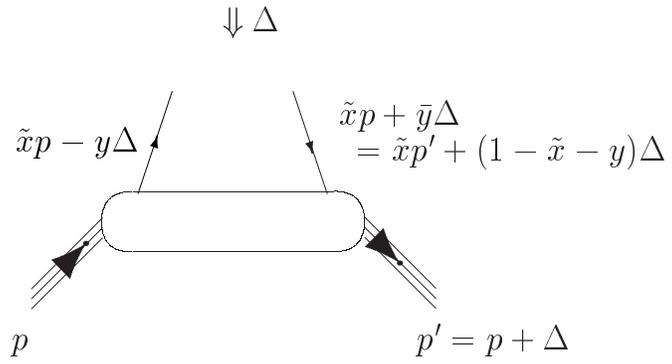}}
\vspace{-9cm}
\caption[]{\small Diagram to denote the arguments $\tilde x$, $y$ and
  $\Delta^2$ of the double distributions.}
\label{fig:double}
\end{figure}

\begin{itemize}
\item When \( x>\xi  \), both quark propagators represent quarks, whereas for \( x<-\xi  \)
both represent antiquarks. In these regions, the OFPD's are the generalizations
of the usual parton distributions from DIS. Actually, in the forward direction,
the OFPD's \( H \) and \( \tilde{H} \) respectively reduce to the quark density
distribution \( q(x) \) and the quark helicity distribution \( \Delta q(x) \)
: 
\begin{equation}
\label{eq:dislimit}
H^{q}(x,0,0)\, =\, q(x)\; ,\hspace {2cm}\tilde{H}^{q}(x,0,0)\, =\, \Delta q(x)\; ,
\end{equation}
 where \( q \) and \( \Delta q \) are defined by the Fourier integrals on
the light cone \cite{Jaf96} \footnotemark{}: 
\begin{eqnarray}
q(x)\,  & = & \, {{p^{+}}\over {4\pi }}\, \int dy^{-}e^{ixp^{+}y^{-}}\langle p|\bar{\psi }(0)\gamma .n\psi (y)|p\rangle {\Bigg |}_{y^{+}=\vec{y}_{\perp }=0}\; ,\label{eq:dis1} \\
\Delta q(x)\,  & = & \, {{p^{+}}\over {4\pi }}\, \int dy^{-}e^{ixp^{+}y^{-}}\langle pS_{\Vert }|\bar{\psi }(0)\gamma .n\gamma _{5}\psi (y)|pS_{\Vert }\rangle {\Bigg |}_{y^{+}=\vec{y}_{\perp }=0}\; .\label{eq:dis} 
\end{eqnarray}
 In Eqs.~(\ref{eq:dis1}, \ref{eq:dis}), \( p \) represents the initial nucleon
momentum in the DIS process and \( S_{\Vert } \) is the longitudinal nucleon
spin projection. From Eqs.~(\ref{eq:dis1}, \ref{eq:dis}) the quark distributions
for negative momentum fractions are related to the antiquark distributions as
\( q(-x)=-\, \bar{q}(x) \) and \( \Delta q(-x)=+\, \Delta \bar{q}(x) \). 
\item In the region \( -\xi <x<\xi  \), one quark propagator represents a quark and
the other one an antiquark. In this region, the OFPD's behave like a meson distribution
amplitude. 
\end{itemize}
\footnotetext{ Note the typing errors in Eqs.~(120, 121) of Ref.~\cite{vcsrev},
where the factors \( 2\pi  \) should be replaced by \( 4\pi  \). }An alternative,
but equivalent, parametrization of the matrix elements of the bilocal operators
has been proposed by Radyushkin \cite{Rady96} by expressing the longitudinal
momentum fractions with respect to the initial nucleon momentum \( p \) instead
of the average momentum \( P \) of Fig.~\ref{fig:handbags}. In this notation,
the active quark has momentum \( k^{+}=Xp^{+} \) and \( \Delta ^{+}=-\zeta p^{+} \).
The two variables \( X \) and \( \zeta  \) are related to the previously introduced
momentum fractions \( x \) and \( \xi  \) by: 
\begin{equation}
\label{eq:nfpd}
x\equiv {{X-\zeta /2}\over {1-\zeta /2}}\; ,\hspace {0.5cm}\xi \equiv {{\zeta /2}\over {1-\zeta /2}}\; .
\end{equation}
 In terms of the variables \( X \), \( \zeta  \) and \( t \), so-called `non-forward
parton distributions' (NFPD's) \( f^{q}(X,\zeta ,t) \) were introduced by Radyushkin
\cite{Rady96}. They are equivalent to the OFPD's and are related to them by
\begin{eqnarray}
H^{q}(x,\xi ,t)\,  & = & \, {{1}\over {1+\xi }}\, \left\{ \, f^{q}(X,\zeta ,t)\, \right\} \; ,\hspace {1cm}{\mathrm{f}or}\hspace {0.5cm}\xi \leq x\leq 1\; ,\nonumber \\
H^{q}(x,\xi ,t)\,  & = & \, {{1}\over {1+\xi }}\, \left\{ \, f^{q}(X,\zeta ,t)\; -\; f^{\bar{q}}(\zeta -X,\zeta ,t)\, \right\} \; ,\hspace {1cm}{\mathrm{f}or}\hspace {0.5cm}-\xi \leq x\leq \xi \; ,\nonumber \\
H^{q}(x,\xi ,t)\,  & = & \, {{1}\over {1+\xi }}\, \left\{ \, -\, f^{\bar{q}}(\zeta -X,\zeta ,t)\, \right\} \; ,\hspace {1cm}{\mathrm{f}or}\hspace {0.5cm}-1\leq x\leq -\xi \; ,\label{eq:relofpdnfpd} 
\end{eqnarray}
 where \( f^{q} \) are the quark and \( f^{\bar{q}} \) the anti-quark NFPD's
respectively. \\
 \indent The ``double'' nature of the skewed parton distributions according
to whether \( 0\leq x\leq \xi  \) (\( 0\leq X\leq \zeta  \)) or \( \xi \leq x\leq 1 \)
(\( \zeta \leq X\leq 1 \)) was made more transparent by Radyushkin by expressing
the quark momenta connected to the lower blob of Fig.~\ref{fig:handbags} in
a fraction \( \tilde{x} \) of the initial nucleon momentum \( p \) and a fraction
\( y \) of the momentum transfer \( \Delta  \) as shown in Fig.~\ref{fig:double}.
The relation with the previously introduced momentum fraction \( X \) is readily
seen to be \( X=\tilde{x}+\zeta y \). In terms of the variables \( \tilde{x},y,t \),
Radyushkin then introduced \cite{Rady96,Rady98} the so-called double distributions
\( F^{q} \), \( K^{q} \) for quarks and \( F^{\bar{q}} \), \( K^{\bar{q}} \)
for the anti-quarks, according to: 
\begin{eqnarray}
 &  & \langle p^{'}|\bar{\psi }(0)\; \gamma ^{+}\; \psi (z)|p\rangle {\Bigg |}_{z^{+}=\vec{z}_{\perp }=0}\nonumber \\
 &  & =\bar{N}(p^{'})\gamma ^{+}N(p)\; \cdot \; \int _{0}^{1}d\tilde{x}\, \int _{0}^{1-\tilde{x}}dy\; \left\{ e^{-i\left( \tilde{x}p^{+}\, -\, y\Delta ^{+}\right) z^{-}}F^{q}(\tilde{x},y,t)\; -\; e^{i\left( \tilde{x}p^{+}\, +\, \bar{y}\Delta ^{+}\right) z^{-}}F^{\bar{q}}(\tilde{x},y,t)\right\} \nonumber \\
 &  & +\; \bar{N}(p^{'})i\sigma ^{+\kappa }{{\Delta _{\kappa }}\over {2m_{N}}}N(p)\nonumber \\
 &  & \hspace {.5cm}\times \int _{0}^{1}d\tilde{x}\, \int _{0}^{1-\tilde{x}}dy\; \left\{ e^{-i\left( \tilde{x}p^{+}\, -\, y\Delta ^{+}\right) z^{-}}K^{q}(\tilde{x},y,t)\; -\; e^{i\left( \tilde{x}p^{+}\, +\, \bar{y}\Delta ^{+}\right) z^{-}}K^{\bar{q}}(\tilde{x},y,t)\right\} \; ,\label{eq:doubledef} 
\end{eqnarray}
 with both \( \tilde{x} \) and \( y \) positive and \( \tilde{x}+y\leq 1 \).
The double distributions are related to the NFPD's through integrals over \( y \)
\cite{Rady96,Rady98} : 
\begin{eqnarray}
 &  & f^{q}(X,\zeta ,t)\, =\, \int _{0}^{\bar{X}/\bar{\zeta }}dy\; \; F^{q}(X-\zeta y,y,t)\; ,\hspace {1cm}{\mathrm{f}or}\hspace {1cm}X\geq \zeta \; ,\nonumber \\
 &  & f^{q}(X,\zeta ,t)\, =\, \int _{0}^{X/\zeta }dy\; \; F^{q}(X-\zeta y,y,t)\; ,\hspace {1cm}{\mathrm{f}or}\hspace {1cm}X\leq \zeta \; ,\label{eq:double} 
\end{eqnarray}
 with \( \bar{X}\equiv 1-X \) and \( \bar{\zeta }\equiv 1-\zeta  \). Relations
similar to Eq.~(\ref{eq:double}) hold for the anti-quark distributions \( f^{\bar{q}} \).
Analogously to Eq.~(\ref{eq:doubledef}), one defines the helicity dependent
double distributions \( \tilde{F} \) and \( \tilde{K} \) through the matrix
element of a bilocal axial vector operator. \\
 \indent Besides coinciding with the quark distributions at vanishing momentum
transfer, the skewed parton distributions have interesting links with other
nucleon structure quantities. The first moments of the OFPD's are related to
the elastic form factors of the nucleon through model independent sum rules
\cite{Ji97} . By integrating Eq.~(\ref{eq:qsplitting}) over \( x \), one
gets the following relations for one quark flavor : 
\begin{eqnarray}
 &  & \int _{-1}^{+1}dx\; H^{q}(x,\xi ,t)\, =\, F_{1}^{q}(t)\; ,\hspace {0.5cm}\int _{-1}^{+1}dx\; E^{q}(x,\xi ,t)\, =\, F_{2}^{q}(t)\; ,\label{eq:vecsumrule} \\
 &  & \int _{-1}^{+1}dx\; \tilde{H}^{q}(x,\xi ,t)\, =\, g_{A}^{q}(t)\; ,\hspace {0.5cm}\int _{-1}^{+1}dx\; \tilde{E}^{q}(x,\xi ,t)\, =\, h_{A}^{q}(t)\; .\label{eq:axvecsumrule} 
\end{eqnarray}
 The elastic form factors for one quark flavor on the RHS of Eqs.~(\ref{eq:vecsumrule},
\ref{eq:axvecsumrule}) have to be related to the physical ones. Restricting
to the \( u,d \) and \( s \) quark flavors, the Dirac form factors are expressed
as 
\begin{equation}
\label{eq:vecff}
F_{1}^{u/p}\, =\, 2F_{1}^{p}+F_{1}^{n}+F_{1}^{s}\; ,\hspace {0.5cm}F_{1}^{d/p}\, =\, 2F_{1}^{n}+F_{1}^{p}+F_{1}^{s}\; ,
\end{equation}
 where \( F_{1}^{p} \) and \( F_{1}^{n} \) are the proton and neutron electromagnetic
form factors respectively. The strange form factor is given by \( F_{1}^{s/p}\equiv F_{1}^{s} \),
but since it is small and not so well known we set it to zero in the following
numerical evaluations. Relations similar to Eq.~(\ref{eq:vecff}) hold for the
Pauli form factors \( F_{2}^{q} \). For the axial vector form factors one uses
the isospin decomposition~: 
\begin{equation}
\label{eq:axff}
g_{A}^{u/p}\, =\, {1\over 2}g_{A}+{1\over 2}g_{A}^{0}\; ,\hspace {0.5cm}g_{A}^{d/p}\, =\, -{1\over 2}g_{A}+{1\over 2}g_{A}^{0}\; .
\end{equation}
 The isovector axial form factor \( g_{A} \) is known from experiment, with
\( g_{A}(0)\approx 1.267 \) \cite{PDG98}. For the unknown isoscalar axial
form factor \( g_{A}^{0} \), we use the quark model relation : \( g_{A}^{0}(t)=3/5\, g_{A}(t) \).
For the pseudoscalar form factor \( h_{A}^{q} \), we have relations similar
to Eq.~(\ref{eq:axff}). \\
 \indent The second moment of the OFPD's is relevant for the nucleon spin structure.
It was shown in Ref.~\cite{Ji97} that there exists a (color) gauge-invariant
decomposition of the nucleon spin~: 
\begin{equation}
{1\over 2}\, =\, J_{q}\, +\, J_{g}\; ,
\label{eq:totalspin}
\end{equation}
 where \( J_{q} \) and \( J_{g} \) are respectively the total quark and gluon
angular momentum. The second moment of the unpolarized OFPD's at \( t=0 \)
gives 
\begin{equation}
\label{eq:dvcsspin}
J_{q}\, =\, {1\over 2}\, \int _{-1}^{+1}dx\, x\, \left[ H^{q}(x,\xi ,t=0)+E^{q}(x,\xi ,t=0)\right] \; ,
\end{equation}
 and this relation is independent of \( \xi  \). The quark angular momentum
\( J_{q} \) decomposes as 
\begin{equation}
\label{eq:spindecomp}
J_{q}={1\over 2}\Delta \Sigma +L_{q}\; ,
\end{equation}
 where \( \Delta \Sigma /2 \) and \( L_{q} \) are respectively the quark spin
and orbital angular momentum. As \( \Delta \Sigma  \) is measured through polarized
DIS experiments, a measurement of the sum rule of Eq.~(\ref{eq:dvcsspin}) in
terms of the OFPD's, provides a model independent way to determine the quark
orbital contribution to the nucleon spin and thus from 
Eq.~(\ref{eq:totalspin}) the gluon contribution.

\subsection{Modelization of the skewed parton distributions}

\label{sec:modelofpd}

Ultimately one wants to extract the skewed parton distributions from the data
but, in order to evaluate electroproduction observables, we need a first guess
for them. We shall restrict our considerations to the near forward direction
because this kinematical domain is the closest to inclusive DIS, which we want
to use as a guide. Since \( E \) and \( \tilde{E} \) are always multiplied
by the momentum transfer, this means that the amplitudes are dominated by \( H \)
and \( \tilde{H} \) which we discuss first.

\subsubsection{\protect\protect\( \xi \protect \protect \)-independent ansatz for \protect\protect\( H\protect \protect \)
and \protect\protect\( \tilde{H}\protect \protect \)}

The simplest ansatz, already used in our previous works \cite{ourprl,vcsrev},
is to write \( H \) and \( \tilde{H} \) as a product of a form factor and
a quark distribution function, neglecting any \( \xi  \) dependence, which
yields : 
\begin{equation}
\label{eq:factunpol}
H^{u/p}(x,\xi ,t)=u(x)\, {F_{1}^{u/p}(t)}/2\; ,\hspace {0.5cm}H^{d/p}(x,\xi ,t)=d(x)\, {F_{1}^{d/p}(t)}\; ,\hspace {0.5cm}H^{s/p}(x,\xi ,t)=0\; ,
\end{equation}
 where \( u(x),d(x) \) and \( s(x) \) are the unpolarized quark distributions.
Obviously the ansatz of Eq.~(\ref{eq:factunpol}) satisfies both Eq.~(\ref{eq:dislimit})
and the sum rule of Eq.~(\ref{eq:vecsumrule}), if one uses the valence quark
distributions normalized as: 
\begin{equation}
{1\over 2}\int _{0}^{+1}dx\; u_{V}(x)=\int _{0}^{+1}dx\; d_{V}(x)=1\; .
\end{equation}
 For all the calculations presented in this paper, we shall use the parametrization
MRST98 \cite{MRST98} for the quark distributions.

To model \( \tilde{H}^{q} \), we need the corresponding polarized quark distributions.
For this we follow the recent work of Ref.\cite{Leader98}, where a next to
leading order QCD analysis of inclusive polarized deep-inelastic lepton-nucleon
scattering was performed and which yields an excellent fit of the world data.
In this analysis, the input polarized densities (at a scale \( Q_{0}^{2} \)
= 1 GeV\( ^{2} \)) are given by : 
\begin{eqnarray}
 &  & \Delta u_{V}(x,Q_{0}^{2})=\eta _{u}\, A_{u}\; x^{0.250}\; u_{V}(x,Q_{0}^{2})\; ,\nonumber \\
 &  & \Delta d_{V}(x,Q_{0}^{2})=\eta _{d}\, A_{d}\; x^{0.231}\; d_{V}(x,Q_{0}^{2})\; ,\nonumber \\
 &  & \Delta \bar{q}(x,Q_{0}^{2})=\eta _{\bar{q}}\, A_{S}\; x^{0.576}\; S(x,Q_{0}^{2})\; ,\label{eq:poldistr} 
\end{eqnarray}
 where we assume a SU(3) symmetric sea, i.e. \( \Delta \bar{q}\equiv \Delta \bar{u} \)
= \( \Delta \bar{d} \) = \( \Delta \bar{s} \) and \( S \) represents the
total sea. On the rhs of Eqs.~(\ref{eq:poldistr}), the normalization factors
\( A_{u},A_{d},A_{S} \) are determined so that the first moments of the polarized
densities are given by \( \eta _{u},\eta _{d},\eta _{\bar{q}} \) respectively.
For the valence quark densities, \( \eta _{u} \) and \( \eta _{d} \) were
fixed in Ref.\cite{Leader98} by the octet hyperon \( \beta  \) decay constants
which yield : 
\begin{equation}
\eta _{u}\; =\; \int _{0}^{+1}dx\; \Delta u_{V}(x)\; \approx 0.918\; ,\hspace {1.cm}\eta _{d}\; =\; \int _{0}^{+1}dx\; \Delta d_{V}(x)\; \approx -0.339\; .
\end{equation}
 The first moment of the polarized sea quark density was determined by the fit
of Ref.\cite{Leader98} which gives \( \eta _{\bar{q}}\approx -0.054 \). \\
 \indent Our ansatz for \( \tilde{H}^{q} \) is to multiply the polarized quark
distributions of Eq.~(\ref{eq:poldistr}) by the axial form factor, neglecting
any \( \xi  \)-dependence and sea contribution, which yields 
\begin{equation}
\label{eq:factpol}
\tilde{H}^{u/p}(x,\xi ,t)=\Delta u_{V}(x)\, {g_{A}^{u/p}(t)}/{g_{A}^{u/p}(0)}\; ,\hspace {0.5cm}\tilde{H}^{d/p}(x,\xi ,t)=\Delta d_{V}(x)\, {g_{A}^{d/p}(t)}/{g_{A}^{d/p}(0)}\; .
\end{equation}
 One can check that Eq.~(\ref{eq:dislimit}) is verified by construction and
that, using the (\ref{eq:axff}) and the quark model relation to evaluate the
axial form factors \( g_{A}^{u/p} \) and \( g_{A}^{d/p} \), the sum rule of
Eq.~(\ref{eq:axvecsumrule}) is satisfied within 10 \%.

\subsubsection{\protect\protect\( \xi \protect \protect \)-dependent ansatz for \protect\protect\( H\protect \protect \)
and \protect\protect\( \tilde{H}\protect \protect \)}

\indent To generate the dependence on \( \xi  \), we start from the double
distributions introduced in Eq.~(\ref{eq:doubledef}). As shown in Fig.~\ref{fig:double},
the momentum flow in the double distributions is decomposed in a part along
the initial nucleon momentum \( p \) and a part along the momentum transfer
\( \Delta  \) . Therefore, a reasonable guess for the helicity independent
double distribution \( F(\tilde{x},y,t) \) was proposed in Ref.\cite{Rady98},
as a product of a quark distribution function \( q(\tilde{x}) \), describing
the momentum flow along \( p \), and an asymptotic ``meson-like'' distribution
amplitude \( \sim y(1-\tilde{x}-y) \), describing the momentum flow along \( \Delta  \).
Properly normalized, this yields : 
\begin{equation}
\label{eq:ddmodel}
F^{q}(\tilde{x},y,t)\; =\; F_{1}^{q}(t)/F_{1}^{q}(0)\; q(\tilde{x})\; 6\, {{y(1-\tilde{x}-y)}\over {(1-\tilde{x})^{3}}}\; ,
\end{equation}
 where similarly to Eq.~(\ref{eq:factunpol}), the \( t \)-dependent part is
given by the form factor \( F_{1} \). Similarly for the helicity dependent
double distribution, this ansatz yields : 
\begin{equation}
\label{eq:ddhelmodel}
\tilde{F}^{q}(\tilde{x},y,t)\; =\; g_{A}^{q}(t)/g_{A}^{q}(0)\; \Delta q(\tilde{x})\; 6\, {{y(1-\tilde{x}-y)}\over {(1-\tilde{x})^{3}}}\; ,
\end{equation}
 Starting from Eq.(\ref{eq:ddmodel}), the OFPD \( H^{q} \) is then constructed
by using Eqs.~(\ref{eq:relofpdnfpd}) and (\ref{eq:double}). Analogously, the
OFPD \( \tilde{H}^{q} \) is constructed starting from Eq.~(\ref{eq:ddhelmodel}).

\begin{figure}[ht]
\epsfxsize=15 cm
\epsfysize=21.5cm
\centerline{\epsffile{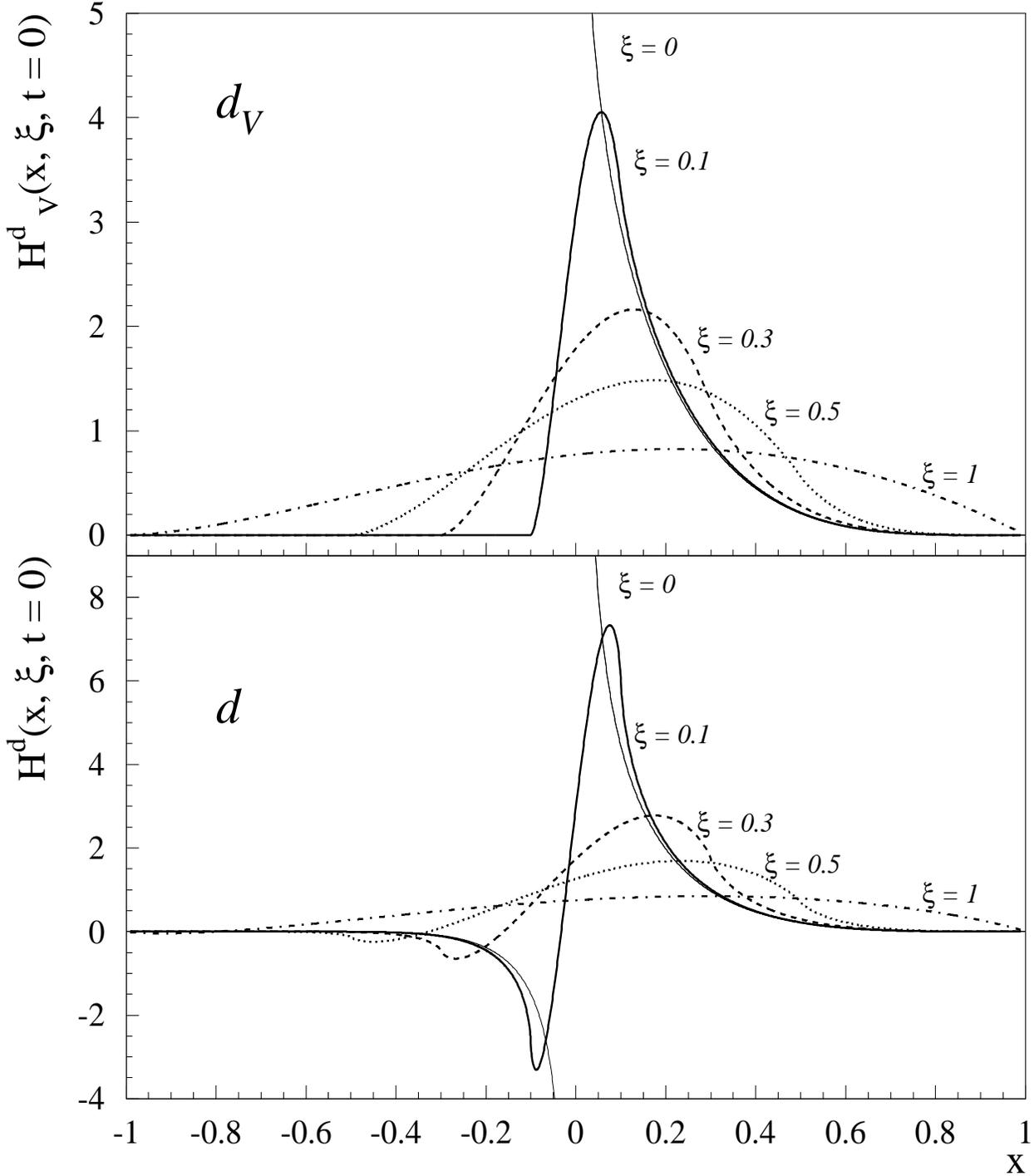}}
\vspace{-0.75cm}
\caption[]{\small $\xi$ dependence of the OFPD $H^d$ at $t = 0$ 
using the ansatz (based on the MRST98 \cite{MRST98} quark
distributions) as described in the text. 
Upper panel~: valence down quark OFPD, lower panel~: total down quark
OFPD. The thin lines ($\xi = 0$) correspond with the ordinary
$d$-quark distributions (MRST98 parametrization at $Q^2$ = 2 GeV$^2$).}
\label{fig:ofpdxi}
\end{figure}

As an example, we show on Fig.~\ref{fig:ofpdxi} the dependence on \( \xi  \)
predicted by this model for the valence \( d \)-quark and total \( d \)-quark
skewed distributions.

\subsubsection{The skewed distributions \protect\protect\( E\protect \protect \) and \protect\protect\( \tilde{E}\protect \protect \)}

\indent The modelization of \( E \) and \( \tilde{E} \), which correspond
to helicity flip amplitudes, is more difficult as we don't have the DIS constraint
for the \( x \)-dependence in the forward limit. However, as already pointed
out, \( E \) and \( \tilde{E} \) are multiplied by a momentum transfer and
therefore their contribution is suppressed at small \( t \). Nevertheless it
was argued by the authors of Ref.\cite{FPS98} that the pion exchange, which
contributes to the region \( -\xi \leq x\leq \xi  \) of \( \tilde{E} \), may
be non negligible at small \( t \) due to the proximity of the pion pole at
\( t=m_{\pi }^{2} \). We follow therefore the suggestion of Refs.\cite{FPS98,FPPS99}
and evaluate \( \tilde{E} \) assuming it is entirely due to the pion pole.
(By contrast we continue to neglect the contributions due to \( E \) since
there is no dangerous pole in this case). According to this hypothesis, since
the pion exchange is isovector, we must have 
\begin{equation}
\label{eq:etildeisov}
\tilde{E}^{u/p}=-\tilde{E}^{d/p}={1\over 2}\; \tilde{E}_{\pi -pole}\; .
\end{equation}
 The \( t \) -dependence of \( \tilde{E}_{\pi -pole}(x,\xi ,t) \) is fixed
by the sum rule (\ref{eq:axvecsumrule}) pseudoscalar form factor \( h_{A}(t) \)
: 
\begin{equation}
\label{eq:axvecsumrulepipole}
\int _{-1}^{+1}dx\; \tilde{E}^{(3)}(x,\xi ,t)\, =\, h_{A}(t)\longrightarrow {{g_{A}\; \left( 2M_{N}\right) ^{2}}\over {-t+m_{\pi }^{2}}}\; ,
\end{equation}
 where \( \tilde{E}^{(3)}\equiv \tilde{E}^{u/p}-\tilde{E}^{d/p} \) denotes
the isovector OFPD and where we have approximated the pseudoscalar form factor
by its pion pole dominance expression.

In the region \( -\xi \leq x\leq \xi  \), the quark and antiquark couple to
the pion field of the nucleon. Therefore, this coupling should be proportional
to the pion distribution amplitude. For the later we take the asymptotic form,
which is well supported now by experiment (see discussion in section \ref{sec:dvcskperp}).
Expressing the quark's longitudinal momentum fraction \( z \) in the pion in
the symmetric range \( -1\leq z\leq 1 \), the asymptotic distribution amplitude
\( \Phi _{as} \) is given by \( \Phi _{as}(z)=3/4 \) \( (1-z^{2}) \), and
is normalized as \( \int _{-1}^{+1}dz\; \Phi _{as}(z)=1 \). The light-cone
momentum fractions of the quark and antiquark in the pion which couples to
the lower blobs in Figs.~\ref{fig:handbags}, \ref{fig:factmeson}, are respectively
given by \( (1+x/\xi )/2 \) and \( (1-x/\xi )/2 \). Therefore, \( \tilde{E}_{\pi -pole} \)
is finally modelled as 
\begin{equation}
\label{eq:etildepipole}
\tilde{E}_{\pi -pole}\; =\; \Theta \left( -\xi \leq x\leq \xi \right) \; h_{A}(t)\; {1\over \xi }\; \Phi _{as}\left( {x\over \xi }\right) \; ,
\end{equation}
 which satisfies the sum rule of Eq.~(\ref{eq:axvecsumrulepipole}).

As there is no dangerous pole in the case of \( E \), it seems safe to neglect
its contribution to the amplitude at small \( t \), which of course does not
mean that \( E \) itself is small. In this respect it is amusing to note that,
if one adopts for \( E \) an ansatz similar to the one for \( H \), Eq.~(\ref{eq:factunpol}),
that is 
\begin{equation}
\label{eq:factunpol2}
E^{u/p}(x,\xi ,t)=u(x)\, {F_{2}^{u/p}(t)}/2\; ,\hspace {0.5cm}E^{d/p}(x,\xi ,t)=d(x)\, {F_{2}^{d/p}(t)}\; ,\hspace {0.5cm}E^{s/p}(x,\xi ,t)=0\; ,
\end{equation}
 and if one uses this ansatz to evaluate the total spin carried by the quarks
through the sum rule of Eq.~(\ref{eq:dvcsspin}), one get the results shown
in Table \ref{table1}. We can see that, even though there is little theoretical
basis for the ansatz of Eq.~(\ref{eq:factunpol2}), we get the reasonable result
that 43\% of the nucleon spin in carried by the quarks. This is consistent with
a recent QCD sum rule estimate \cite{JiBal97} (at a low scale) where the gluons
were found to contribution to half of the nucleon spin. 
Note that only the information
from the unpolarized parton distributions has been used to evaluate 
the quark contribution to the spin of the nucleon. \\

\indent  By using the decomposition of the total quark spin \( J_{q} \) as
in Eq.~(\ref{eq:spindecomp}) and by using the most recent values measured at
SMC \cite{SMC98} for the quark helicity distributions \( \Delta u \), \( \Delta d \)
and \( \Delta s \), one can extract the quark orbital angular momentum contribution
to the nucleon spin. Using the factorized, \( \xi  \)-independent ansatz, the
orbital contribution is estimated in Table \ref{table1} to yield about 15\%
of the nucleon spin. \\

\subsubsection{Other models of the skewed distributions}

\indent Some model calculations of the skewed distributions exist. A bag model
evaluation of the skewed distributions \cite{Ji97c} found (at a low scale)
that they depend weakly on \( \xi  \). However, a calculation in the chiral
quark soliton model \cite{Petrov98} found (also at a low scale) a strong dependence
on \( \xi  \) and exhibits in particular fast ``crossovers'' at \( |x|=\xi  \).
The chiral quark soliton model is based on the large-\( N_{c} \) picture of
the nucleon as a heavy semiclassical system whose \( N_{c} \) valence quarks
are bound by a self-consistent pion field. The crossovers in this model can
be interpreted as being due to meson-exchange type contributions in the region
\( -\xi \leq x\leq \xi  \), which are not present in the bag model calculation
and which are also not contained in the phenomenological ansatz of Eqs.~(\ref{eq:ddmodel},
\ref{eq:ddhelmodel}) for the double distributions (see Ref.\cite{Poly99} for
a detailed discussion of these meson-exchange type contributions to the skewed
quark distributions). It was noticed in Ref.\cite{Petrov98} that this crossover
behavior at \( |x|=\xi  \) could lead to an enhancement of observables, which
will be interesting to quantify. \\

\indent As a last remark, as we are mainly interested in this paper in giving
estimates for electroproduction reactions in the valence region at only moderately
large values of \( Q^{2} \) (\( 1\sim 20 \)~GeV\( ^{2} \)), which is where
experiments of exclusive electroproduction reactions can be performed currently
or in the near future, we neglect here the scale dependence of the skewed distributions.
At very high values of \( Q^{2} \), the scale dependence should be considered.
This scale dependence is described by generalized evolution equations that are
under intensive investigation in the literature.

\section{Leading order amplitudes for DVCS and hard electroproduction of mesons}

\label{sec:sec3}

\subsection{DVCS}

Factorization proofs for DVCS have been given by several authors \cite{JiOs98,Collins98}.
They give the theoretical underpinning which allows to express the leading order
DVCS amplitude in terms of OFPD's. Using the parametrization of Eq.~(\ref{eq:qsplitting}),
the leading order DVCS tensor \( H^{\mu \nu }_{L.O.\, DVCS} \) follows from
the two handbag diagrams of Fig.~\ref{fig:handbags} as 
\begin{eqnarray}
 &  & H^{\mu \nu }_{L.O.\, DVCS}\nonumber \\
 & = & {1\over 2}\, \left[ \tilde{p}^{\mu }n^{\nu }+\tilde{p}^{\nu }n^{\mu }-g^{\mu \nu }\right] \; \int _{-1}^{+1}dx\left[ {1\over {x-\xi +i\epsilon }}+{1\over {x+\xi -i\epsilon }}\right] \nonumber \\
 &  & \hspace {.5cm}\times \left[ H^{p}_{DVCS}(x,\xi ,t)\; \bar{N}(p^{'})\gamma .nN(p)+\; E^{p}_{DVCS}(x,\xi ,t)\; \bar{N}(p^{'})i\sigma ^{\kappa \lambda }{{n_{\kappa }\Delta _{\lambda }}\over {2m_{N}}}N(p)\right] \nonumber \\
 & + & \; {1\over 2}\, \left[ -i\varepsilon ^{\mu \nu \kappa \lambda }\tilde{p}_{\kappa }n_{\lambda }\right] \; \int _{-1}^{+1}dx\left[ {1\over {x-\xi +i\epsilon }}-{1\over {x+\xi -i\epsilon }}\right] \nonumber \\
 &  & \hspace {.5cm}\times \left[ \tilde{H}^{p}_{DVCS}(x,\xi ,t)\bar{N}(p^{'})\gamma .n\gamma _{5}N(p)+\tilde{E}^{p}_{DVCS}(x,\xi ,t)\bar{N}(p^{'})\gamma _{5}{{\Delta \cdot n}\over {2m_{N}}}N(p)\right] \; ,\label{eq:dvcsampl} 
\end{eqnarray}
 with \( \varepsilon _{0123}=+1 \). We refer to Ref.\cite{vcsrev} for the
formalism to calculate DVCS observables starting from the DVCS tensor of Eq.~(\ref{eq:dvcsampl}).
\\
 \indent In the DVCS on the proton, the OFPD's enter in the combination 
\begin{equation}
H^{p}_{DVCS}(x,\xi ,t)\, =\, {4\over 9}H^{u/p}\, +\, {1\over 9}H^{d/p}\, +\, {1\over 9}H^{s/p}\; ,
\end{equation}
 and similarly for \( \tilde{H} \), \( E \) and \( \tilde{E} \). \\
 \indent For the \( \pi ^{0} \) pole contribution to the DVCS amplitude, which
contributes to the function \( \tilde{E}^{p}_{DVCS} \), the convolution integral
in Eq.~(\ref{eq:dvcsampl}) can be worked out analytically. By using Eqs.~(\ref{eq:etildeisov},
\ref{eq:etildepipole}) one obtains : 
\begin{equation}
\int _{-1}^{+1}dx\left[ {1\over {x-\xi +i\epsilon }}-{1\over {x+\xi -i\epsilon }}\right] \; \tilde{E}^{p}_{DVCS}(x,\xi ,t)\; =\; -\, {1\over {2\xi }}\; h_{A}(t)\; .
\end{equation}
\\
 \indent The leading order DVCS amplitude of Eq.~(\ref{eq:dvcsampl}), is exactly
gauge invariant with respect to the virtual photon, i.e. \( q_{\nu }\, H^{\mu \nu }_{L.O.\, DVCS}=0 \).
However, electromagnetic gauge invariance is violated by the real photon except
in the forward direction. In fact \( q^{'}_{\mu }\, H^{\mu \nu }_{L.O.DVCS}\sim \Delta _{\perp } \).
This violation of gauge invariance is a higher twist effect of order \( 1/Q^{2} \)
compared to the leading order term \( H^{\mu \nu }_{L.O.\, DVCS} \). So in
the limit \( Q^{2}\rightarrow \infty  \) it is innocuous but for actual experiments
it matters. Actually for any cross section estimate one needs to choose a gauge
and this explicit gauge dependence for nonzero angles is unpleasant. In the
absence of a dynamical gauge invariant higher twist calculation of the DVCS
amplitude, we propose to restore gauge invariance in a heuristic way based on
physical considerations. We propose to introduce the gauge invariant tensor
\( H^{\mu \nu }_{DVCS} \) : 
\begin{equation}
\label{eq:dvcsgaugeinv1}
H^{\mu \nu }_{DVCS}\, =\, H^{\mu \nu }_{L.O.\, DVCS}\, -\, {{a^{\mu }}\over {\left( a\cdot q^{'}\right) }}\; \left( q^{'}_{\lambda }\, H^{\lambda \nu }_{L.O.\, DVCS}\right) \; ,
\end{equation}
 where \( a^{\mu } \) is a four-vector specified below. Obviously \( H^{\mu \nu }_{DVCS} \)
respects gauge invariance for both the virtual and the real photon : 
\begin{equation}
q_{\nu }\, H^{\mu \nu }_{DVCS}=0\; ,\hspace {1cm}q^{'}_{\mu }\, H^{\mu \nu }_{DVCS}=0\; .
\end{equation}
 Furthermore, as 
\begin{equation}
q^{'}_{\lambda }\, H^{\lambda \nu }_{L.O.\, DVCS}=-\left( \Delta _{\perp }\right) _{\lambda }H^{\lambda \nu }_{L.O.\, DVCS}\; ,
\end{equation}
 the gauge restoring term gives zero in the forward direction (\( \Delta _{\perp }=0 \))
which is natural. We choose \( a^{\mu }=\tilde{p}^{\mu } \) because \( \tilde{p}\cdot q^{'} \)
is of order \( Q^{2} \), which gives automatically a gauge restoring term of
order \( O\left( 1/Q^{2}\right)  \). This choice for \( a^{\mu } \) is furthermore
motivated by the fact that in the derivation of the leading order amplitude
of Eq.~(\ref{eq:dvcsampl}), only the \( \tilde{p}^{\mu } \) components at
the electromagnetic vertices are retained \cite{Ji97}. The above arguments
lead to the following DVCS amplitude: 
\begin{equation}
\label{eq:dvcsgaugeinv}
H^{\mu \nu }_{DVCS}\, =\, H^{\mu \nu }_{L.O.\, DVCS}\, +\, {{\tilde{p}^{\mu }}\over {\left( \tilde{p}\cdot q^{'}\right) }}\; \left( \Delta _{\perp }\right) _{\lambda }\, H^{\lambda \nu }_{L.O.\, DVCS}\; ,
\end{equation}
 which is now gauge invariant with respect to both the virtual and real photons.
We will illustrate the influence of this gauge invariance prescription, by comparing
DVCS observables calculated with Eq.~(\ref{eq:dvcsgaugeinv}) and with the leading
order formula of Eq.~(\ref{eq:dvcsampl}).

\subsection{Hard electroproduction of mesons}

\begin{figure}[ht]
\epsfxsize=15 cm
\epsfysize=20. cm
\centerline{\epsffile{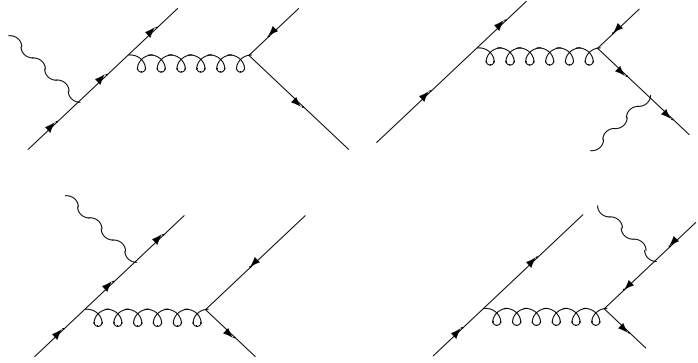}}
\vspace{-11cm}
\caption[]{\small Leading order diagrams to hard meson electroproduction.}
\label{fig:4diaghard}
\end{figure}

The OFPD's reflect the structure of the nucleon independently of the reaction
which probes the nucleon. In this sense, they are universal quantities and can
also be accessed, in different flavor combinations, through the hard exclusive
electroproduction of mesons - \( \pi ^{0,\pm },\eta ,...,\rho ^{0,\pm },\omega ,\phi ,... \)
- (see Fig.~\ref{fig:factmeson}) for which a QCD factorization proof was given
in Refs.~\cite{Rady96b,Collins97}. According to Ref.~\cite{Collins97}, the
factorization applies when the virtual photon is longitudinally polarized because
in this case, the end-point contributions in the meson wave function are power
suppressed. Furthermore, it was shown that the cross section for a transversely
polarized photon is suppressed by 1/\( Q^{2} \) compared to a longitudinally
polarized photon.\\
 Collins et al. \cite{Collins97} also showed that leading order PQCD predicts
that the longitudinally polarized vector meson channels (\( \rho ^{0,\pm }_{L} \),
\( \omega _{L} \), \( \phi _{L} \)) are sensitive only to the unpolarized
OFPD's (\( H \) and \( E \)) whereas the pseudo-scalar channels (\( \pi ^{0,\pm },\eta ,... \))
are sensitive only to the polarized OFPD's (\( \tilde{H} \) and \( \tilde{E} \)).
It was shown in Ref.\cite{Diehl99} that the leading twist contribution to exclusive
electroproduction of transversely polarized vectors mesons vanishes at all orders
in perturbation theory. In comparison to meson electroproduction reactions,
we recall that DVCS depends at the same time on \textit{both} the unpolarized
(\( H \) and \( E \)) and polarized (\( \tilde{H} \) and \( \tilde{E} \))
OFPD's. \\
 \indent According to the above discussion, we give predictions only for the
meson electroproduction cross section by a longitudinal virtual photon. The
longitudinal \( \gamma ^{*}_{L}+p\rightarrow M+p \) two-body cross section
\( d\sigma _{L}/dt \) is 
\begin{equation}
\label{eq:mescross}
{{d\sigma _{L}}\over {dt}}={1\over {16\pi \left( s-m_{N}^{2}\right) \Lambda (s,-Q^{2},m_{N}^{2})}}\; {1\over 2}\sum _{h_{N}}\sum _{h^{'}_{N}}|{\mathcal{M}}^{L}(\lambda _{M}=0,h^{'}_{N};h_{N})|^{2}\; ,
\end{equation}
 where \( h_{N} \), \( h^{'}_{N} \) are the initial and final nucleon helicities
and where the standard kinematic function \( \Lambda (x,y,z) \) is defined
by 
\begin{equation}
\Lambda (x,y,z)=\sqrt{x^{2}+y^{2}+z^{2}-2xy-2xz-2yz}\; .
\end{equation}
 which gives \( \Lambda (s,-Q^{2},m_{N}^{2})=2m_{N}|\vec{q}_{L}| \), where
\( |\vec{q}_{L}| \) is the virtual photon momentum in the lab system. As the
way to extract \( d\sigma _{L}/dt \) from the fivefold electroproduction cross
section is a matter of convention, all results for \( d\sigma _{L}/dt \) in
this paper are given with the choice of the flux factor of Eq.~(\ref{eq:mescross}).
\\
 \indent In Eq.~(\ref{eq:mescross}), \( {\mathcal{M}}^{L} \) is the amplitude
for the production of a meson with \( \lambda _{M}=0 \) by a longitudinal photon.
In the valence region, the leading order amplitude is given by the hard scattering
diagrams of Fig.~\ref{fig:4diaghard} (\( T_{H} \) part in Fig.~\ref{fig:factmeson}).
The amplitudes \( {\mathcal{M}}^{L} \) for \( \rho ^{0}_{L} \) and \( \pi ^{0} \)
electroproduction were calculated in Refs.~\cite{ourprl} (see also Ref.~\cite{Mank98})
for which the following gauge invariant expressions were found : 
\begin{eqnarray}
 &  & {\mathcal{M}}^{L}_{\rho ^{0}_{L}}=\; -ie\, {4\over 9}\, {1\over {Q}}\; \left[ \, \int _{0}^{1}dz{{\Phi _{\rho }(z)}\over z}\right] \, {1\over 2}\, \int _{-1}^{+1}dx\left[ {1\over {x-\xi +i\epsilon }}+{1\over {x+\xi -i\epsilon }}\right] \nonumber \\
 &  & \hspace {2.cm}\times (4\pi \alpha _{s})\; \left\{ H^{p}_{\rho ^{0}_{L}}(x,\xi ,t)\bar{N}(p^{'})\gamma .nN(p)+E^{p}_{\rho ^{0}_{L}}(x,\xi ,t)\bar{N}(p^{'})i\sigma ^{\kappa \lambda }{{n_{\kappa }\Delta _{\lambda }}\over {2m_{N}}}N(p)\right\} ,\label{eq:rhoampl} 
\end{eqnarray}
\begin{eqnarray}
 &  & {\mathcal{M}}^{L}_{\pi ^{0}}=\; -ie\, {4\over 9}\, {1\over {Q}}\; \left[ \int _{0}^{1}dz{{\Phi _{\pi }(z)}\over z}\right] \, {1\over 2}\, \int _{-1}^{+1}dx\left[ {1\over {x-\xi +i\epsilon }}+{1\over {x+\xi -i\epsilon }}\right] \nonumber \\
 &  & \hspace {2.cm}\times (4\pi \alpha _{s})\; \left\{ \tilde{H}^{p}_{\pi ^{0}}(x,\xi ,t)\bar{N}(p^{'})\gamma .n\gamma _{5}N(p)+\tilde{E}^{p}_{\pi ^{0}}(x,\xi ,t)\bar{N}(p^{'})\gamma _{5}{{\Delta \cdot n}\over {2m_{N}}}N(p)\right\} ,\label{eq:piampl} 
\end{eqnarray}
 where \( \Phi _{\rho }(z) \) and \( \Phi _{\pi }(z) \) are the \( \rho  \)
and \( \pi  \) distribution amplitudes (DA) respectively. The factor 4/9 is
a color factor corresponding with one-gluon exchange. From Eqs.~(\ref{eq:rhoampl},
\ref{eq:piampl}), one sees that the leading order longitudinal amplitude for
meson electroproduction behaves as \( 1/Q \). As the phase space factor in
the cross section Eq.~(\ref{eq:mescross}) behaves as \( 1/Q^{4} \) at fixed
\( x_{B} \) and fixed \( t \), this leads to a \( 1/Q^{6} \) behavior of
\( d\sigma _{L}/dt \) at large \( Q^{2} \) and at fixed \( x_{B} \) and fixed
\( t \). \\
 \indent According to the considered reaction, the proton OFPD's enter in different
combinations due to the charges and isospin factors. For the electroproduction
of \( \rho ^{0} \), which corresponds to the quark state (with spin \( S=1 \))
\begin{equation}
\label{eq:rhowf}
|\rho ^{0}\rangle \; =\; {1\over \sqrt{2}}\left\{ |u\bar{u}\rangle \; -\; |d\bar{d}\rangle \right\} \; ,
\end{equation}
 the corresponding skewed distribution is given by: 
\begin{equation}
H^{p}_{\rho ^{0}_{L}}(x,\xi ,t)\; =\; {1\over {\sqrt{2}}}\left\{ {2\over 3}H^{u/p}\, +\, {1\over 3}H^{d/p}\right\} \; .
\end{equation}
 For the longitudinal electroproduction of the \( \omega _{L} \) and \( \phi _{L} \)
vector mesons, the amplitudes have an expression analogous to Eq.~(\ref{eq:rhoampl})
in terms of \( H^{p}_{\omega _{L}} \) and \( H^{p}_{\phi _{L}} \). By considering
the \( \omega  \) and \( \phi  \) mesons as pure quark states (with \( S=1 \))
\begin{equation}
\label{eq:omphiwavef}
|\omega \rangle \; =\; {1\over \sqrt{2}}\left\{ |u\bar{u}\rangle \; +\; |d\bar{d}\rangle \right\} \; ,\hspace {1cm}|\phi \rangle \; =\; |s\bar{s}\rangle \; ,
\end{equation}
 one then obtains for the corresponding OFPD's : 
\begin{eqnarray}
H^{p}_{\omega _{L}}(x,\xi ,t)\;  & = & \; {1\over {\sqrt{2}}}\left\{ {2\over 3}H^{u/p}\, -\, {1\over 3}H^{d/p}\right\} \; ,\\
H^{p}_{\phi _{L}}(x,\xi ,t)\;  & = & \; -\; {1\over 3}H^{s/p}\; .
\end{eqnarray}
 Besides the longitudinal electroproduction amplitude Eq.~(\ref{eq:piampl})
for \( \pi ^{0} \), one again has analogous expressions for other neutral pseudoscalar
mesons such as the \( \eta  \). Starting from the quark states (with spin S
= 0) for pseudoscalar mesons and neglecting any mixing for the \( \eta  \)~:
\begin{eqnarray}
|\pi ^{0}\rangle \;  & = & \; {1\over \sqrt{2}}\left\{ |u\bar{u}\rangle \; -\; |d\bar{d}\rangle \right\} \; ,\label{eq:piostate} \\
|\eta \rangle \;  & = & \; {1\over \sqrt{6}}\left\{ |u\bar{u}\rangle \; +\; |d\bar{d}\rangle \; -\; 2\; |s\bar{s}\rangle \right\} \; ,\label{eq:etastate} 
\end{eqnarray}
 one obtains for the polarized OFPD's : 
\begin{eqnarray}
\tilde{H}^{p}_{\pi ^{0}}(x,\xi ,t)\;  & = & \; {1\over {\sqrt{2}}}\left\{ {2\over 3}\tilde{H}^{u/p}\, +\, {1\over 3}\tilde{H}^{d/p}\right\} \; ,\\
\tilde{H}^{p}_{\eta }(x,\xi ,t)\;  & = & \; {1\over {\sqrt{6}}}\left\{ {2\over 3}\tilde{H}^{u/p}\, -\, {1\over 3}\tilde{H}^{d/p}\, +\, {2\over 3}\tilde{H}^{s/p}\right\} \; .
\end{eqnarray}
\\
 \indent For the charged meson channels \( \rho ^{\pm } \) and \( \pi ^{\pm } \),
we obtain from \( |\rho ^{+}>=-|u\bar{d}> \) and \( |\rho ^{-}>=|d\bar{u}> \)
for the corresponding longitudinal electroproduction amplitudes (which have
been given previously for the \( \rho ^{\pm } \) in Ref.\cite{Mank99a} and
for the \( \pi ^{\pm } \) in Refs.\cite{Mank99b,FPPS99}) : 
\begin{eqnarray}
 &  & {\mathcal{M}}^{L}_{\rho ^{\pm }}=\; -ie\, {4\over 9}\, {1\over {Q}}\; \left[ \, \int _{0}^{1}dz{{\Phi _{\rho }(z)}\over z}\right] \, \nonumber \\
 &  & \hspace {1.5cm}\times {1\over 2}\, \int _{-1}^{+1}dx\left[ \left\{ \begin{array}{c}
-2/3\\
-1/3
\end{array}\right\} {1\over {x-\xi +i\epsilon }}+\left\{ \begin{array}{c}
1/3\\
2/3
\end{array}\right\} {1\over {x+\xi -i\epsilon }}\right] \nonumber \\
 &  & \hspace {1.5cm}\times (4\pi \alpha _{s})\; \left\{ H^{(3)}(x,\xi ,t)\bar{N}(p^{'})\gamma .nN(p)+E^{(3)}(x,\xi ,t)\bar{N}(p^{'})i\sigma ^{\kappa \lambda }{{n_{\kappa }\Delta _{\lambda }}\over {2m_{N}}}N(p)\right\} \; ,\label{eq:rhopmampl} 
\end{eqnarray}
\begin{eqnarray}
 &  & {\mathcal{M}}^{L}_{\pi ^{\pm }}=\; -ie\, {4\over 9}\, {1\over {Q}}\; \left[ \int _{0}^{1}dz{{\Phi _{\pi }(z)}\over z}\right] \, \nonumber \\
 &  & \hspace {1.5cm}\times {1\over 2}\, \int _{-1}^{+1}dx\left[ \left\{ \begin{array}{c}
-2/3\\
-1/3
\end{array}\right\} {1\over {x-\xi +i\epsilon }}+\left\{ \begin{array}{c}
1/3\\
2/3
\end{array}\right\} {1\over {x+\xi -i\epsilon }}\right] \nonumber \\
 &  & \hspace {1.5cm}\times (4\pi \alpha _{s})\; \left\{ \tilde{H}^{(3)}(x,\xi ,t)\bar{N}(p^{'})\gamma .n\gamma _{5}N(p)+\tilde{E}^{(3)}(x,\xi ,t)\bar{N}(p^{'})\gamma _{5}{{\Delta \cdot n}\over {2m_{N}}}N(p)\right\} \; ,\label{eq:pipmampl} 
\end{eqnarray}
 where the \( u \)- and \( d \)-quark charges appear in front of the direct
and crossed terms and where the following isovector forms for the corresponding
OFPD's enter : 
\begin{eqnarray}
H^{(3)}(x,\xi ,t)\;  & = & \; H^{u/p}\, -\, H^{d/p}\; ,\\
\tilde{H}^{(3)}(x,\xi ,t)\;  & = & \; \tilde{H}^{u/p}\, -\, \tilde{H}^{d/p}\; .
\end{eqnarray}
 In all of the above we have only given the expressions for the OFPD's \( H \)
and \( \tilde{H} \), but exactly the same isospin relations are valid for the
OFPD's \( E \) and \( \tilde{E} \). In particular, the isovector OFPD \( \tilde{E}^{(3)} \)
provides a prominent contribution to the charged pion electroproduction amplitude
because it contains the \( t \)-channel pion pole contribution, given by Eq.~(\ref{eq:etildepipole}).
As seen before for the \( \pi ^{0} \) pole contribution in case of the DVCS,
the convolution integral for the charged pion pole contribution in Eq.~(\ref{eq:pipmampl})
can also be worked out analytically. In the case of the \( \pi ^{+} \) electroproduction
amplitude, one has~: 
\begin{equation}
\label{eq:pipmpipole}
\int _{-1}^{+1}dx\left[ {{-e_{u}}\over {x-\xi +i\epsilon }}+{{-e_{d}}\over {x+\xi -i\epsilon }}\right] \; \tilde{E}_{\pi -pole}(x,\xi ,t)\; =\; {3\over {2\xi }}\, \left( e_{u}-e_{d}\right) \; h_{A}(t)\; .
\end{equation}
\\
\indent In the amplitudes for longitudinal meson electroproduction Eqs.(\ref{eq:rhopmampl},
\ref{eq:pipmampl}), the meson distribution amplitudes \( \Phi  \) enter. For
the pion, recent data \cite{Gro98} for the \( \pi ^{0}\gamma ^{*}\gamma  \)
transition form factor up to \( Q^{2} \) = 9 GeV\( ^{2} \), which will be
briefly discussed in section \ref{sec:dvcskperp}, support the asymptotic form
: 
\begin{equation}
\label{eq:piondistr}
\Phi _{\pi }(z)=\sqrt{2}f_{\pi }\; 6z(1-z)\; ,
\end{equation}
 with \( f_{\pi } \) = 0.0924 GeV from the pion weak decay. With this asymptotic
DA for the pion, the charged pion pole contribution to the amplitude \( {\mathcal{M}}^{L}_{\pi ^{\pm }} \)
can be worked out by using Eq.~(\ref{eq:pipmpipole}), where we insert the pion
pole formula of Eq.~(\ref{eq:axvecsumrulepipole}) for the induced pseudoscalar
FF \( h_{A}(t) \). By using the PCAC relation \( g_{A}/f_{\pi }=g_{\pi NN}/m_{N} \),
where \( g_{\pi NN} \) is the \( \pi NN \) coupling constant, we finally obtain
for the pion pole part of the amplitude \( {\mathcal{M}}^{L}_{\pi ^{\pm }} \)~:
\begin{equation}
{\mathcal{M}}^{L}_{\pi ^{+}}\left( \pi ^{+}-{\mathrm{p}ole}\right) \;
=\; ie\, \sqrt{2}\, Q\, F_{\pi }\left( Q^{2}\right) \; {{g_{\pi
      NN}}\over {-t+m_{\pi }^{2}}}\; \bar{N}(p^{'})\gamma _{5}N(p)\;,
\label{eq:lopipole}
\end{equation}
where \( F_{\pi} \) represents the pion electromagnetic FF. 
The leading order pion pole amplitude is obtained by using in 
Eq.~(\ref{eq:lopipole}) the asymptotic pion FF \( F_{\pi}^{as} \), 
which is given by (see also section \ref{sec:kperpmeson})~: 
\begin{equation}
F_{\pi }^{as}\left( Q^{2}\right) \; =\; {{16\pi \alpha _{s}\, f_{\pi }^{2}}\over {Q^{2}}}\; .
\end{equation}
\\
\indent For the \( \eta  \), the transition form factor has also been measured
in Ref.\cite{Gro98}, which also supports an asymptotic shape for the \( \eta  \)
distribution amplitude : 
\begin{equation}
\label{eq:etadistr}
\Phi _{\eta }(z)=f_{\eta }\; 6z(1-z)\; .
\end{equation}
 For the normalization, we adopt the value used in Ref.\cite{Gro98}, which
in the notation of Eq.~(\ref{eq:etadistr}) is given by : \( f_{\eta }\approx  \)
0.138 GeV. \\
 \indent For the vector mesons, no experimental determination of the DA exists
besides its normalization. Both recent updated QCD sum rule analyses \cite{Bal96,Bakulev98}
and a calculation in the instanton model of the QCD vacuum \cite{Poly98} favor
a DA for the longitudinally polarized \( \rho  \) meson that is rather close
to its asymptotic form. The calculations in both models differ however in the
deviations from the asymptotic form. In all calculations for vector meson electroproduction
shown below, we will also use the asymptotic DA for the vector mesons \footnote{
Because we want to use the same convention for the DA (Eq.~(\ref{eq:vectormda}))
for all vector mesons, we changed our definition of \( f_{\rho } \) compared
to our earlier work, i.e. \( f_{\rho } \) of this work corresponds with \( \sqrt{2}f_{\rho } \)
used in Refs.\cite{ourprl,vcsrev}. 
}~: 
\begin{equation}
\label{eq:vectormda}
\Phi _{V}(z)=f_{V}\; 6z(1-z)\; ,
\end{equation}
 with \( f_{\rho }\approx  \) 0.216 GeV, \( f_{\omega }\approx  \) 0.195 GeV
and \( f_{\phi }\approx  \) 0.237 GeV determined from the electromagnetic decay
\( V\rightarrow e^{+}e^{-} \). \\
 \indent When evaluating the leading order meson electroproduction amplitudes
at relatively low scales, we use the value of the strong coupling which was
also used in the QCD sum rule analysis of the vector meson distribution amplitudes
of Ref.\cite{Bal96} : \( \alpha _{s}(\mu  \) = 1 GeV) \( \approx  \) 0.56.
At high values of \( Q^{2} \), the running of the coupling has to be considered.
However, the average virtuality of the exchanged gluon in the leading order
meson electroproduction amplitudes can be considerably less than the external
\( Q^{2} \), which is therefore not the ``optimal'' choice for the renormalization
scale. In the next section, we will study the inclusion of the transverse momentum
dependence in the considered hard scattering processes and will then be able
to adapt the renormalization scale to the average gluon virtuality. The corresponding
strong coupling constant will then be calculated within the convolution integral,
through an IR finite expression, which reduces to the standard running coupling
at larger scales.

\section{Corrections to leading order amplitude : intrinsic transverse momentum dependence
and soft overlap contribution}

\label{sec:sec4}

A systematic study of higher twist corrections to hard exclusive processes is
beyond the scope of the present work. Our limited goal is to model here two
important mechanisms which give rise to power corrections to the leading amplitude.

First, we will consider the intrinsic (or primordial) transverse momentum dependence
of hard exclusive processes. When one neglects the intrinsic transverse momentum
of the active quark, the hard exclusive electroproduction amplitudes can be
written as a one-dimensional convolution of a hard scattering operator and a
non-perturbative soft quantity which depends \textit{\emph{only}} on the longitudinal
momentum fractions of the quark. This neglect of the parton intrinsic transverse
momentum is exact only up to corrections of order \( 1/Q^{2} \).

Second, for the meson electroproduction reactions, which contain a one-gluon
exchange, soft `overlap' mechanisms can compete with the leading order amplitude
at lower \( Q^{2} \) values. In the case of meson form factors (FF) both corrections
have been studied in details by several groups. It turns out that including
these corrections allows a quantitative interpretation of the available data
for the FF at \( Q^{2} \) values down to a few GeV\( ^{2} \). We will briefly
review these corrections for the \( \pi ^{0}\gamma ^{*}\gamma  \) transition
FF and the pion electromagnetic FF for \( Q^{2} \) values in the range \( 1\sim 20 \)
GeV\( ^{2} \). Our aim is to introduce the notations and to set the stage to
calculate these corrections for the hard electroproduction. We will see that
the DVCS is analogous to the \( \pi ^{0}\gamma ^{*}\gamma  \) FF because the
leading order diagrams involve no gluon exchange. The leading order diagrams
for meson electroproduction, which proceed through a one-gluon exchange mechanism,
are analogous to those for the pion electromagnetic FF.

\subsection{Intrinsic transverse momentum dependence in \protect\protect\( \pi ^{0}\gamma ^{*}\gamma \protect \protect \)
FF and in DVCS}

\label{sec:dvcskperp}

\begin{figure}[ht]
\epsfxsize=16 cm
\epsfysize=22.cm
\centerline{\epsffile{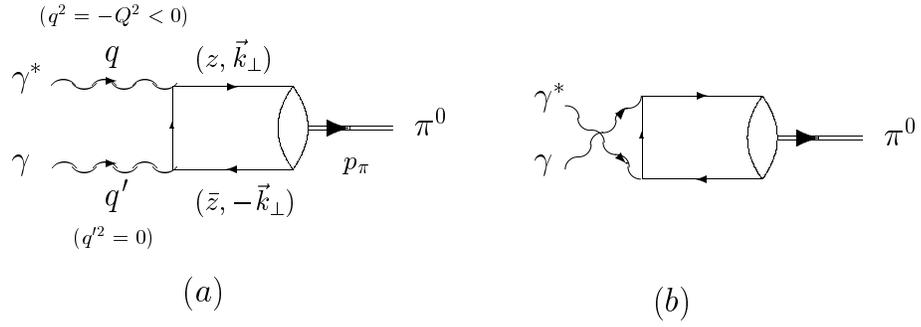}}
\vspace{-8.5cm}
\caption[]{\small Leading order direct (a) and crossed (b) diagrams to the 
$\pi^0 \gamma^* \gamma$ transition form factor.}
\label{fig:piogaga}
\end{figure}

\begin{figure}[ht]
\epsfxsize=15 cm
\epsfysize=20.cm
\centerline{\epsffile{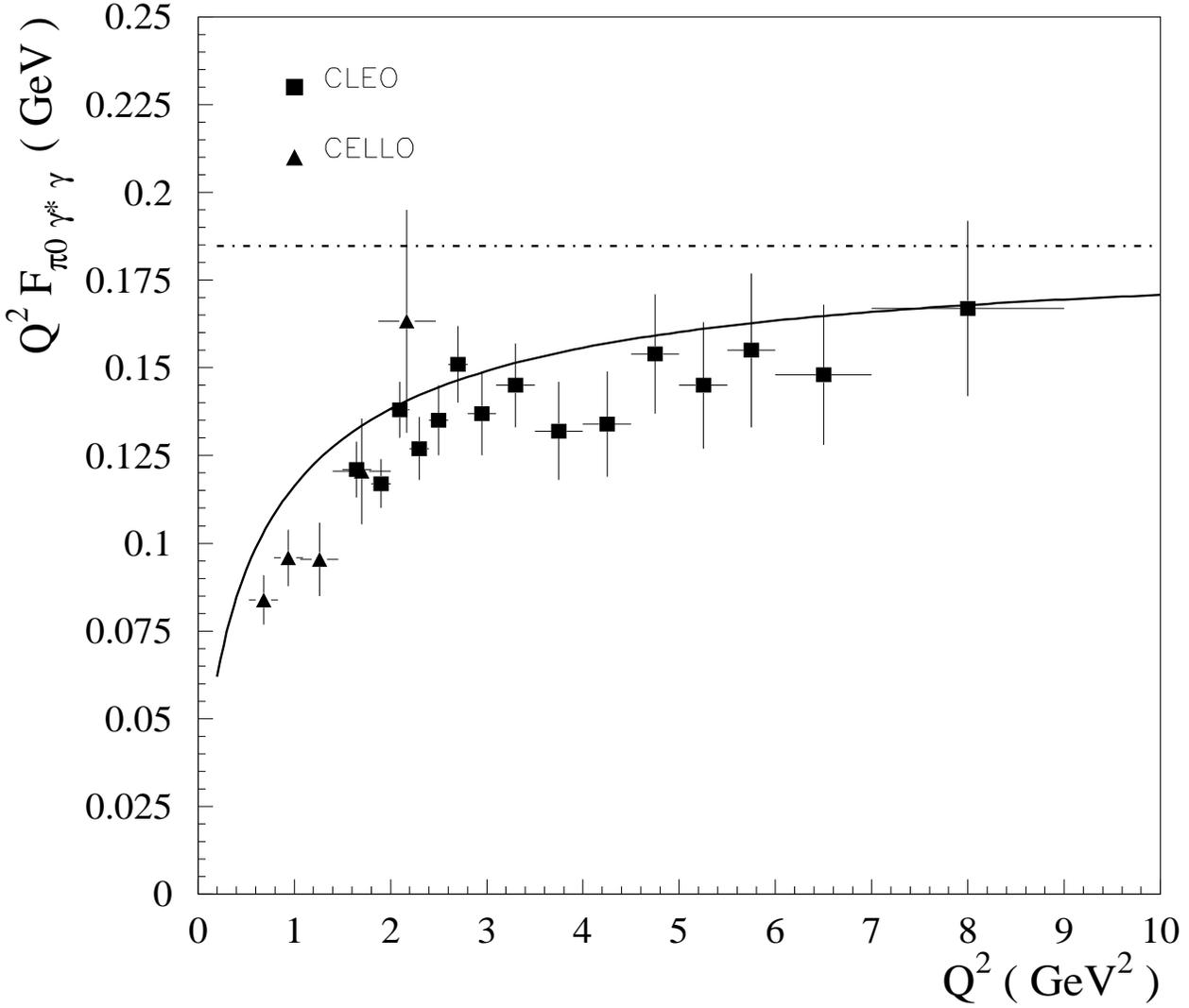}}
\vspace{-2cm}
\caption[]{\small Results for the $\pi^0 \gamma^* \gamma$ transition
  form factor of the leading order PQCD prediction 
(dashed-dotted line) compared with the prediction including 
transverse momentum dependence (full line). 
Data are from CELLO \cite{CELLO} and CLEO \cite{Gro98}.}
\label{fig:pioff}
\end{figure}

The calculation of the \( \pi ^{0}\gamma ^{*}\gamma  \) transition FF will
serve as our starting point to calculate the corrections due to intrinsic transverse
momentum dependence for the case of DVCS. The leading order contribution to
the \( \pi ^{0}\gamma ^{*}\gamma  \) transition FF is given by the diagrams
of Fig.~\ref{fig:piogaga} which involve a one-dimensional convolution integral
over the quark's longitudinal momentum fraction \( z \). The corrections to
the leading order amplitude have been estimated by various groups (see Ref.\cite{Musatov97}
for a recent comparative discussion). We start here from the method used in
in Refs.\cite{Kroll96} which is based on the modified factorization approach
of Refs.\cite{Sterman92} and which we will extend to the case of hard electroproduction
reactions. \\
 \indent In this modified factorization approach, the amplitude corresponding
to the two diagrams of Fig.~\ref{fig:piogaga} is expressed as a three-fold
convolution integral over the quark's longitudinal momentum fraction \( z \)
and its relative transverse momentum \( \vec{k}_{\perp } \)of the hard scattering
operator \( T_{H} \) and the pion light-cone wavefunction \( \Psi _{\pi }(z,\vec{k}_{\perp }) \)
: 
\begin{equation}
\label{eq:piogaga}
F_{\pi ^{0}\gamma ^{*}\gamma }(Q^{2})\; =\; \int _{0}^{1}\, dz\, \int \, {{d^{2}\vec{k}_{\perp }}\over {16\pi ^{3}}}\; \Psi _{\pi }(z,\vec{k}_{\perp })\; T_{H}(z,\vec{k}_{\perp };Q^{2})\; .
\end{equation}
 The pion wavefunction \( \Psi _{\pi }(z,\vec{k}_{\perp }) \) represents the
amplitude to find a pion in a valence \( q\bar{q} \) state, where one quark
has longitudinal momentum fraction \( z \) and relative transverse momentum
\( \vec{k}_{\perp } \).The pion light-cone wavefunction is defined by the following
bilocal quark matrix element at equal light-cone time \( (y^{+}=0) \) : 
\begin{equation}
\label{eq:wfmekperp}
\Psi _{\pi }(z,\vec{k}_{\perp })\; =\; {1\over {\sqrt{6}}}\; \int dy^{-}\, e^{i\left( zp_{\pi }^{+}\right) y^{-}}\, \int d^{2}\vec{y}_{\perp }\, e^{-i\vec{k}_{\perp }\cdot \vec{y}_{\perp }}\; \langle 0|\bar{d}(0)\, \gamma ^{+}\gamma ^{5}\, u(y)|\pi ^{+}\left( p_{\pi }\right) \rangle {\Bigg |}_{y^{+}=0}\; ,
\end{equation}
 where we have written the isospin structure for a positively charged pion moving
with large momentum \( p_{\pi } \) and where the normalization factor \( 1/\sqrt{6}=1/\sqrt{2N_{c}} \)
corresponds to the Brodsky-Lepage convention \cite{BrodLep}. The hard scattering
operator \( T_{H} \) in Eq.~(\ref{eq:piogaga}) is calculated from the diagrams
of Fig.~\ref{fig:piogaga} by keeping the \( \vec{k}_{\perp } \) dependence
in the intermediate quark propagators. Its evaluation yields~: 
\begin{equation}
\label{eq:thkperp}
T_{H}(z,\vec{k}_{\perp };Q^{2})\; =\; \left( {1\over {3\, \sqrt{2}}}\right) \; 2\sqrt{6}\; \left\{ {1\over {zQ^{2}\, +\, \vec{k}_{\perp }^{2}}}\; +\; {1\over {\bar{z}Q^{2}\, +\, \vec{k}_{\perp }^{2}}}\right\} \; ,
\end{equation}
 where the factor \( 1/(3\sqrt{2}) \) is due to the squared quark charges in
the \( \pi ^{0} \) state (Eq.~(\ref{eq:piostate})). When the \( \vec{k}_{\perp } \)
dependence is neglected in \( T_{H} \) the transverse momentum integral in
Eq.~(\ref{eq:piogaga}) acts only on \( \Psi _{\pi }(z,\vec{k}_{\perp }) \)
and one finds back the usual expression for the \( \pi ^{0}\gamma ^{*}\gamma  \)
FF in terms of the pion distribution amplitude \( \Phi _{\pi }(z) \), which
depends only on \( z \) : 
\begin{equation}
\label{eq:piogaga2}
F_{\pi ^{0}\gamma ^{*}\gamma }(Q^{2})\; \stackrel{Q^{2}\rightarrow \infty }{\longrightarrow }\; \int _{0}^{1}\, dz\; {{\Phi _{\pi }(z)}\over {2\sqrt{6}}}\; T_{H}(z;Q^{2})\; .
\end{equation}
 The pion distribution amplitude \( \Phi _{\pi } \) is defined as 
\begin{equation}
\label{eq:normpsi}
\int ^{\mu _{F}}{{d^{2}\vec{k}_{\perp }}\over {16\pi ^{3}}}\; \Psi _{\pi }(z,\vec{k}_{\perp })\; =\; {1\over {2\sqrt{6}}}\; \Phi _{\pi }(z,\mu _{F})\; ,
\end{equation}
 where we have indicated explicitely the factorization scale \( \mu _{F} \).
For \( \mu _{F} \) much larger than the average value of the quark transverse
momentum, \( \Phi _{\pi } \) depends only weakly on \( \mu _{F} \) and we
will neglect this dependence in the following. Using the asymptotic DA Eq.~(\ref{eq:piondistr})
for \( \Phi _{\pi } \), one obtains from Eq.~(\ref{eq:piogaga2}) the well
known leading order PQCD result: 
\begin{equation}
\label{eq:piogagaasy}
F_{\pi ^{0}\gamma ^{*}\gamma }(Q^{2})\; \stackrel{Q^{2}\rightarrow \infty }{\longrightarrow }\; {{2\, f_{\pi }}\over {Q^{2}}}\; .
\end{equation}
  \indent
To investigate the corrections to the \( \pi ^{0}\gamma ^{*}\gamma  \) FF when
\( Q^{2} \) is not asymptotically large, one needs an ansatz for the non-perturbative
wavefunction \( \Psi _{\pi }(z,\vec{k}_{\perp }) \) in order to perform the
integral of Eq.~(\ref{eq:piogaga}) over both \( z \) and \( \vec{k}_{\perp } \).
For the soft meson wavefunction, we follow Ref.\cite{Kroll96} and adopt a gaussian
form for the dependence on \( \vec{k}_{\perp } \) : 
\begin{equation}
\label{eq:kperpwf}
\Psi _{\pi }(z,\vec{k}_{\perp })\; =\; {{\Phi _{\pi }(z)}\over {2\sqrt{6}}}\; {{8\pi ^{2}}\over {\sigma \, z\bar{z}}}\; \exp {\left\{ -{1\over 2}\; {{\vec{k}_{\perp }^{2}}\over {\sigma \, z\bar{z}}}\right\} }\; ,
\end{equation}
 which satisfies Eq.~(\ref{eq:normpsi}). The parameter \( \sigma  \) in Eq.~(\ref{eq:kperpwf})
is related to the average squared transverse momentum \( \langle \vec{k}_{\perp }^{2}\rangle  \)
of the quarks in the meson. For the pion, this free parameter has been fixed
in Ref.\cite{Brodsky82} through the axial anomaly which leads to the condition
\begin{equation}
\label{eq:axanom}
\int _{0}^{1}\, dz\, \Psi _{\pi }(z,\vec{k}_{\perp }=0)\; =\; {{\sqrt{3}}\over {f_{\pi }}}\; .
\end{equation}
 Using the asymptotic DA for \( \Phi _{\pi } \) in the gaussian ansatz Eq.~(\ref{eq:kperpwf})
for \( \Psi _{\pi } \), the condition Eq.~(\ref{eq:axanom}) yields 
\begin{equation}
\label{eq:sigmapi}
\sigma \; =\; 8\, \pi ^{2}\, f_{\pi }^{2}\; \approx \; 0.67\; {\mathrm{G}eV}^{2}\; .
\end{equation}
 Furthermore using the gaussian ansatz for \( \Psi _{\pi } \) with an asymptotic
DA, one finds that the probability \( P_{q\bar{q}} \) of the valence quark
Fock state of the meson, which is defined as 
\begin{equation}
\label{eq:probval}
P_{q\bar{q}}\; =\; \int _{0}^{1}\, dz\, \int \, {{d^{2}\vec{k}_{\perp }}\over {16\pi ^{3}}}\; \arrowvert \Psi (z,\vec{k}_{\perp })\arrowvert ^{2}\; ,
\end{equation}
 is given by \( P^{\pi }_{q\bar{q}}=2\pi ^{2}f_{\pi }^{2}/\sigma  \) = 1/4.
For the average squared transverse momentum defined by 
\begin{equation}
\langle \vec{k}_{\perp }^{2}\rangle \; =\; {1\over {P_{q\bar{q}}}}\; \int _{0}^{1}\, dz\, \int \, {{d^{2}\vec{k}_{\perp }}\over {16\pi ^{3}}}\; \vec{k}_{\perp }^{2}\; \arrowvert \Psi (z,\vec{k}_{\perp })\arrowvert ^{2}\; ,
\end{equation}
 one obtains an average quark transverse momentum of \( \sqrt{\langle \vec{k}_{\perp }^{2}\rangle } \)
= \( \sqrt{\sigma /5}\approx  \) 0.366 GeV. \\
 \indent In Fig.~\ref{fig:pioff} we show the \( Q^{2} \) dependence of the
\( \pi ^{0}\gamma ^{*}\gamma  \) FF evaluated with the asymptotic DA for the
pion. The CLEO data \cite{Gro98} which extend to the highest \( Q^{2} \) values
\( (\sim  \) 9 GeV\( ^{2}) \), clearly approach the leading order PQCD result
of Eq.~(\ref{eq:piogagaasy}). The correction to the 1/\( Q^{2} \) scaling
behavior in the range 1 - 10 GeV\( ^{2} \) is rather well described - as found
in Refs.~\cite{Kroll96} - by the calculation including the intrinsic \( \vec{k}_{\perp } \)
dependence and using the gaussian ansatz of Eq.~(\ref{eq:kperpwf}) for the
pion wavefunction. This \( \vec{k}_{\perp } \) dependence results in a reduction
of the \( \pi ^{0}\gamma ^{*}\gamma  \) FF compared to the leading order result
when \( Q^{2} \) decreases. At a value \( Q^{2}\approx  \) 3 GeV\( ^{2} \),
this reduction is about 20 \%. \\
 \indent We now study the intrinsic transverse momentum dependence for DVCS,
which is a higher twist correction to the leading order DVCS amplitude of Eq.~(\ref{eq:dvcsampl}).
Our strategy is to evaluate the power corrections due to the intrinsic transverse
momentum dependence only for the leading twist terms in Eq.~(\ref{eq:dvcsampl})
proportional to \( H,\tilde{H},E \) and \( \tilde{E} \). Guided by the leading
order DVCS amplitude, we expect the contribution from these four OFPD's to be
dominant and do not consider other higher twist OFPD's in this work. Starting
from the handbag diagrams of Fig.~\ref{fig:handbags} for DVCS, we evaluate
the amplitude by keeping the transverse momentum dependence of the quarks propagating
between the two electromagnetic vertices, as was discussed above for the \( \pi ^{0}\gamma ^{*}\gamma  \)
FF. In analogy to Eq.~(\ref{eq:piogaga}) for the \( \pi ^{0}\gamma ^{*}\gamma  \)
FF, the calculation of the DVCS amplitude leads to a three-fold convolution
integral over the longitudinal and transverse components of \( k \) (see Fig.~\ref{fig:handbags}
for the definition). \\
 \indent Let us first consider the generalization of Eq.~(\ref{eq:qsplitting})
- representing the lower blob in Fig.~\ref{fig:handbags} - when the momentum
\( k \) in Fig.~\ref{fig:handbags} has both a longitudinal component \( (k^{+}=xP^{+}) \)
and a transverse component \( (\vec{k}_{\perp }) \). We parametrize the bilocal
quark operators matrix element at equal light-cone time \( (y^{+}=0) \) as
\begin{eqnarray}
 &  & {P^{+}}\, \int dy^{-}\, e^{i\left( xP^{+}\right) y^{-}}\, \int d^{2}\vec{y}_{\perp }\, e^{-i\vec{k}_{\perp }\cdot \vec{y}_{\perp }}\; \langle p^{'}|\bar{\psi }(-{y\over 2})\, \gamma ^{+}\, \psi ({y\over 2})|p\rangle {\Bigg |}_{y^{+}=0}\nonumber \\
 &  & =\; H^{q}(x,\vec{k}_{\perp },\xi ,t)\; \bar{N}(p^{'})\gamma ^{+}N(p)\; +\; E^{q}\; {\mathrm{t}erm}\; ,\label{eq:qsplittingveckperp} 
\end{eqnarray}
 and 
\begin{eqnarray}
 &  & {P^{+}}\, \int dy^{-}e^{i\left( xP^{+}\right) y^{-}}\, \int d^{2}\vec{y}_{\perp }\, e^{-i\vec{k}_{\perp }\cdot \vec{y}_{\perp }}\; \langle p^{'}|\bar{\psi }(-{y\over 2})\, \gamma ^{+}\gamma _{5}\, \psi ({y\over 2})|p\rangle {\Bigg |}_{y^{+}=0}\nonumber \\
 &  & =\; \tilde{H}^{q}(x,\vec{k}_{\perp },\xi ,t)\; \bar{N}(p^{'})\gamma ^{+}\gamma _{5}N(p)\; +\; \tilde{E}^{q}\; {\mathrm{t}erm}\; ,\label{eq:qsplittingaxveckperp} 
\end{eqnarray}
 where we have introduced \( \vec{k}_{\perp } \)-dependent OFPD's \( H^{q}(x,\vec{k}_{\perp },\xi ,t),E^{q},\tilde{H}^{q} \)
and \( \tilde{E}^{q} \). Remark that in writing down Eqs.~(\ref{eq:qsplittingveckperp},
\ref{eq:qsplittingaxveckperp}), we use a gauge as discussed in section \ref{sec:bilocal},
for which the gauge link between the quark fields is equal to one. \\
 \indent Similarily to Eq.~(\ref{eq:normpsi}), where integrating the meson
wavefunction over \( \vec{k}_{\perp } \) leads to the meson distribution amplitude,
we find back the OFPD \( H(x,\xi ,t) \) introduced in Eq.~(\ref{eq:qsplitting})
by integrating the \( \vec{k}_{\perp } \)-dependent distributions \( H(x,\vec{k}_{\perp },\xi ,t) \),
that is 
\begin{equation}
\label{eq:ofpdkperp}
H^{q}(x,\xi ,t)\; =\; \int \, {{d^{2}\vec{k}_{\perp }}\over {8\pi ^{3}}}\; H^{q}(x,\vec{k}_{\perp },\xi ,t)\; ,
\end{equation}
 and similarly for \( E,\tilde{H} \) and \( \tilde{E} \). Indeed, one readily
verifies that by integrating both sides of Eqs.~(\ref{eq:qsplittingveckperp},
\ref{eq:qsplittingaxveckperp}) over \( \vec{k}_{\perp } \) and by using Eq.(\ref{eq:ofpdkperp}),
one finds back the leading twist parametrization of Eq.~(\ref{eq:qsplitting})
for the lower blob in Fig.~\ref{fig:handbags}. \\
 \indent Next we must evaluate the hard scattering operator for the handbag
diagrams of Fig.~\ref{fig:handbags}. We keep the \( \vec{k}_{\perp } \) dependence
only in the denominators of the (massless) quark propagators where it is most
important. The quarks propagating between the electromagnetic vertices in Fig.~\ref{fig:handbags}
have four-momenta \( (k+q-\Delta /2) \) in the direct diagram and \( (k-q+\Delta /2) \)
in the crossed diagram respectively. Therefore, the transverse momenta in the
propagators are \( (\vec{k}_{\perp }-\vec{\Delta }_{\perp }/2) \) for the direct
diagram and \( (\vec{k}_{\perp }+\vec{\Delta }_{\perp }/2) \) for the crossed
diagram. This leads to the following generalization of the leading order DVCS
amplitude of Eq.~(\ref{eq:dvcsampl}): 
\begin{eqnarray}
 &  & H^{\mu \nu }_{DVCS}(\vec{k}_{\perp })\nonumber \\
 &  & =\; {1\over 2}\left[ \tilde{p}^{\mu }n^{\nu }+\tilde{p}^{\nu }n^{\mu }-g^{\mu \nu }\right] \; \int _{-1}^{+1}dx\; \int \, {{d^{2}\vec{k}_{\perp }}\over {8\pi ^{3}}}\; \left[ H^{p}_{DVCS}(x,\vec{k}_{\perp },\xi ,t)\; \bar{N}(p^{'})\gamma .nN(p)\; +\; E\; {\mathrm{t}erm}\right] \nonumber \\
 &  & \hspace {6.5cm}\times \left[ {1\over {x-\xi _{-}+i\epsilon }}\; +\; {1\over {x+\xi _{+}-i\epsilon }}\right] \nonumber \\
 &  & +\; {1\over 2}\, \left[ -i\varepsilon ^{\mu \nu \kappa \lambda }\tilde{p}_{\kappa }n_{\lambda }\right] \; \int _{-1}^{+1}dx\; \int \, {{d^{2}\vec{k}_{\perp }}\over {8\pi ^{3}}}\; \left[ \tilde{H}^{p}_{DVCS}(x,\vec{k}_{\perp },\xi ,t)\; \bar{N}(p^{'})\gamma .n\gamma _{5}N(p)\; +\; \tilde{E}\; {\mathrm{t}erm}\right] \nonumber \\
 &  & \hspace {6.5cm}\times \left[ {1\over {x-\xi _{-}+i\epsilon }}\; -\; {1\over {x+\xi _{+}-i\epsilon }}\right] ,\label{eq:dvcskperp} 
\end{eqnarray}
 with 
\begin{equation}
\label{eq:xiperpdef}
\xi _{\pm }\; =\; \xi \; +\; {{2\xi }\over {Q^{2}}}\left( \vec{k}_{\perp }\pm {{\vec{\Delta }_{\perp }}\over 2}\right) ^{2}\; .
\end{equation}
 One readily sees from Eqs.~(\ref{eq:dvcskperp}, \ref{eq:xiperpdef}) that
by neglecting the transverse momenta in the quark denominators compared to \( Q^{2} \)
and by using Eq.~(\ref{eq:ofpdkperp}), one finds back the leading order DVCS
amplitude of Eq.~(\ref{eq:dvcsampl}). \\
 \indent In order to estimate the DVCS amplitude of Eq.~(\ref{eq:dvcskperp}),
one needs to model the \( \vec{k}_{\perp } \)-dependent OFPD's. We will model
here only the OFPD's \( H(x,\vec{k}_{\perp },\xi ,t) \) and \( \tilde{H} \),
as they give the dominant contribution to the DVCS amplitude in the near forward
direction. Therefore we will neglect the contribution from the OFPD's \( E \)
and \( \tilde{E} \).

To generate the \( \vec{k}_{\perp } \)-dependent OFPD's, we start from the
double distributions which we model in the near forward direction by a gaussian
ansatz by analogy with what was done in Eq.~(\ref{eq:kperpwf}) for the pion
wavefunction. This yields for the helicity independent double distribution \( F^{q} \)
: 
\begin{eqnarray}
F^{q}(\tilde{x},y,\vec{k}_{\perp },t)\; =\; F^{q}_{1}(t)/F^{q}_{1}(0)\; q(\tilde{x})\;  &  & 6\, {{y(1-\tilde{x}-y)}\over {(1-\tilde{x})^{3}}}\; \nonumber \\
 &  & \times \, \left[ {{4\pi ^{2}}\over {\sigma \, y(1-\tilde{x}-y)}}\; \exp {\left\{ -{1\over 2}\; {{\vec{k}_{\perp }^{2}}\over {\sigma \, y(1-\tilde{x}-y)}}\right\} }\right] \; .\label{eq:ddmodelkperp} 
\end{eqnarray}
 For the helicity dependent double distribution \( \tilde{F}^{q} \), our ansatz
yields 
\begin{eqnarray}
\tilde{F}^{q}(\tilde{x},y,\vec{k}_{\perp },t)\; =\; g_{A}^{q}(t)/g_{A}^{q}(0)\; \Delta q(\tilde{x})\;  &  & 6\, {{y(1-\tilde{x}-y)}\over {(1-\tilde{x})^{3}}}\; \nonumber \\
 &  & \times \, \left[ {{4\pi ^{2}}\over {\sigma \, y(1-\tilde{x}-y)}}\; \exp {\left\{ -{1\over 2}\; {{\vec{k}_{\perp }^{2}}\over {\sigma \, y(1-\tilde{x}-y)}}\right\} }\right] \; .\label{eq:ddhelmodelkperp} 
\end{eqnarray}
 When integrating Eq.~(\ref{eq:ddmodelkperp}) over \( \vec{k}_{\perp } \),
one finds back the double distribution \( F^{q} \) of Eq.~(\ref{eq:ddmodel}),
i.e. Eq.~(\ref{eq:ddmodelkperp}) satisfies the normalization 
\begin{equation}
\label{eq:ddkperp}
F^{q}(\tilde{x},y,t)\; =\; \int \, {{d^{2}\vec{k}_{\perp }}\over {8\pi ^{3}}}\; F^{q}(\tilde{x},y,\vec{k}_{\perp },t)\; .
\end{equation}
 In the same way, when integrating Eq.~(\ref{eq:ddhelmodelkperp}) over \( \vec{k}_{\perp } \),
one finds back the helicity dependent double distribution \( \tilde{F}^{q} \)
of Eq.~(\ref{eq:ddhelmodel}).

\begin{figure}[ht]
\epsfxsize=15 cm
\epsfysize=20.cm
\centerline{\epsffile{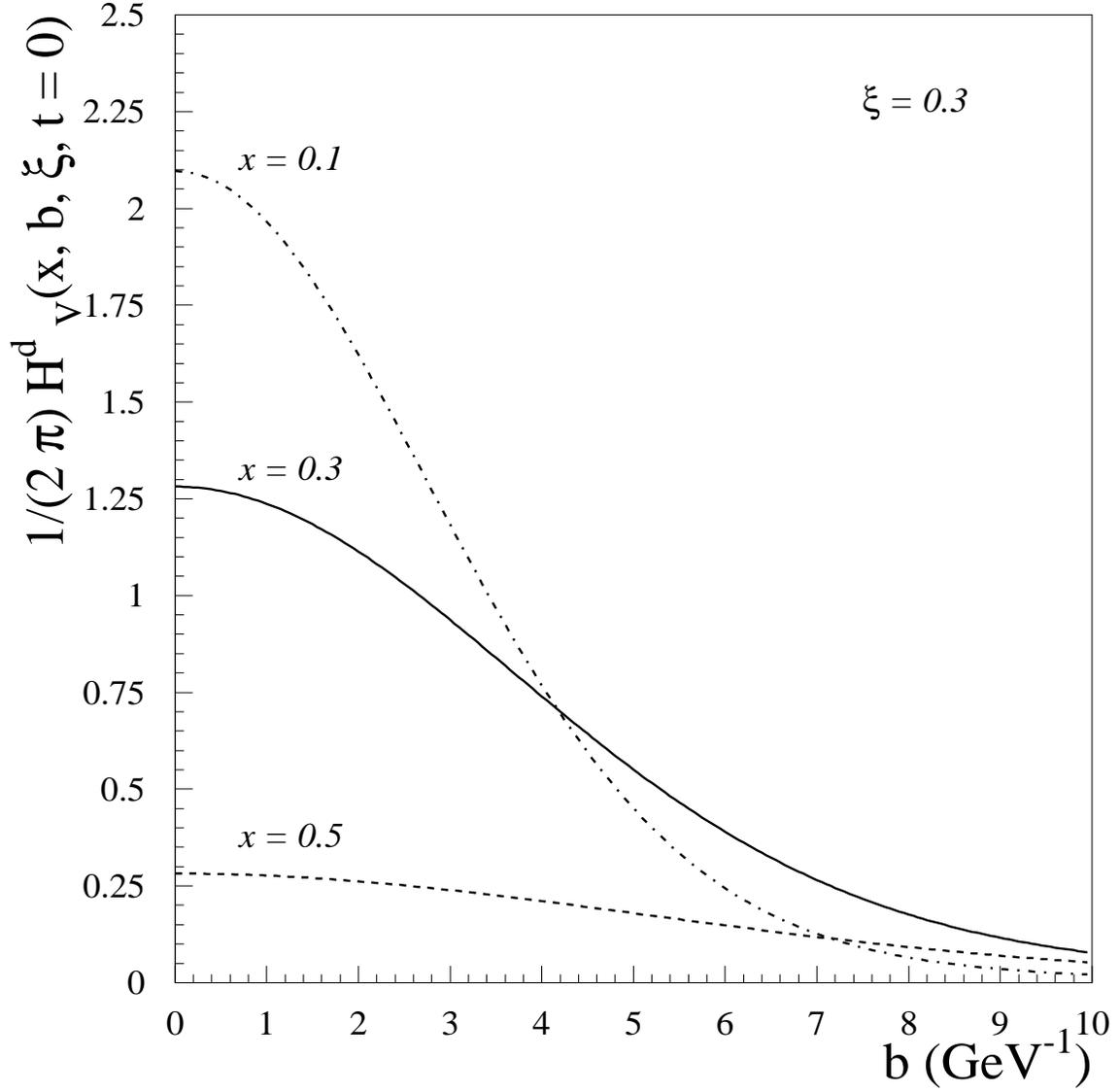}}
\vspace{-2cm}
\caption[]{\small Impact parameter (b) dependence of the OFPD $H^d_V$ 
for the valence down quark, using the model ansatz (based on the MRST98
quark distributions) as described in the text.}
\label{fig:ofpdimpact}
\end{figure}

The gaussian ansatz of Eqs.~(\ref{eq:ddmodelkperp}, \ref{eq:ddhelmodelkperp})
depends upon one free parameter \( \sigma  \) which is related to the average
squared transverse momentum of the quarks in the nucleon. We take here the same
value as determined for the pion in Eq.~(\ref{eq:sigmapi}). Having modelled
the \( \vec{k}_{\perp } \)-dependent double distributions, the \( \vec{k}_{\perp } \)-dependent
OFPD's \( H \) and \( \tilde{H} \) are then obtained from Eq.~(\ref{eq:ddmodelkperp})
using the \( \vec{k}_{\perp } \) dependent analogues of Eqs.~(\ref{eq:relofpdnfpd})
and (\ref{eq:double}). \\
 \indent When calculating the \( \vec{k}_{\perp } \)-dependence for the hard
meson electroproduction reactions in section \ref{sec:kperpmeson}, we will
find it convenient to introduce the OFPD's in impact parameter space. They are
obtained by taking the Fourier transform of the \( \vec{k}_{\perp } \)-dependent
OFPD's with respect to \( \vec{k}_{\perp } \) : 
\begin{equation}
\label{eq:ofpdimpact}
H^{q}(x,\vec{b},\xi ,t)\; =\; \int \, {{d^{2}\vec{k}_{\perp }}\over {\left( 2\pi \right) ^{2}}}\; e^{i\vec{k}_{\perp }\cdot \vec{b}}\; H^{q}(x,\vec{k}_{\perp },\xi ,t)\; ,
\end{equation}
 where \( \vec{b} \) plays the role of a transverse distance between the active
quarks. Using Eq.~(\ref{eq:ofpdkperp}), one readily verifies that the ordinary
OFPD's correspond to a zero transverse distance : 
\begin{equation}
{1\over {2\pi }}\; H^{q}(x,\vec{b}=0,\xi ,t)\; =\; H^{q}(x,\xi ,t)\; .
\end{equation}
 We show in Fig.~\ref{fig:ofpdimpact} the impact parameter dependence for the
valence down quark OFPD \( H^{d}_{V} \) using the ansatz of Eq.~(\ref{eq:ddmodelkperp}).
Note that the values at the origin (\( b=0 \)) in Fig.~\ref{fig:ofpdimpact}
can be directly read off from the corresponding ordinary OFPD displayed in Fig.~\ref{fig:ofpdxi}.

\subsection{Intrinsic transverse momentum dependence in pion electromagnetic FF and in
hard meson electroproduction reactions}

\label{sec:kperpmeson}

\begin{figure}[ht]
\epsfxsize=15 cm
\epsfysize=20.cm
\centerline{\epsffile{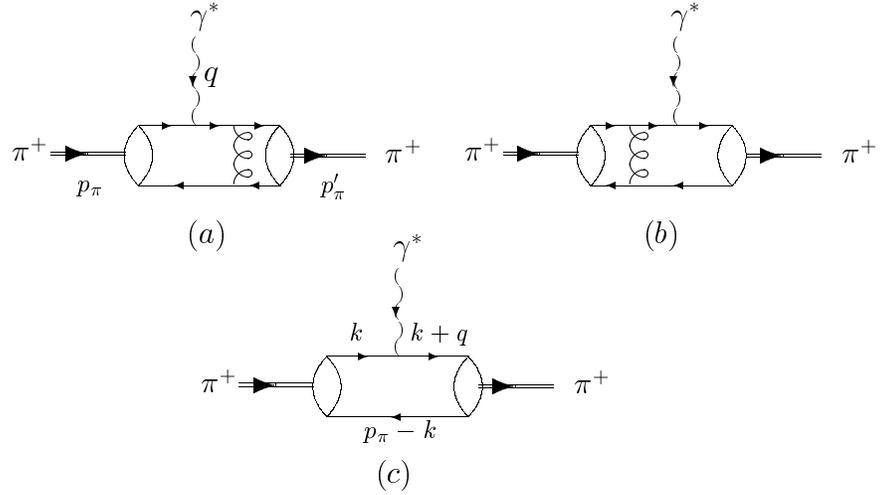}}
\vspace{-5cm}
\caption[]{\small Leading order direct (a) and crossed (b) diagrams to the 
$\pi$ electromagnetic form factor involving one-gluon exchange. 
Soft overlap contribution (c) to the $\pi$ form factor. 
There are analogous diagrams where the virtual photon couples 
to the lower quark line, which are not shown here.}
\label{fig:piemffdiag}
\end{figure}

Having discussed the intrinsic transverse momentum dependence of the DVCS amplitude,
we now turn to the hard meson electroproduction reactions. As before, where
we have made the parallel between the \( \pi ^{0}\gamma ^{*}\gamma  \) FF and
DVCS, we will find it very instructive to study the intrinsic transverse momentum
dependence of the hard meson electroproduction amplitudes by comparison with
the charged pion electromagnetic FF. In contrast to the \( \pi ^{0}\gamma ^{*}\gamma  \)
FF, the leading order contribution to the pion electromagnetic FF proceeds through
one-gluon exchange as shown in Figs.~\ref{fig:piemffdiag} a, b.

To calculate the form factor, we consider the process in a frame (e.g. Breit
frame) where the initial (final) pion moves with large momentum along the positive
(negative) \( z \)-direction and where the momentum transfered by the virtual
photon points along the negative \( z \)-direction. We denote the longitudinal
momentum fractions of the quarks in the initial (final) pion by \( x \) (\( y \))
and their relative transverse momenta by \( \vec{k}_{\perp } \) (\( \vec{l}_{\perp } \)).
The expression of the pion electromagnetic form factor including the transverse
momentum dependencies is given by the six-fold convolution integral : 
\begin{equation}
\label{eq:piemff}
F_{\pi }(Q^{2})\; =\; \int _{0}^{1}\, dy\, \int \, {{d^{2}\vec{l}_{\perp }}\over {16\pi ^{3}}}\; \int _{0}^{1}\, dx\, \int \, {{d^{2}\vec{k}_{\perp }}\over {16\pi ^{3}}}\; \Psi _{\pi }(y,\vec{l}_{\perp })\; T_{H}(y,\vec{l}_{\perp },x,\vec{k}_{\perp };Q^{2})\; \Psi _{\pi }(x,\vec{k}_{\perp })\; .
\end{equation}
 The hard scattering operator in Eq.~(\ref{eq:piemff}) can be easily calculated
by the diagrams of Figs.~\ref{fig:piemffdiag} a, b and two other diagrams where
the virtual photon couples to the lower quark line. It has been shown in Ref.\cite{Li94}
that the most important transverse momentum dependence in \( T_{H} \) comes
from the gluon propagator. The reason is that if one neglects the transverse
momentum dependence, the gluon virtuality then depends quadratically on the
quark longitudinal momentum fractions (\( x \) and \( y \)). In contrast,
the virtuality of intermediate quark lines depends only linearly upon the longitudinal
momentum fractions. Therefore the gluon propagator weights most heavily the
end-point regions and it is important in these regions to include the transverse
momentum dependence to get a consistent PQCD calculation. We keep therefore
the transverse momentum dependence of \( T_{H} \) only in the gluon propagator,
which simplifies considerably the integral of Eq.~(\ref{eq:piemff}). Using
the symmetry of the integrand in Eq.~(\ref{eq:piemff}) under \( x\leftrightarrow \bar{x} \)
and \( y\leftrightarrow \bar{y} \), this yields for \( T_{H} \) : 
\begin{equation}
\label{eq:piemff2}
T_{H}(y,\vec{l}_{\perp },x,\vec{k}_{\perp };Q^{2})\; =\; {4\over 9}\; (4\pi \alpha _{s})\; {{\left( 2\sqrt{6}\right) ^{2}}\over 2}\; {1\over {xyQ^{2}\; +\; (\vec{k}_{\perp }-\vec{l}_{\perp })^{2}}}\; ,
\end{equation}
 where the factor (4/9) is a color factor due to one-gluon exchange. Neglecting
the transverse momenta dependence in \( T_{H} \) and using asymptotic DA's,
one obtains the well known leading order PQCD result for \( F_{\pi } \) : 
\begin{equation}
\label{eq:piemff3}
F_{\pi }(Q^{2})\; \stackrel{Q^{2}\rightarrow \infty }{\longrightarrow }\; {{16\pi \, \alpha _{s}\left( Q^{2}\right) \, f_{\pi }^{2}}\over {Q^{2}}}\; .
\end{equation}
 The calculation of the full six-fold convolution integral of Eq.~(\ref{eq:piemff})
can be simplified by noting that the hard scattering operator \( T_{H} \) of
Eq.~(\ref{eq:piemff2}) depends only on the transverse momentum difference \( \vec{m}_{\perp }\equiv \vec{k}_{\perp }-\vec{l}_{\perp } \).
Therefore, one has advantage to express Eq.~(\ref{eq:piemff}) in impact parameter
space. The pion wavefunction in impact parameter space is obtained from \( \Psi _{\pi }(x,\vec{k}_{\perp }) \)
by the Fourier transform 
\begin{equation}
\label{eq:wavefimpact}
\Psi _{\pi }(x,\vec{b})\; =\; \int \, {{d^{2}\vec{k}_{\perp }}\over {(2\pi )^{2}}}\; e^{i\, \vec{k}_{\perp }\cdot \vec{b}}\; \Psi _{\pi }(x,\vec{k}_{\perp })\; ,
\end{equation}
 where \( \vec{b} \) represents the transverse separation between the quarks
in the pion. For the gaussian ansatz of Eq.~(\ref{eq:kperpwf}), \( \Psi _{\pi }(x,\vec{b}) \)
is given by : 
\begin{equation}
\label{eq:impactwf}
\Psi _{\pi }(x,\vec{b})\; =\; 4\pi \; {{\Phi _{\pi }(x)}\over {2\sqrt{6}}}\; \exp {\left\{ -{1\over 2}\; \sigma \, b^{2}\, x\bar{x}\right\} }\; .
\end{equation}
 The hard scattering operator in impact parameter space is given by 
\begin{eqnarray}
T_{H}(x,y,\vec{b};Q^{2})\;  & = & \; \int \, {{d^{2}\vec{m}_{\perp }}\over {(2\pi )^{2}}}\; e^{i\, \vec{m}_{\perp }\cdot \vec{b}}\; T_{H}(x,y,\vec{m}_{\perp };Q^{2})\nonumber \\
 & = & \; {4\over 3}\; \left( 16\pi \alpha _{s}(\mu _{R}^{2})\right) \; {1\over {2\pi }}\; K_{0}(\sqrt{xy}\, b\, Q)\; ,\label{eq:thardpiimpact} 
\end{eqnarray}
 where the strong coupling is evaluated at the renormalization scale \( \mu _{R} \)
to be specified shortly, and where the modified Bessel function \( K_{0} \)
in Eq.~(\ref{eq:thardpiimpact}) is obtained by Fourier transforming the gluon
propagator (in Eq.~(\ref{eq:piemff2})) with spacelike four-momentum. As \( \Psi _{\pi } \)
and \( T_{H} \) depend only on the magnitude (\( b \)) of \( \vec{b} \),
Eq.~(\ref{eq:piemff}) for the pion FF reduces to a three-fold convolution integral
: 
\begin{equation}
\label{eq:piemffimpact}
F_{\pi }(Q^{2})\; =\; \int _{0}^{1}\, dx\, \int _{0}^{1}\, dy\, \int _{0}^{\infty }\, {{db\; b}\over {8\pi }}\; \Psi _{\pi }(x,b)\; T_{H}(x,y,b;Q^{2})\; \Psi _{\pi }(y,b)\; .
\end{equation}
 To calculate the hard scattering operator of Eq.~(\ref{eq:thardpiimpact}),
one has to specify the renormalization scale \( \mu _{R} \) in the strong coupling
constant. Here we follow the recent work of Ref.\cite{Stefanis98} and take
it as \( \mu _{R}^{2}={{xy}\over 2}Q^{2} \), which represents in a sense the
average gluon virtuality in the leading order diagrams for the pion FF. Near
the endpoints, where the quark longitudinal momenta vanish (\( x\approx 0 \)
or \( y\approx 0 \)), the gluon virtuality becomes very small and one leaves
the region where the PQCD result for the running coupling can be used. An often
used practise is to freeze the coupling at low gluon virtuality at some finite
value (\( \sim 0.5 \)). Instead of this ad-hoc procedure, the pion FF was evaluated
in Ref.~\cite{Stefanis98} using the infrared (IR) finite analytic model for
\( \alpha _{s} \) of Refs.~\cite{Shirkov97,Milton97}. In Ref.\cite{Shirkov97},
Shirkov and Solovtsov have deduced from the asymptotic freedom expression for
\( \alpha _{s} \) and by imposing \( Q^{2} \) analyticity, the following IR
finite result for the (one-loop) strong coupling : 
\begin{equation}
\label{eq:irfinalphas}
\alpha _{s}^{an}(Q^{2})\; =\; {{4\pi }\over {\beta _{0}}}\; \left[ {1\over {\ln Q^{2}/\Lambda _{QCD}^{2}}}\; +\; {{\Lambda _{QCD}^{2}}\over {\Lambda _{QCD}^{2}-Q^{2}}}\right] \; ,
\end{equation}
 where \( \beta _{0}=11-2/3N_{f} \) and \( \Lambda _{QCD} \) is the QCD scale
parameter. The first term in Eq.~(\ref{eq:irfinalphas}) is the standard (one-loop)
asymptotic freedom expression and the second term ensures that the coupling
has no ghost pole at \( Q^{2}=\Lambda _{QCD}^{2} \). At \( Q^{2}=0 \), the
coupling constant takes on the finite value \( \alpha _{s}(Q^{2}=0)=4\pi /\beta _{0} \)
(\( \approx  \) 1.40 for \( N_{f} \) = 3), which depends only upon group symmetry
factors. Therefore, Eq.~(\ref{eq:irfinalphas}) provides a coupling that can
be used from low to high values of \( Q^{2} \) without any adjustable parameters
other than \( \Lambda _{QCD} \). The value of \( \Lambda _{QCD} \) in the
one-loop case can be fixed from \( \alpha _{s}(M_{\tau }^{2})\approx 0.34 \)
and results in \( \Lambda _{QCD}^{an}\approx 280 \) MeV \cite{Shirkov97} -
where the superscript refers to the analytic model of Eq.~(\ref{eq:irfinalphas})
(it was noted in Ref.\cite{Shirkov97} that this value of \( \Lambda _{QCD}^{an} \)
corresponds with \( \Lambda ^{\overline{MS}}_{QCD}\approx  \) 230 MeV when
using the \( \overline{MS} \) scheme in the usual renormalization group solution).
As an example one obtains from Eq.~(\ref{eq:irfinalphas}) \( \alpha _{s}^{an}(1 \)
GeV\( ^{2} \)) \( \approx  \) 0.43. Having fixed \( \Lambda _{QCD} \), the
convolution integral of Eq.~(\ref{eq:piemffimpact}) for the pion form factor
can now be evaluated with the IR finite strong coupling of Eq.~(\ref{eq:irfinalphas}),
as proposed recently in Ref.\cite{Stefanis98}. \\
 \indent The result is shown in Fig.~\ref{fig:piemff}, where the leading order
PQCD expression of Eq.~(\ref{eq:piemff3}) for the pion form factor (dotted
line) is compared with the result including the intrinsic transverse momentum
dependence (dashed-dotted line). As is seen from Fig.~\ref{fig:piemff}, the
leading order PQCD result is approached only at very large \( Q^{2} \). The
correction including the transverse momentum dependence gives a substantial
suppression at lower \( Q^{2} \) (about a factor of two around \( Q^{2}\approx  \)
5 GeV\( ^{2} \)). At these lower \( Q^{2} \) values, the inclusion of the
transverse momentum dependence renders the PQCD calculation internally consistent
in the sense that the dominant contributions come from regions in which the
coupling is relatively small. In addition to the reduction due to the transverse
momentum dependence there is an additional suppression due to the Sudakov effect
which ensures that at large \( Q^{2} \), soft gluon exchange between quarks
is suppressed due to gluonic radiative corrections. At the intermediate values
of \( Q^{2} \) shown in Fig.~\ref{fig:piemff}, the exponential reduction due
to the Sudakov form factor was calculated and found to be small provided one
has already taken into account the intrinsic transverse momentum dependence,
as was also noted in Ref.\cite{Kroll96}. Only at very large values of \( Q^{2} \),
the Sudakov form factor takes over and yields an additional reduction compared
with the one due to the intrinsic transverse momentum dependence. As we give
in this paper only predictions for low and intermediate \( Q^{2} \) values,
we will not consider the Sudakov effect.\\
 \indent We now transpose the result for the pion electromagnetic FF and study
the transverse momentum dependence in hard meson electroproduction amplitudes.
Keeping the transverse momentum dependence of the hard scattering operator only
in the gluon propagators, we obtain for the longitudinal electroproduction amplitude
of \( \rho ^{0}_{L} \) : 
\begin{eqnarray}
 &  & {\mathcal{M}}^{L}_{\rho ^{0}_{L}}(\vec{k}_{\perp })\nonumber \\
 &  & =\; -ie\, {4\over 9}\, {1\over {Q}}\; \left( {{Q^{2}}\over {2\xi }}\right) \int _{0}^{1}dz\; \int \, {{d^{2}\vec{l}_{\perp }}\over {16\pi ^{3}}}\; (2\sqrt{6})\; \Psi _{\rho }(z,\, \vec{l}_{\perp }\, +\, z\, \vec{\Delta }_{\perp })\nonumber \\
 &  & \times \int _{-1}^{+1}dx\; \int \, {{d^{2}\vec{k}_{\perp }}\over {8\pi ^{3}}}\; \left[ H^{p}_{\rho ^{0}_{L}}(x,\vec{k}_{\perp },\xi ,t)\; \bar{N}(p^{'})\gamma .nN(p)\; +\; E\; {\mathrm{t}erm}\right] \; \left( 4\pi \alpha _{s}(\mu _{R}^{2})\right) \nonumber \\
 &  & \times {1\over 2}\, \left[ {1\over {\bar{z}\left( x-\xi \right) {{Q^{2}}\over {2\xi }}\, -\, \left( \vec{k}_{\perp }-\vec{l}_{\perp }-{{\vec{\Delta }_{\perp }}\over 2}\right) ^{2}+i\epsilon }}\, +\, {1\over {z\left( x+\xi \right) {{Q^{2}}\over {2\xi }}\, +\, \left( \vec{k}_{\perp }-\vec{l}_{\perp }-{{\vec{\Delta }_{\perp }}\over 2}\right) ^{2}-i\epsilon }}\right] \, .\label{eq:rhoamplkperp} 
\end{eqnarray}
 When neglecting the transverse momenta dependencies \( \vec{k}_{\perp } \)
and \( \vec{l}_{\perp } \) in the gluon propagators in Eq.~(\ref{eq:rhoamplkperp})
and by using the normalization conditions for the \( \rho  \) wavefunction
(analogous as Eq.~(\ref{eq:normpsi}) for the pion) and the normalization condition
for the OFPD's (Eq.~(\ref{eq:ofpdkperp})), we find back the leading order \( \rho ^{0}_{L} \)
electroproduction amplitude of Eq.~(\ref{eq:rhoampl}). \\
 \indent The evaluation of Eq.~(\ref{eq:rhoamplkperp}) requires to perform
a six-fold convolution integral in analogy with Eq.~(\ref{eq:piemff}) for the
pion electromagnetic FF. To make the calculation tractable, we will limit ourselves
here to the forward direction where \( \vec{\Delta }_{\perp }=0 \). In this
case, the gluon propagators depend only on the transverse momentum difference
\( \vec{k}_{\perp }-\vec{l}_{\perp } \) and one can use the impact parameter
space, as for the pion form factor. Using Eq.~(\ref{eq:wavefimpact}) for the
wavefunction and Eq.~(\ref{eq:ofpdimpact}) for the OFPD, Eq.~(\ref{eq:rhoamplkperp})
for the hard \( \rho ^{0}_{L} \) electroproduction amplitude becomes in the
forward direction (\( \vec{\Delta }_{\perp }=0 \)) : 
\begin{eqnarray}
 &  & {\mathcal{M}}^{L}_{\rho ^{0}_{L}}(\vec{k}_{\perp })\nonumber \\
 &  & =\; -ie\, {4\over 9}\, {1\over {Q}}\, \left( {{Q^{2}}\over {2\xi }}\right) \; \int _{0}^{1}dz\; \int _{0}^{\infty }\, {{db\; b}\over {4\pi }}\; (2\sqrt{6})\; \Psi _{\rho }(z,\, b)\nonumber \\
 &  & \hspace {.5cm}\times \int _{-1}^{+1}dx\; \left[ H^{p}_{\rho ^{0}_{L}}(x,b,\xi ,t)\; \bar{N}(p^{'})\gamma .nN(p)\, +\, E\; {\mathrm{t}erm}\right] \, \left( 4\pi \alpha _{s}(\mu _{R}^{2})\right) \, {1\over 2}\, T_{H}(x,\xi ,z,b;Q^{2})\; ,\label{eq:rhoamplimpact} 
\end{eqnarray}
 where \( T_{H}(x,\xi ,z,b;Q^{2}) \) represents the hard gluon propagators
in impact parameter space. It is given by 
\begin{eqnarray}
 &  & T_{H}(x,\xi ,z,b;Q^{2})\nonumber \\
 &  & =\; {1\over {2\pi }}\int _{0}^{\infty }dm_{\perp }\; m_{\perp }\, J_{0}(b\, m_{\perp })\; \left[ {1\over {z\left( x-\xi \right) {{Q^{2}}\over {2\xi }}\, -\, m_{\perp }^{2}+i\epsilon }}\; +\; {1\over {z\left( x+\xi \right) {{Q^{2}}\over {2\xi }}\, +\, m_{\perp }^{2}-i\epsilon }}\right] \nonumber \\
 &  & =\; {1\over {2\pi }}\, \left\{ -i{\pi \over 2}\, H_{0}^{(1)}\left( \sqrt{z\, (x-\xi )/(2\xi )}\, Q\, b\right) \; +\; K_{0}\left( \sqrt{z\, (x+\xi )/(2\xi )}\, Q\, b\right) \right\} \, ,\hspace {.25cm}{\mathrm{f}or}\hspace {.25cm}x\geq \xi \; ,\label{eq:gluonpropmeson1} \\
 &  & =\; {1\over {2\pi }}\, \left\{ -\; K_{0}\left( \sqrt{z\, (\xi -x)/(2\xi )}\, Q\, b\right) \; +\; K_{0}\left( \sqrt{z\, (x+\xi )/(2\xi )}\, Q\, b\right) \right\} \, ,\hspace {.25cm}{\mathrm{f}or}\hspace {.25cm}0<x<\xi \; ,\label{eq:gluonpropmeson2} 
\end{eqnarray}
 where the analytical results are obtained through a contour integration in
the complex plane. The modified Bessel function \( K_{0} \) originates from
Fourier transforming a gluon propagator with spacelike momentum which appears
in the direct diagram when \( 0<x<\xi  \) and in the crossed diagram. The Hankel
function \( H_{0}^{(1)} \) in Eq.~(\ref{eq:gluonpropmeson1}) originates from
Fourier transforming a gluon propagator with timelike momentum, which appears
in the direct diagram when \( x>\xi  \). In Eqs.~(\ref{eq:gluonpropmeson1},
\ref{eq:gluonpropmeson2}), the analytical expressions of \( T_{H} \) are given
for positive values of \( x \). The expression of \( T_{H} \) for negative
values of \( x \) is directly seen to be minus the expression for positive
values of \( x \). \\
 \indent For the estimates of meson electroproduction, which include the transverse
momentum dependence, we use the IR finite strong coupling of Eq.~(\ref{eq:irfinalphas}),
as in the case of the pion form factor. The scale \( \mu _{R} \) at which the
strong coupling \( \alpha _{s} \) in Eq.~(\ref{eq:rhoamplimpact}) is evaluated,
is given by the largest mass scale in the hard scattering, which we can take
as 
\begin{equation}
\mu _{R}^{2}={\mathrm{m}ax}\left( |x-\xi |\, z\, {{Q^{2}}\over {2\xi }},\; {1\over {b^{2}}}\right) \; ,
\end{equation}
 for the direct diagram and 
\begin{equation}
\mu _{R}^{2}={\mathrm{m}ax}\left( |x+\xi |\, z\, {{Q^{2}}\over {2\xi }},\; {1\over {b^{2}}}\right) \; ,
\end{equation}
 for the crossed diagram.

\subsection{Soft overlap mechanism for pion electromagnetic FF and for
  meson electroproduction reactions}

\label{sec:overlap}

\begin{figure}[ht]
\epsfxsize=15 cm
\epsfysize=20.cm
\centerline{\epsffile{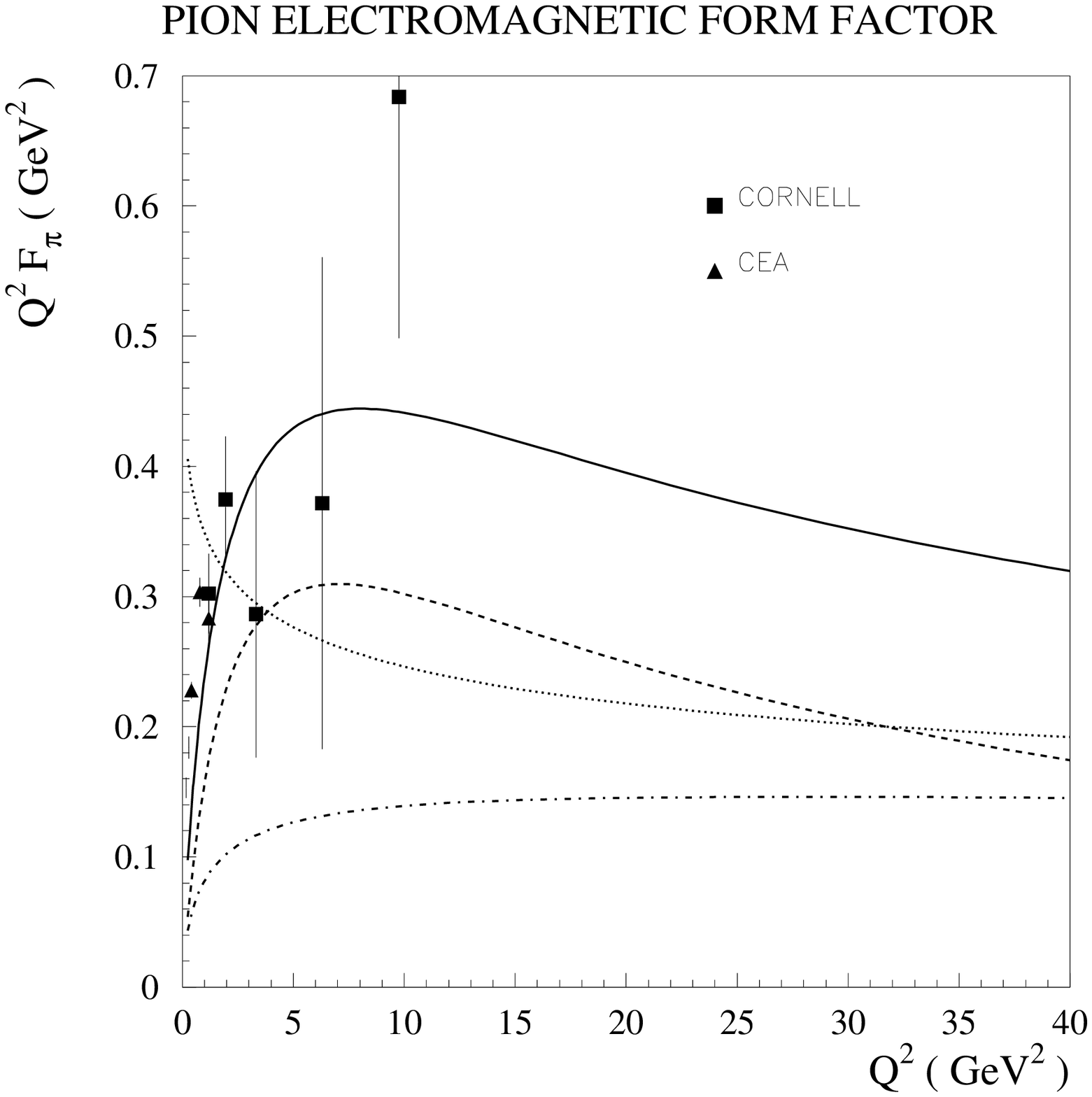}}
\vspace{-2cm}
\caption[]{\small Results for the $\pi$ electromagnetic form factor of
the leading order PQCD prediction without (dotted line) and with 
(dashed-dotted line ) inclusion of the corrections due to 
intrinsic transverse momentum dependence. 
The dashed curve shows the result for the soft overlap contribution of
Fig.~\ref{fig:piemffdiag}c and the total result (full line) 
is the sum of the dashed and dashed-dotted lines. 
The ``data'' points from Refs.\cite{Brown73} (CEA) and \cite{Bebek78}
(Cornell) were obtained indirectly from pion electroproduction.}
\label{fig:piemff}
\end{figure}

\begin{figure}[ht]
\epsfxsize=18 cm
\epsfysize=23.cm
\centerline{\epsffile{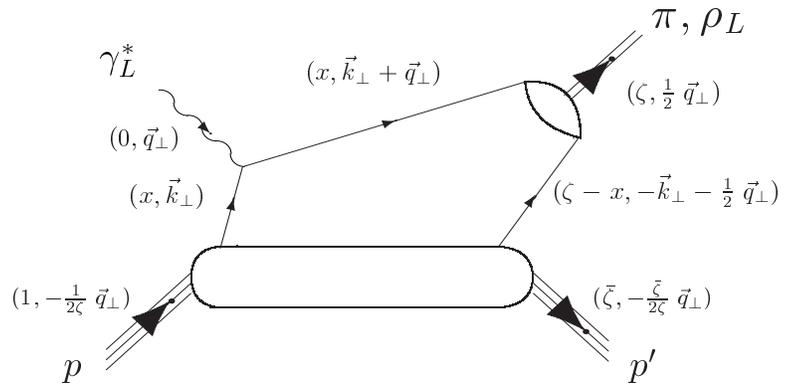}}
\vspace{-9cm}
\caption[]{\small Soft overlap contribution to the meson 
electroproduction amplitude.}
\label{fig:overlap}
\end{figure}

For hard exclusive reactions that proceed through one-gluon exchange at leading
order, soft `overlap' mechanisms can compete with the leading order amplitude
at intermediate values of \( Q^{2} \). We will estimate the soft overlap contribution
to meson electroproduction by analogy with the soft overlap to the pion electromagnetic
FF. \\
 \indent The overlap contribution to the pion electromagnetic FF is shown in
Fig.~\ref{fig:piemffdiag}c. Its calculation is most easily performed in the
frame where the spacelike virtual photon has only a transversal momentum \( \vec{q}_{\perp } \)
(i.e. \( q^{+}=0 \) and \( Q^{2}=\vec{q}_{\perp }^{\, 2} \)). Although the
final result for the FF cannot depend on the reference frame in which
one performs the calculations, the calculation however in 
a frame where $q^+ \neq 0$ is more complicated  due to additional 
\( Z \)-graph contributions (see e.g. Ref.~\cite{Sawicki} for a
comprehensive discussion).  
\newline
\indent
Indeed, in a frame where $q^+ \neq 0$, 
one obtains two types of contributions to the FF. The first 
corresponds to the light-cone time ordering in the diagram of 
Fig.~\ref{fig:piemffdiag}c, where the initial pion consists of
partons with light-cone
momenta \( k^{+} \) and \( (p_{\pi }^{+}-k^{+}) \) before the
interaction, and where the final pion consists of an active parton 
with momentum \( (k^{+}+q^{+}) \) and a spectator parton  
of momentum \( (p_{\pi }^{+}-k^{+}) \).
In this case, the non-perturbative objects at both meson sides 
are the pion light-cone wavefunctions \( \Psi _{\pi} \) as defined before. 
\newline
\indent
The second type of contribution 
corresponds to the situation where the virtual photon 
creates a quark-antiquark pair with light-cone momenta 
\( (q^{+}-k^{+}) \) and \( k^{+} \) prior (in light-cone time) 
to the interaction with the bound state. In contrast to the
first type of contribution, these so-called \( Z \)-graph
contribution \textit{cannot}
be expressed as the product of two pion light-cone wavefunctions. 
Indeed, it involves  the amplitude for a parton to split 
into a meson and another parton, which is a different 
non-perturbative object than a meson light-cone wavefunction. 
However, as pair creation from a photon with zero momentum
\( q^{+} \) = 0 is not possible for light-cone time ordered diagrams, these
pair creation or \( Z \)-graph contributions vanish when \( q^{+}=0 \).
Therefore, it is very convenient to estimate the soft overlap
contribution to the pion FF in the frame \( q^{+}=0 \) 
where only the first contribution survives,
and which is given by the Drell-Yan formula \cite{DrellYan} as a product of two
meson light-cone wavefunctions : 
\begin{equation}
\label{eq:piemffsoft1}
F_{\pi }^{soft}(Q^{2})\; =\; \int _{0}^{1}\, dz\, \int \, {{d^{2}\vec{k}_{\perp }}\over {16\pi ^{3}}}\; \Psi _{\pi }\left( z,\vec{k}_{\perp }+\bar{z}\vec{q}\right) \; \Psi _{\pi }\left( z,\vec{k}_{\perp }\right) \; .
\end{equation}
In the arguments of the soft meson wavefunctions in Eq.~(\ref{eq:piemffsoft1}),
the relative transverse momenta enter. Using the gaussian ansatz of Eq.~(\ref{eq:kperpwf})
for the soft pion wave function \( \Psi _{\pi } \) in Eq.~(\ref{eq:piemffsoft1}),
one obtains : 
\begin{equation}
\label{eq:piemffsoft2}
F_{\pi }^{soft}(Q^{2})\; =\; {{4\pi ^{2}}\over \sigma }\; \int _{0}^{1}\, dz\, \left( {{\Phi _{\pi }(z)}\over {2\sqrt{6}}}\right) ^{2}\, {1\over {z\bar{z}}}\; \exp \left\{ -{{Q^{2}\, \bar{z}^{2}}\over {4\sigma z\bar{z}}}\right\} \; .
\end{equation}
 The soft overlap contribution of Eq.~(\ref{eq:piemffsoft2}) to the
 pion electromagnetic FF is shown in Fig.~\ref{fig:piemff}, 
calculated with the asymptotic distribution
amplitude for \( \Phi _{\pi } \). It is seen from Fig.~\ref{fig:piemff} that
although the soft overlap contribution drops faster than \( 1/Q^{2} \), one
has to go to rather large values of \( Q^{2} \) before the perturbative one-gluon
exchange contribution dominates. In the intermediate \( Q^{2} \) range shown
in Fig.~\ref{fig:piemff}, the soft overlap contribution makes up more than
half of the total form factor. One also sees from Fig.~\ref{fig:piemff} that
the sum of the soft overlap and the perturbative one-gluon exchange contributions,
corrected for transverse momentum dependence, is compatible with the existing
data. When comparing to the reported data for the pion electromagnetic FF, one
has to be aware however that these sparse data are rather imprecise at the
larger \( Q^{2} \) values, as they are obtained indirectly from pion electroproduction
experiments through a (model-dependent) extrapolation to the pion pole. \\
 \indent We now calculate the soft overlap 
contribution to longitudinal electroproduction of mesons, which is shown
in Fig.~\ref{fig:overlap}. As we are interested in this paper in the
study of meson electroproduction reactions at a non-zero 
transfer \( \Delta  \), the choice of frame to suppress
\( Z \)-graph contributions completely is not as obvious as in the form factor
case. Among the natural choices, one has now the possibility 
to choose a frame where \( q^{+} \) = 0 
or a frame where \( \Delta ^{+} \) = 0. As we are concerned in this work with
the kinematical regime of large \( Q^{2} \) and small momentum
transfer \( t = \Delta^2 \), it seems natural to estimate the overlap 
contribution in a frame where \( q^{+} \) = 0, 
which will suppress \( Z \)-graph contributions at the virtual photon vertex. 
Remark that the choice of frame $q^+ = 0$ is different from the one 
used before in the calculation of the leading order amplitude for hard
 meson electroproduction. We will use this different choice of frame
 only to provide an estimate for the soft overlap contribution and
 will therefore show the results for the leading 
order and overlap amplitudes separately. 
To add the leading order and overlap amplitudes coherently, 
one would have to perform a Lorentz boost.   
\newline
\indent
Making the choice \( q^{+} \) = 0, we show 
in Fig.~\ref{fig:overlap} the leading \( (+,\perp ) \)
components of the external and quark momenta in the overlap diagram, 
where the + components refer here to the initial nucleon momentum, 
to keep a close analogy to the overlap calculation for the pion FF. 
In the present work, we will only evaluate the contribution 
from the region \( 0\leq x\leq \zeta  \) in Fig.~\ref{fig:overlap}, 
in which case the meson vertex can be parametrized by the 
usual meson wavefunction. In order to estimate also the contribution 
outside this region, when \( x > \zeta \), we
would again need to know how to parametrize the nonperturbative object 
 where a quark splits into a quark and a meson. We postpone to future work 
the investigation of this problem.
\newline
\indent
In the region \( 0\leq x\leq \zeta  \) we can parametrize 
the valence quark state of a longitudinally polarised vector meson  as:
\begin{equation}
\label{eq:mesonstate}
|V_{L}\rangle \; =\; {1\over \sqrt{3}}\, \int _{0}^{1}dz\; \int \, {{d^{2}\vec{k}_{\perp }}\over {16\pi ^{3}\; \sqrt{z\bar{z}}}}\; \Psi _{V_{L}}\left( {z,\vec{k}_{\perp }}\right) \; {1\over \sqrt{2}}\, |\; q_{\uparrow }\, \bar{q}_{\downarrow }\; +\; q_{\downarrow }\, \bar{q}_{\uparrow }\; \rangle \; ,
\end{equation}
 where \( 1/\sqrt{3} \) is a color normalization factor  and 
\( q_{\uparrow \downarrow } \) and \( \bar{q}_{\uparrow \downarrow } \) are
the quark and antiquark states in Fig.~\ref{fig:overlap} with
 helicities $\lambda = +1/2$ \( (\uparrow) \) and 
$\lambda = -1/2$ \( (\downarrow) \). 
The quark states are normalized as~\cite{BrodLep}
\begin{equation}
\langle \, q_{\lambda' }(k^{'\,+},\vec{k}^{\, '}_{\perp})\, 
\arrowvert \, q_{\lambda }(k^{+},\vec{k}_{\perp }) \,\rangle 
\;=\; 2 \, k^+ \, (2 \pi)^3 \, \delta \left(k^{' \,+} - k^+ \right) \,
\delta^2 \left( \vec{k}^{\, '}_{\perp} - \vec{k}_{\perp } \right) \,
\delta_{\lambda' \lambda} \;.
\end{equation}
The light-cone valence wavefunction \( \Psi _{V_{L}} \) 
in Eq.~(\ref{eq:mesonstate}), depends upon the quark relative 
light-cone momentum fraction ($z$) 
and the quark relative transverse momentum ($\vec k_\perp$) in the meson. 
This valence wavefunction \( \Psi _{V_{L}} \) is given, in analogy
with Eq.~(\ref{eq:wfmekperp}) for the pion, 
and for a meson with $\vec p_{V \perp} = 0$, 
by the bilocal quark matrix element
at equal light-cone time \( (y^{+}=0) \) : 
\begin{equation}
\label{eq:vecmeswfkperp}
\Psi _{V_{L}}(z,\vec{k}_{\perp })\; =\; {1\over {\sqrt{6}}}\; \int dy^{-}\, e^{i\left( zp_{V}^{+}\right) y^{-}}\, \int d^{2}\vec{y}_{\perp }\, e^{-i\vec{k}_{\perp }\cdot \vec{y}_{\perp }}\; \langle 0|\bar{q}(0)\, \gamma ^{+}\, q(y)|V_{L}\left( p_{V}\right) \rangle \; {\Bigg |}_{y^{+}=0}\;.
\end{equation}
The flavor structure for different vector mesons 
(\( \rho ^{0},\rho ^{\pm },\omega ,\phi  \))
is understood implicitely in Eqs.~(\ref{eq:mesonstate},
\ref{eq:vecmeswfkperp}). \\
 \indent 
We next have to evaluate the lower blob  in the 
overlap diagram of Fig.~\ref{fig:overlap}. As we restrict ourselves to
the contribution from the region \( 0\leq x\leq \zeta  \), this lower
blob corresponds to the amplitude to find a $q \bar q$ state at the
nucleon side. In the calculation of the leading order meson
electroproduction amplitude, when using a frame where the 
initial nucleon's transversal momentum is zero, 
this amplitude is parametrized by the `meson-like' part of the NFPD's.
In the $q^+ = 0$
frame with non-zero transversal momentum $\vec q_\perp$, the relative
quark transverse momentum will appear as an argument in the 
$\vec k_\perp$-dependent NFPD. Therefore,  
in the \( 0\leq x\leq \zeta  \) region the lower blob in Fig.~\ref{fig:overlap}
is~: 
\begin{equation}
\label{eq:mesonoverlap1}
\bar{N}(p^{'})\gamma ^{+}N(p)\, \cdot \, \int _{0}^{\zeta }dx\; \int
\, {{d^{2}\vec{k}_{\perp }}\over {8\pi ^{3}}}\;  
f(x,\vec{k}_{\perp} + {x \over 2 \zeta} \vec q_\perp,\zeta ,t)\; 
{1\over 4}{1\over {\sqrt{x\, (\zeta -x)}\; p^{+}}}\; \arrowvert 
q_{\uparrow }\bar{q}_{\downarrow }\,+\, q_{\downarrow
  }\bar{q}_{\uparrow } \rangle \,.
\end{equation}
In Eq.~(\ref{eq:mesonoverlap1}), \( f \) is the 
\( \vec{k}_{\perp } \)-dependent unpolarized NFPD 
(as we are interested to calculate longitudinally polarized vector
meson electroproduction), which is related 
to the ordinary NFPD  as in Eq.~(\ref{eq:ofpdkperp}).
The term corresponding with the OFPD \( E \) is neglected here. 
Remark that in the infinite momentum limit, where one neglects the
quark transverse momentum compared with its longitudinal momentum, 
Eq.~(\ref{eq:mesonoverlap1}) yields the 
Lorentz structure \( {1\over 4}\, \left( \gamma ^{-}\right)  \)
that appears in the leading order amplitude of Eq.~(\ref{eq:qsplitting}). \\
 \indent We can now calculate the amplitude of the
 soft overlap mechanism  
by taking the matrix element of the electromagnetic
current \( J^{\mu } \) between the vector meson state 
\( |V_{L}\rangle  \) and the \( |q\bar{q}\rangle  \) state at 
the nucleon side of Eq.~(\ref{eq:mesonoverlap1}).
This yields : 
\begin{eqnarray}
&&{\mathcal{M}}_{V_{L},\, soft} \; =\; (-i\, e)\; \bar{N}(p^{'})\gamma ^{+}N(p)\nonumber \\
&&\hspace{.8cm} \times \, \int _{0}^{\zeta }dx\, \int \, {{d^{2}\vec{k}_{\perp
 }}\over {8\pi ^{3}}}\; f(x,\vec{k}_{\perp }+ {x \over 2 \zeta}
 \vec q_\perp,\zeta ,t)\; {1\over 4}{1\over {\sqrt{x\, (\zeta -x)}\;
 p^{+}}}\; \langle \, V_{L}\, \arrowvert \, J\cdot \epsilon \,
 \arrowvert \, q_{\uparrow }\bar{q}_{\downarrow }
\, +\, q_{\downarrow }\bar{q}_{\uparrow }\, \rangle \, ,
\label{eq:mesonoverlap2} 
\end{eqnarray}
 where \( \epsilon  \) is the photon polarization vector. By using the vector
meson state of Eq.~(\ref{eq:mesonstate}), Eq.~(\ref{eq:mesonoverlap2}) can
be worked out as 
\begin{eqnarray}
{\mathcal{M}}_{V_{L},\, soft} &  & \; =\; (-i\, e)\; \bar{N}(p^{'})\gamma ^{+}N(p)\nonumber \\
 &  & \times \, \int _{0}^{\zeta }dx\; \int \, {{d^{2}\vec{k}_{\perp }}\over {8\pi ^{3}}}\; f(x,\vec{k}_{\perp }+ {x \over 2 \zeta} \vec q_\perp,\zeta ,t)\; {1\over \sqrt{6}}\; \Psi _{V_{L}}\left( z={x\over \zeta },\vec{k}_{\perp }+(1-{x\over {2\zeta }})\, \vec{q}_{\perp }\right) \; \nonumber \\
 &  & \times \, {1\over {4\, x\, p^{+}}}\, \left\{ \langle \, q_{\uparrow }(xp^{+},\vec{k}_{\perp }+\vec{q}_{\perp })\, \arrowvert \, J\cdot \epsilon \, \arrowvert \, q_{\uparrow }(xp^{+},\vec{k}_{\perp })\rangle \right. \nonumber \\
 &  & \left. \hspace {1.5cm}\, +\, \langle \, q_{\downarrow }(xp^{+},\vec{k}_{\perp }+\vec{q}_{\perp })\, \arrowvert \, J\cdot \epsilon \, \arrowvert \, q_{\downarrow }(xp^{+},\vec{k}_{\perp })\rangle \right\} \; ,\label{eq:mesonoverlap3} 
\end{eqnarray}
 where the relative quark momentum arguments enter in the vector meson valence
wavefunction and in the NFPD. 
As we only give predictions in this work for the 
longitudinal electroproduction amplitude, we only keep
the good current component \( J^{+} \), i.e. \( J\cdot \epsilon _{L} \) =
\( J^{+}\cdot \epsilon _{L}^{-} \) (where \( \epsilon _{L} \) is the polarization
vector for a longitudinal photon). For the current operator \( J^{+} \), the
quark matrix elements in Eq.~(\ref{eq:mesonoverlap3}) yield : 
\begin{eqnarray}
 &  & \langle \, q_{\lambda' }(xp^{+},\vec{k}_{\perp }+\vec{q}_{\perp })\, \arrowvert \; J^{+}\; \arrowvert \, q_{\lambda }(xp^{+},\vec{k}_{\perp })\rangle \nonumber \\
 & = & \; \bar{u}(xp^{+},\vec{k}_{\perp }+\vec{q}_{\perp };\lambda'
 )\, \gamma ^{+}\, u(xp^{+},\vec{k}_{\perp };\lambda )\; =\; 
\left( 2 \,x\, p^{+}\right) \, \delta _{\lambda \, \lambda' }\; .
\label{eq:mesonoverlap4} 
\end{eqnarray}
 With Eq.~(\ref{eq:mesonoverlap4}), the soft overlap amplitude for production
of a longitudinally polarized vector meson by a longitudinal photon
can  finally be  written as : 
\begin{eqnarray}
&{\mathcal{M}}^{L}_{V_{L}\, ,soft}&\; = \, (-i\, e)\; \epsilon _{L}^{-}\; \bar{N}(p^{'})\gamma ^{+}N(p)\; \nonumber \\
&& \times \, {1\over \sqrt{6}}\, \int _{0}^{\zeta }dx\, \int 
 {{d^{2}\vec{k}_{\perp }}\over {8\pi ^{3}}}\; 
f(x,\vec{k}_{\perp }+{{x}\over {2\zeta }}\vec{q}_{\perp },\zeta
 ,t)\; \Psi _{V_{L}}\left( z={x\over {\zeta }},\, \vec{k}_{\perp } + (1-{{x}\over {2\zeta }})\, \vec{q}_{\perp } \right) \,.\label{eq:mesonoverlap5} 
\end{eqnarray}
In the frame \( q^{+}=0 \) considered here, 
\( q^{\mu } = (0,\vec{q}_{\perp },0) \) and 
\( \epsilon _{L}^{\mu }=(1,0,0,0) \),
 which yields \( \epsilon _{L}^{-}=1/\sqrt{2} \). \\
 \indent To evaluate the soft overlap formula Eq.~(\ref{eq:mesonoverlap5})
for meson electroproduction, we need to model the \( \vec{k}_{\perp } \)-dependent
NFPD for \( x\leq \zeta  \). We do that by  analogy 
with the gaussian ansatz of Eq.~(\ref{eq:kperpwf})
for the soft part of the meson wavefunction, that is: 
\begin{equation}
\label{eq:nfpdkperp}
f(x,\vec k^{\, '}_\perp,\zeta ,t)\; =\; f(x,\zeta ,t)\;
{{4\pi^{2}}\over {\sigma _{N}\, {x\over \zeta }\, \left( 1-{x\over
        \zeta}\right) }}\; \exp \left\{ -\, {{\left(\vec k^{\,
          '}_\perp \right) ^{2}}\over {2\, \sigma _{N}\, {x\over \zeta
        }\, \left( 1-{x\over \zeta }\right) }}\; \right\} \,, 
\;\;\; x \leq \zeta \,,
\end{equation}
where \( \sigma _{N} \) is now related
to the average squared transverse momentum of the quarks in the meson state
at the nucleon side. The ordinary unpolarized NFPD \( f (x, \zeta,
t)\) entering in
Eq.~(\ref{eq:nfpdkperp}) is modelled as discussed before through its definition
Eq.~(\ref{eq:double}) in terms of the double distribution and through the ansatz
of Eq.~(\ref{eq:ddmodel}) for the double distributions. With the ansatz of
Eq.~(\ref{eq:nfpdkperp}) for the NFPD and Eq.~(\ref{eq:kperpwf}) for the soft
meson wavefunction (with parameter \( \sigma _{V_{L}} \) for the longitudinally
polarized vector meson), the integral over transverse momentum in the soft overlap
formula of Eq.~(\ref{eq:mesonoverlap5}) can be worked out analytically as :
\begin{eqnarray}
{\mathcal{M}}^{L}_{V_{L}\; ,soft}\; = &  & \; (-ie)\, {1\over \sqrt{2}}\; \bar{N}(p^{'})\gamma ^{+}N(p)\; \nonumber \\
 &  & \times \; f_{V_{L}}\; {{4\pi ^{2}}\over {\sigma _{N}+\sigma
 _{V_{L}}}}\; \int _{0}^{\zeta }dx\; f^{p}\left( {x,\zeta ,t}\right)
 \; \exp \left\{ -\, {{Q^{2}}\over {2(\sigma _{N}+\sigma _{V_{L}})}}\,
 {{(1 - x/\zeta)}\over {x/\zeta}}\right\} \;,
\label{eq:softgaussian}
\end{eqnarray}
 where \( f_{V_{L}} \) appears in the asymptotic meson distribution amplitude
as given before in Eq.~(\ref{eq:vectormda}). 
Remark that Eq.~(\ref{eq:softgaussian}) is a generalization of the
overlap formula of Eq.~(\ref{eq:piemffsoft2}) for the pion FF using a gaussian
ansatz for the transverse momentum dependence. Note also that 
the overlap amplitude is purely real. In contrast, the leading order
meson electroproduction amplitude, 
including corrections for the intrinsic transverse momentum dependence, 
is complex and is furthermore dominated by its imaginary part. 
Therefore, one can expect to find qualitative differences with respect to the
case of the form factor. 
\newline
\indent
In the actual calculations of this work, 
we use as first guess for \( \sigma _{V_{L}} \) and \( \sigma _{N} \),
the same value as found before (Eq.~(\ref{eq:sigmapi})) for the pion :
\( \sigma _{V_{L}}\approx \sigma _{N}\approx 0.67 \)
GeV\( ^{2} \). In principle, the parameter \( \sigma _{N} \) could be
fixed independently however, if data are available.

\section{Results and discussion}

\label{sec:sec5}

\begin{figure}[ht]
\epsfxsize=11 cm
\epsfysize=6. cm
\centerline{\epsffile{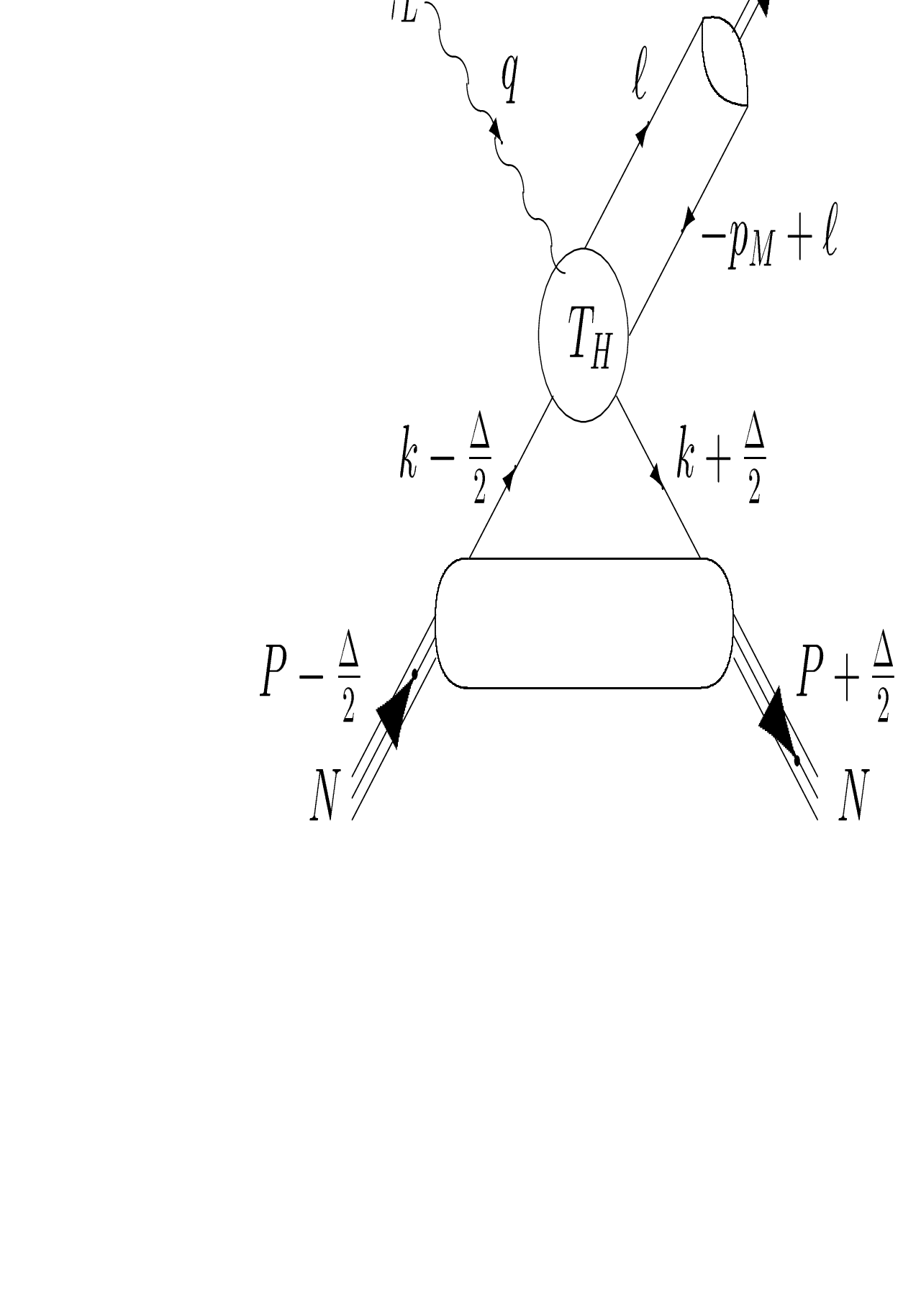}}
\vspace{-1.5cm}
\caption[]{\small Factorization for the leading order hard meson 
electroproduction amplitude.}
\label{fig:factmeson}
\end{figure}

\begin{figure}[h]
\vspace{-2cm}
\hspace{.5cm}
\epsfxsize=15 cm
\epsfysize=22 cm
\centerline{\epsffile{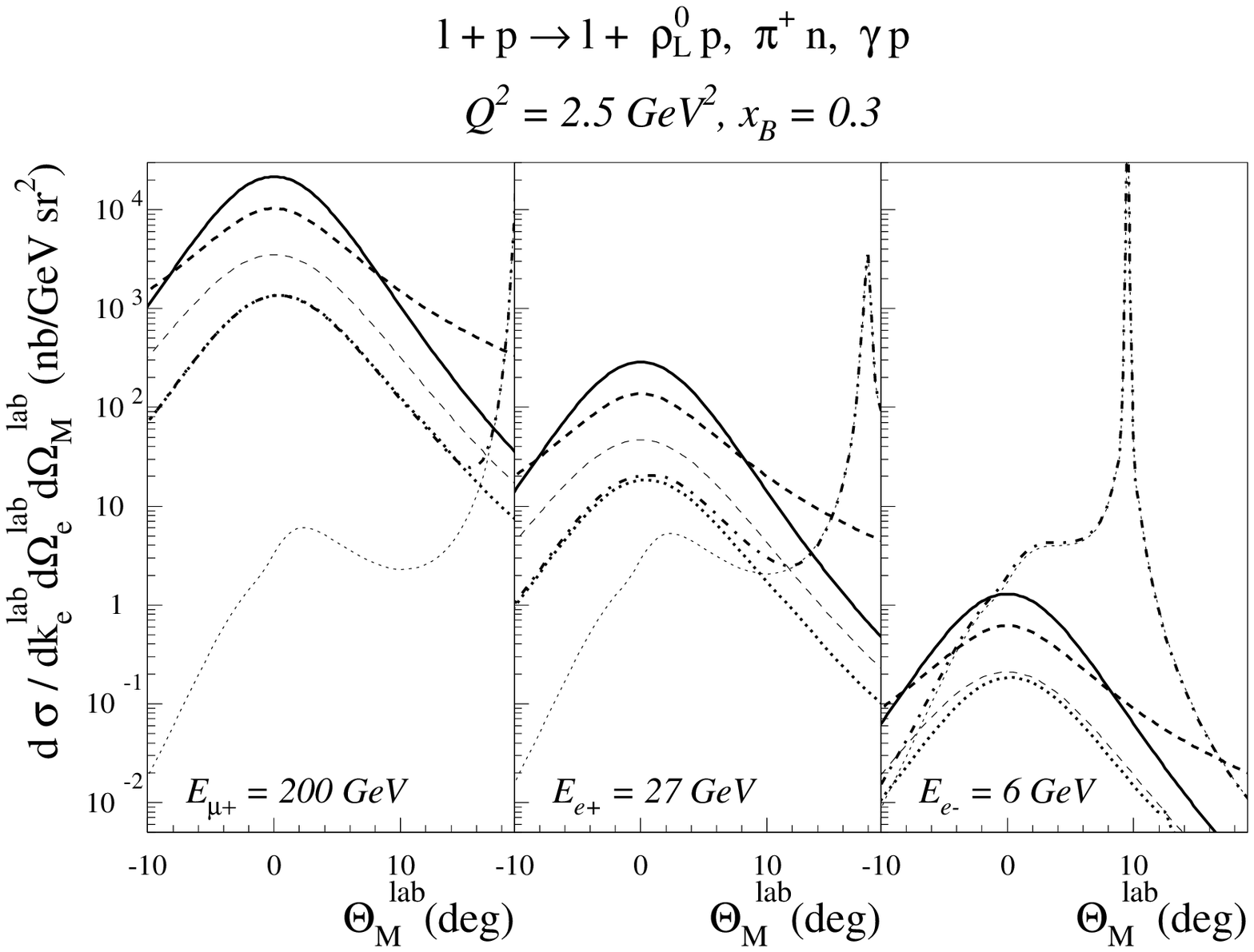}}
\vspace{-5.cm}
\caption[]{\small Comparison between the angular dependence of the 
leading order predictions for 
the $\rho^0_L$ (full lines), $\pi^+$ (thick, upper dashed lines) 
leptoproduction and DVCS (thick dotted lines) 
in-plane cross sections at $Q^2$ = 2.5 GeV$^2$, $x_B$ = 0.3 
and for different beam energies : 
$E_{\mu^+}$ = 200 GeV (COMPASS), $E_{e^+}$ = 27 GeV (HERMES), 
$E_{e^-}$ = 6 GeV (JLab). The BH (thin dotted lines) 
and total $\gamma$ (dashed-dotted lines) cross sections are also
shown. For the $\pi^+$, the result excluding the pion pole is also
shown (thin, lower dashed lines).}
\label{fig:diffkin}
\end{figure}

\begin{figure}[h]
\vspace{-2cm}
\hspace{.5cm}
\epsfxsize=15 cm
\epsfysize=20 cm
\centerline{\epsffile{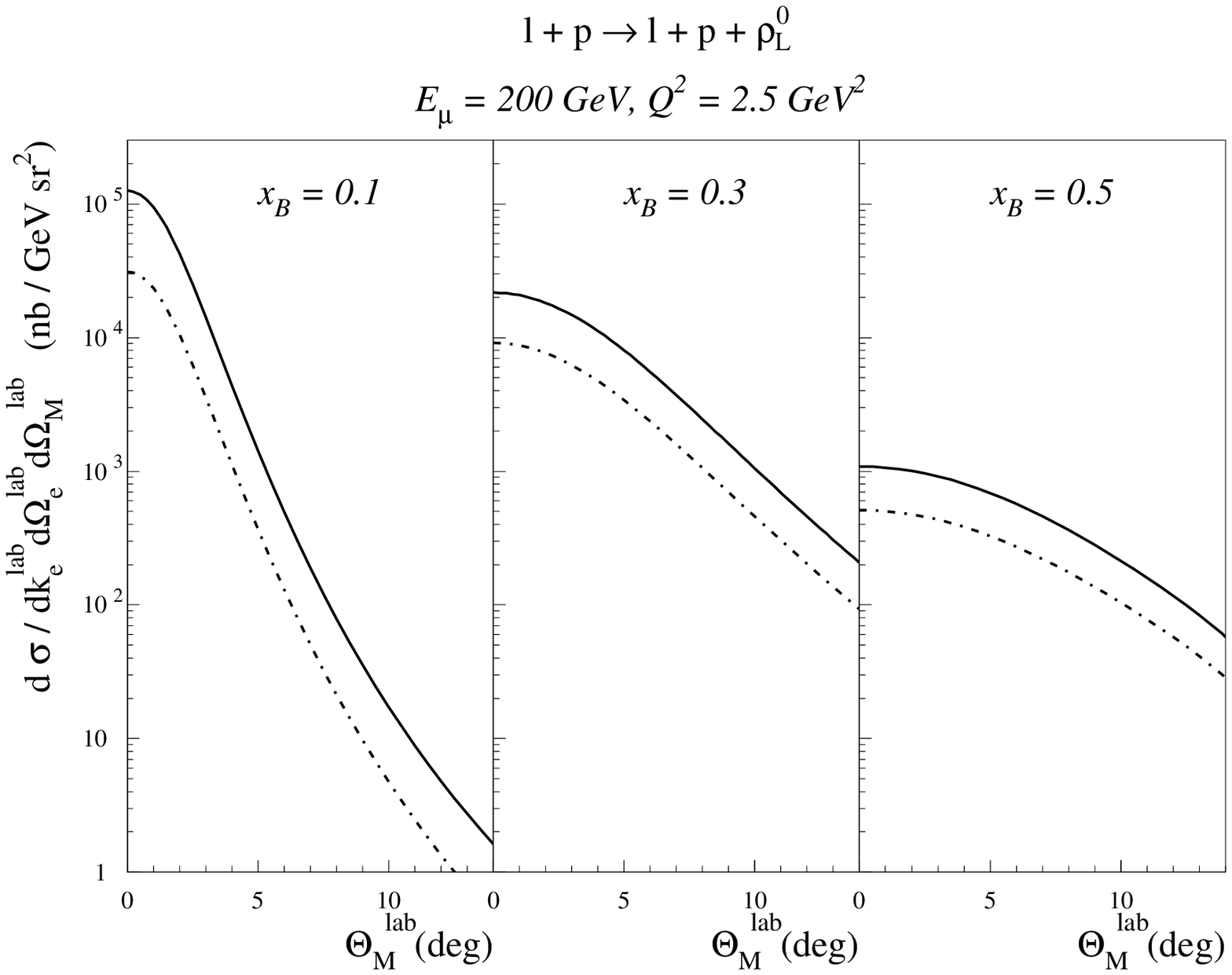}}
\vspace{-4.5cm}
\caption[]{\small Angular dependence of the 
leading order amplitudes for $\rho^0_L$ leptoproduction 
at $E_\mu$ = 200 GeV, $Q^2$ = 2.5 GeV$^2$,  
and for different values of $x_B$. The prediction with a $\xi$-dependent
ansatz for the OFPD's (full lines) are compared with a
$\xi$-independent ansatz for the OFPD's (dashed-dotted lines).}
\label{fig:comparexidep}
\end{figure}

\begin{figure}[h]
\vspace{4.5cm}
\epsfxsize=12 cm
\epsfysize=16 cm
\centerline{\epsffile{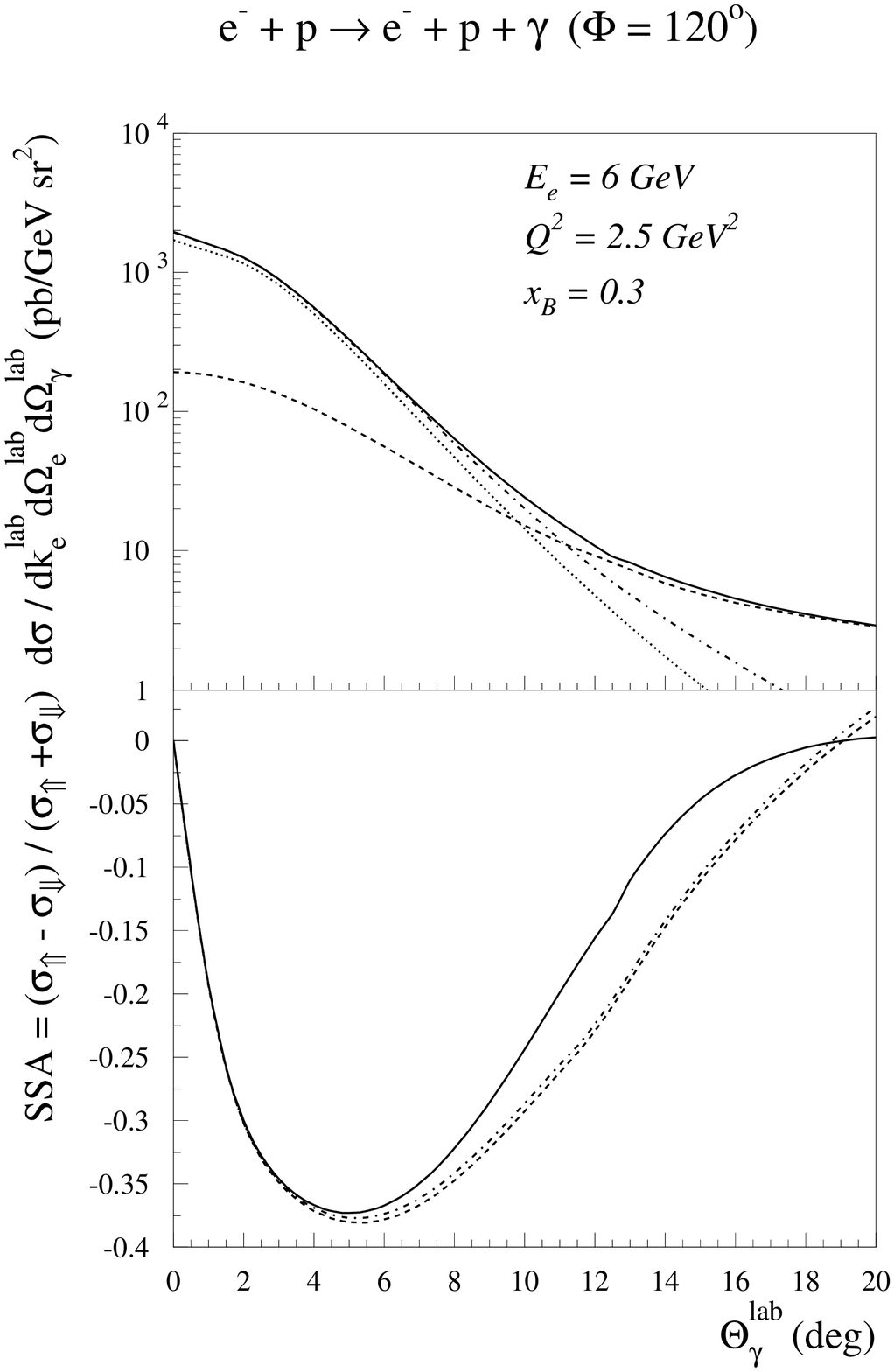}}
\vspace{1.5cm}
\caption[]{\small DVCS at JLab for out-of-plane kinematics ($\phi = 120^o$). 
Upper part shows the differential cross section : 
DVCS including the $\pi^0$ pole contribution (dashed line), BH (dotted
line), total $\gamma$ excluding the $\pi^0$ pole contribution
(dashed-dotted line) and total $\gamma$ including the $\pi^0$ pole
contribution (full line).  
In the lower part, the electron single spin asymmetry is shown 
for the L.O. DVCS amplitude (calculated in radiative gauge) excluding the 
$\pi^0$ pole (dashed line) and including the $\pi^0$ pole (full
line). We also show the result for the gauge invariant 
DVCS amplitude, excluding the $\pi^0$ pole (dashed-dotted line).}
\label{fig:dvcsasymm}
\end{figure}

\begin{figure}[h]
\epsfxsize=15 cm
\epsfysize=20 cm
\centerline{\epsffile{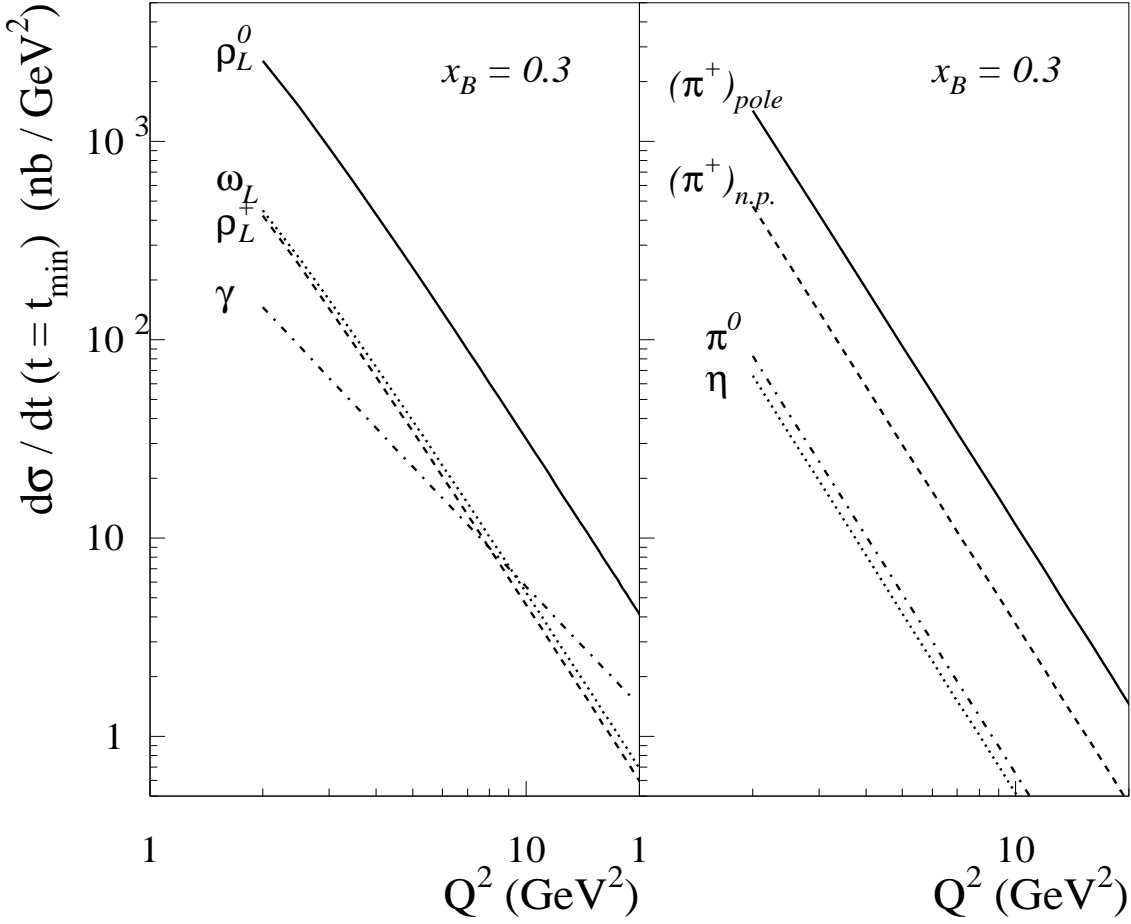}}
\vspace{-3.5cm}
\caption[]{\small Scaling behavior of the leading order predictions
  for the forward differential electroproduction cross section 
$d \sigma_L/d t$ on the proton, for vector mesons (left panel) 
and pseudoscalar mesons (right panel) using a coupling constant frozen
at a scale 1 GeV$^2$. 
For the $\pi^+$ channel, the pion pole contribution
(full line, $(\pi^+)_{pole}$) is shown separately from the 
$\tilde H$ contribution (dashed line, $(\pi^+)_{n.p.}$). 
Also shown is the scaling behavior
of the forward transverse cross section $d \sigma_T/d t$ 
for the leading order DVCS cross section (dashed-dotted line in left panel). }
\label{fig:scaling}
\end{figure}

\begin{figure}[h]
\epsfxsize=15 cm
\epsfysize=20 cm
\centerline{\epsffile{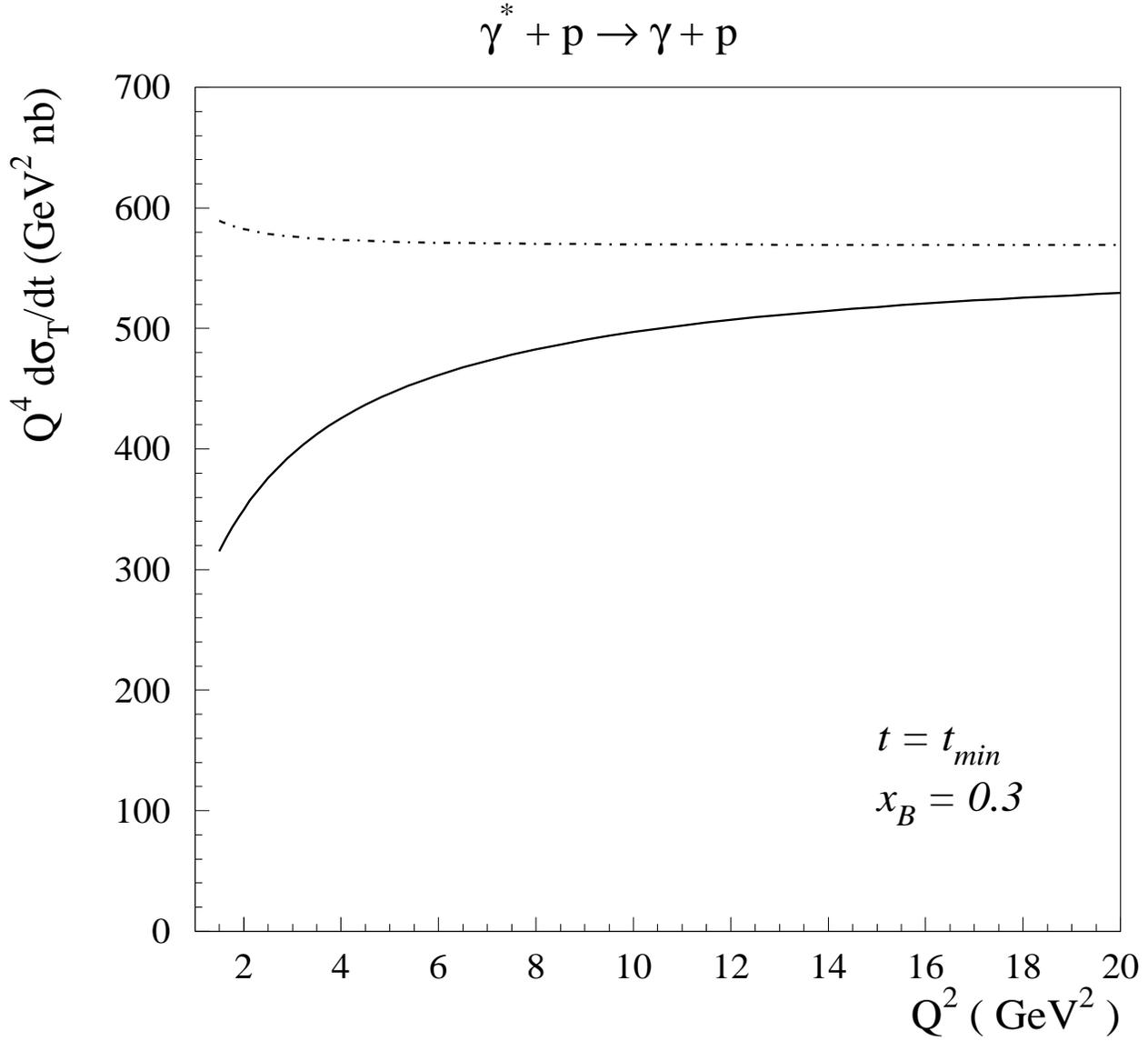}}
\vspace{-2.cm}
\caption[]{\small DVCS transverse cross section 
$d \sigma_T/d t$ (multiplied with the scaling factor $Q^4$) 
in the forward direction ($t = t_{min}$). The 
leading order result is given by the dashed-dotted line  
and the result including the
intrinsic transverse momentum dependence is given by the full line.}
\label{fig:dvcskperp}
\end{figure}

\begin{figure}[h]
\epsfysize=18 cm
\centerline{\epsffile{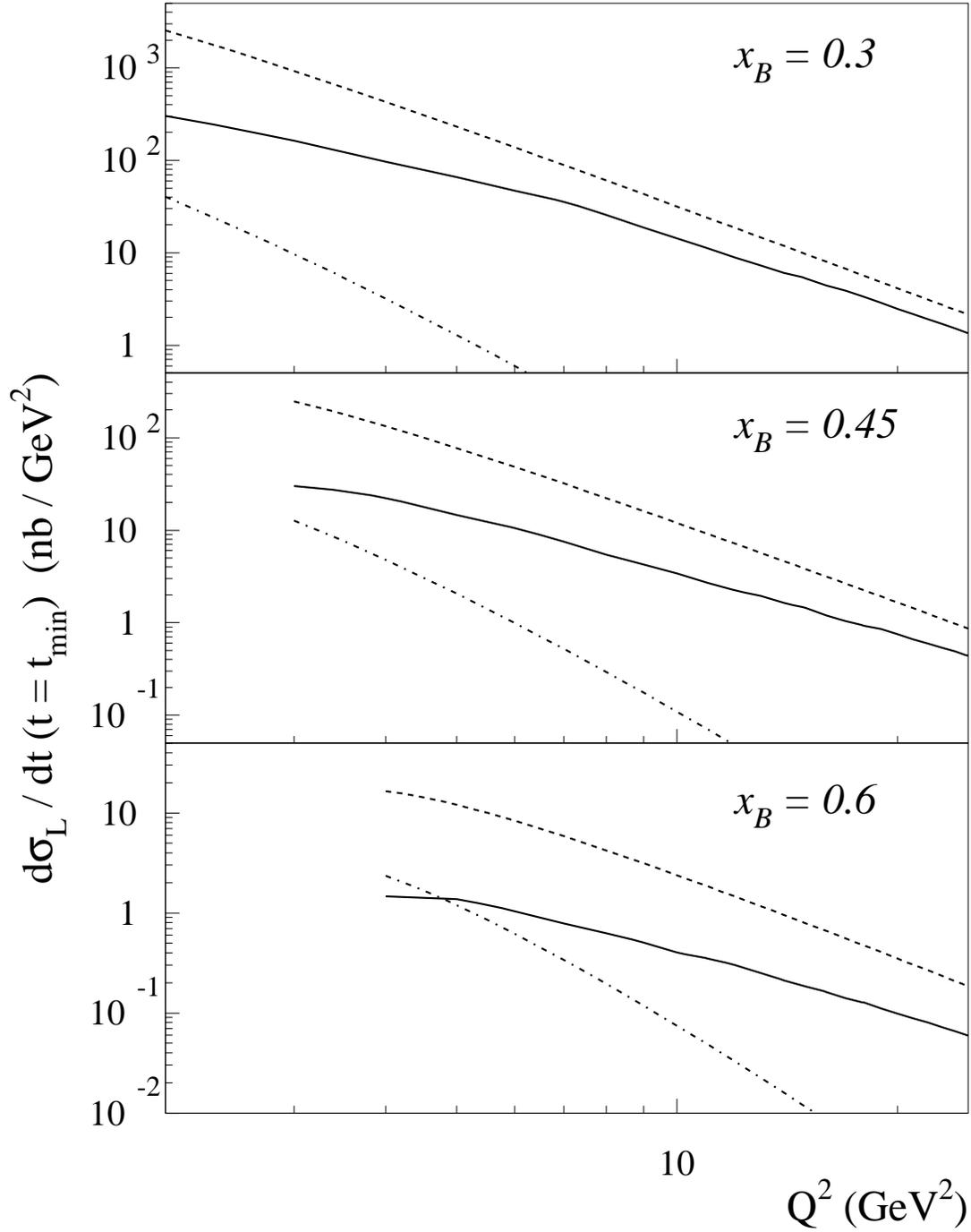}}
\vspace{1.cm}
\caption[]{\small Longitudinal differential cross section 
$d \sigma_L/d t$ in the forward direction ($t = t_{min}$) for 
for $\gamma^*_L p \longrightarrow \rho^0_L p $ . The 
leading order result is given by the dashed line  
and the result including the intrinsic transverse momentum dependence 
is given by the full line.
The calculation of the soft overlap contribution is given by the
dashed-dotted line.}
\label{fig:mesonkperp}
\end{figure}

\begin{figure}[h]
\epsfxsize=14 cm
\epsfysize=20 cm
\centerline{\epsffile{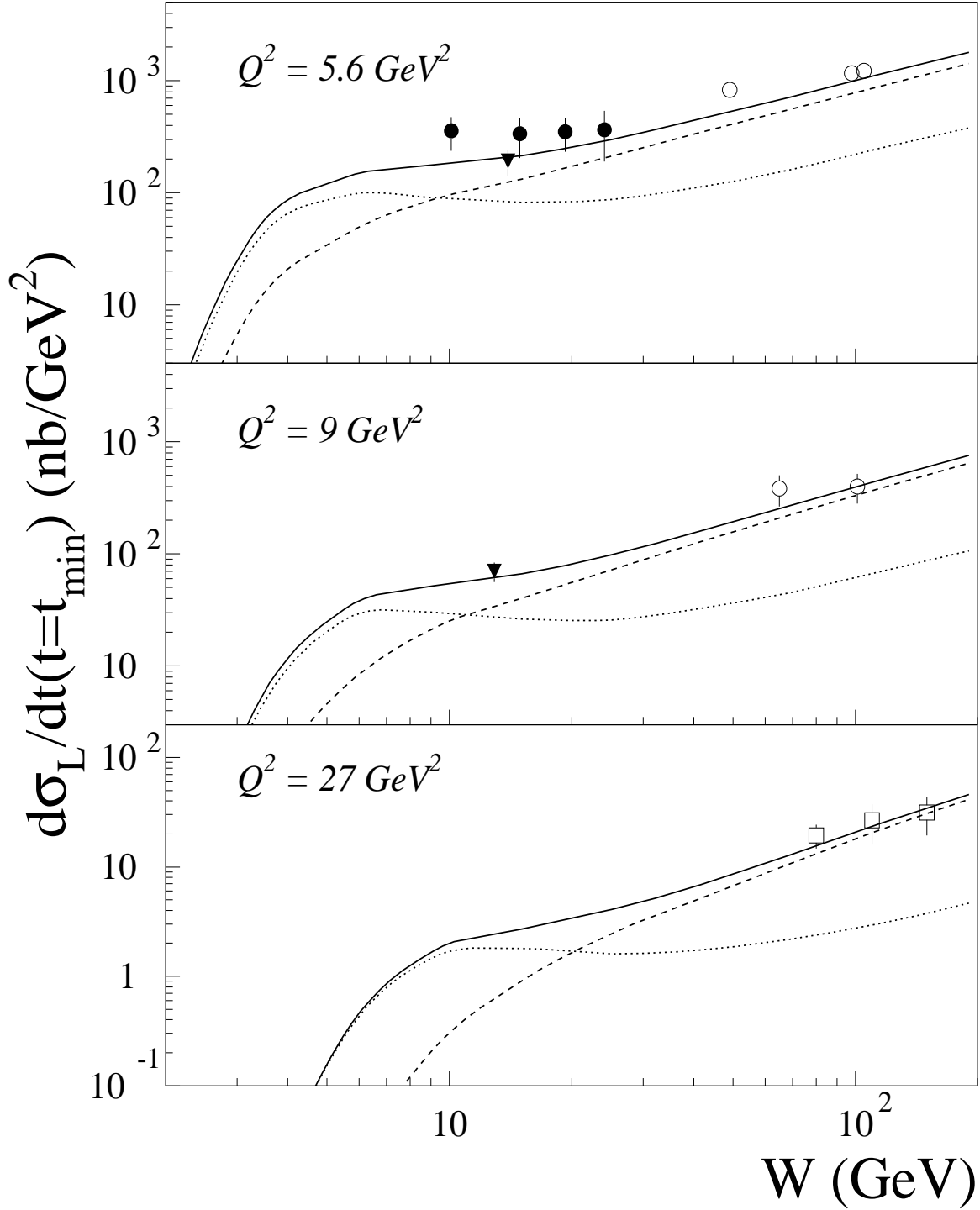}}
\vspace{0cm}
\caption[]{\small Longitudinal forward differential cross section for 
$\rho^0_L$ electroproduction. Calculations compare the quark exchange
mechanism (dotted lines) with the two-gluon
exchange mechanism (dashed lines) and the sum of both (full
lines). Both calculations include the corrections due to intrinsic
transverse momentum dependence as described in the text. 
The data are from NMC (triangles) \cite{Arn94}, 
E665 (solid circles) \cite{Ada97}, 
ZEUS 93 (open circles) \cite{Derrick95} 
and ZEUS 95 (open squares) \cite{Breitweg98}.}
\label{fig:rhotot2}
\end{figure}

\begin{figure}[h]
\epsfxsize=13 cm
\epsfysize=16 cm
\centerline{\epsffile{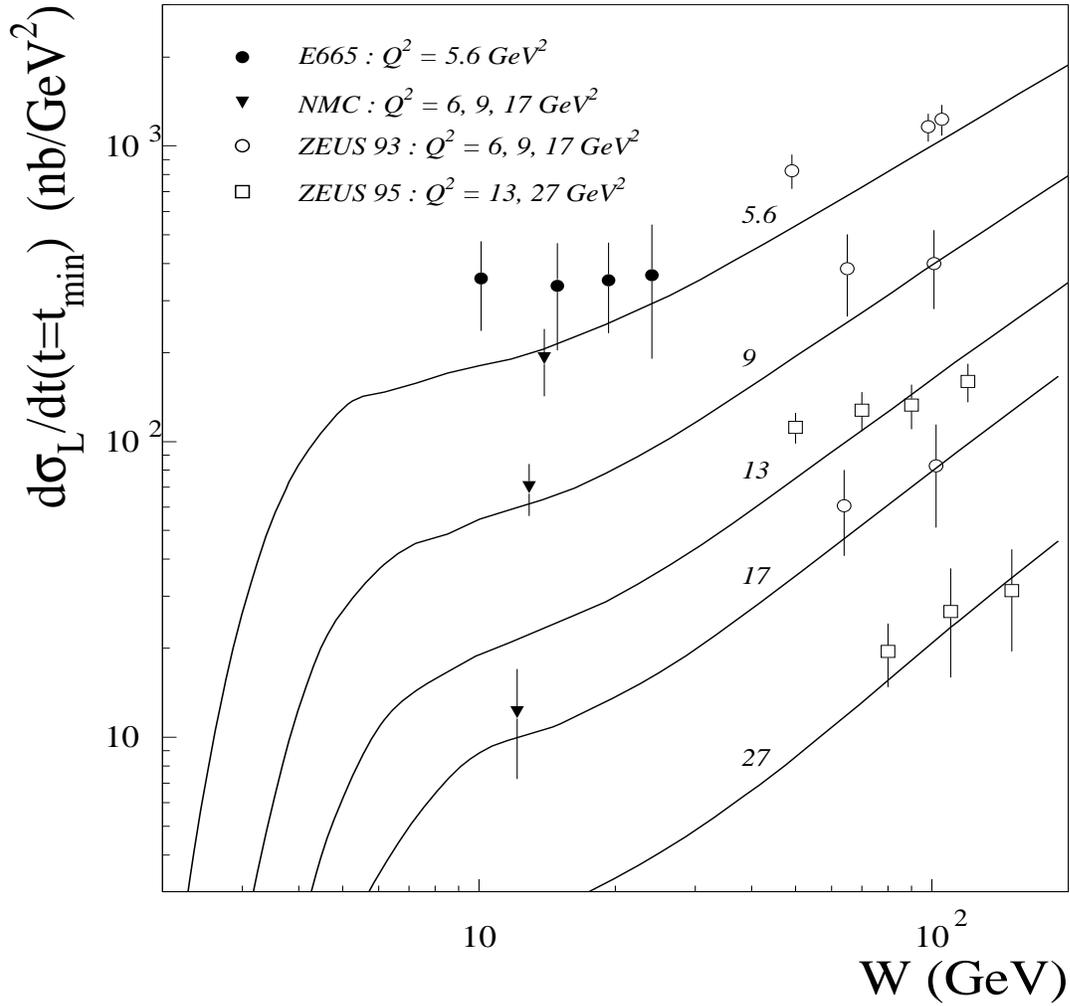}}
\vspace{0cm}
\caption[]{\small Longitudinal forward differential cross section for 
$\rho^0_L$ electroproduction. Calculations show the sum of the quark exchange
and two-gluon exchange mechanisms for different values of $Q^2$ 
(in GeV$^2$) as indicated on the curves. 
The data are from NMC \cite{Arn94}, E665 \cite{Ada97}, 
ZEUS 93 \cite{Derrick95} and ZEUS 95 \cite{Breitweg98}.}
\label{fig:rhotot1}
\end{figure}

\begin{figure}[h]
\epsfxsize=14 cm
\epsfysize=18 cm
\centerline{\epsffile{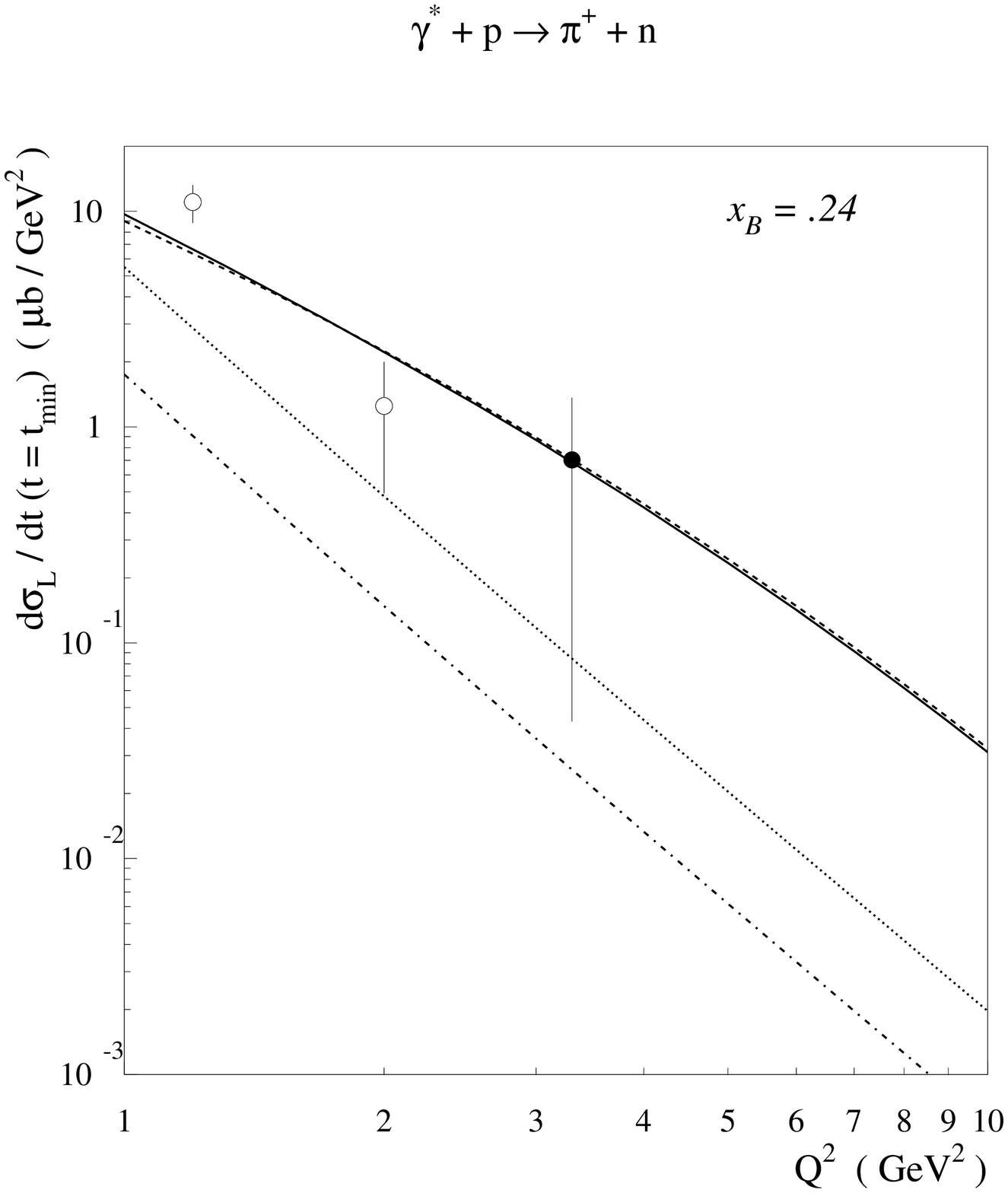}}
\vspace{-.5cm}
\caption[]{\small $Q^2$-dependence 
for the longitudinal forward differential electroproduction cross section 
$d \sigma_L/d t$ for $\gamma^* p \rightarrow \pi^+ n$ at $x_B = 0.24$.
The leading order pion pole contribution
(dotted line) and $\tilde H$ contribution (dashed-dotted line) 
are shown using a running coupling constant with $\Lambda_{QCD}$ = 0.2 GeV. 
Also shown (dashed line) 
is the calculation including the power corrections to the pion
pole contribution (intrinsic transverse momentum dependence 
and overlap contribution to the pion FF), 
as well as the sum of the pion pole and $\tilde H$
contribution including the power corrections (full line). 
The data are from Ref. \cite{Bebek78} : the points
at $Q^2$ = 1.2 and 2.0 GeV$^2$ correspond with $x_B \approx 0.24$,
whereas the point at $Q^2$ = 3.3 GeV$^2$ corresponds with $x_B \approx 0.35$.}
\label{fig:ep_epipn_qdep}
\end{figure}

In this section we present results for several observables for meson electroproduction
and DVCS. We will show the leading order  predictions and  the effects of the  power corrections 
 discussed in the previous section.
In our previous work \cite{ourprl,vcsrev}, we gave predictions for the leading
order amplitudes using a simple \( \xi  \)-independent factorized ansatz for
the OFPD's. Therefore, the \( \xi  \)-dependent ansatz used here allows to study quantitatively the
dependence on the skewedness in the OFPD's. 
\newline
\indent  
We first show in Fig.~\ref{fig:diffkin} the angular and energy dependence
of the leading order meson electroproduction and DVCS 
differential cross sections in
kinematics accessible at JLab, HERMES and COMPASS. For the electroproduction
of photons, one cannot disentangle the DVCS from the Bethe-Heitler (BH) process
where the photon is radiated from an electron line. In our calculations we add
coherently the BH and DVCS amplitudes. From a phenomenological point of view,
it is clear that the best situation to study DVCS occurs when the BH process
is negligible. For fixed \( Q^{2} \) and \( x_{B} \), the only way to favor
the DVCS over the BH is to increase the virtual photon flux and this amounts
to increase the beam energy. This is seen on Fig.~\ref{fig:diffkin} where we
compare separately the BH, the DVCS and the coherent cross section for different
beam energies. According to our estimate, the unpolarized \( (l,l^{'}\gamma ) \)
cross section in the forward region is dominated by the BH process in the few
GeV region. To get a clear dominance of the DVCS process one needs a beam energy
in the 100 GeV range. In the valence region (\( x_{B}\approx 0.3 \)) and for
a \( Q^{2} \) value of \( Q^{2}\approx  \) 2.5 GeV\( ^{2} \), one sees that
at \( E_{\mu } \) = 200 GeV, the DVCS cross section is about two orders of
magnitude larger than the BH in the forward direction. By going to smaller \( x_{B} \),
the BH cross section increases. This is due to the fact that in the BH process
the exchanged photon has 4-momentum \( (q-q') \) which gives a \( 1/t \) behaviour
to the amplitude. The value of \( t \) in the forward direction (\( t_{min} \))
becomes very small for small \( x_{B} \) values. The resulting sharp rise of
the BH process in the forward direction at small \( x_{B} \) puts therefore
a limit on the region where the DVCS can be studied experimentally. At COMPASS
kinematics, \( x_{B}\simeq 0.1 \) seems to be the lower limit. Although the
BH is not a limiting factor at high \( x_{B} \), one cannot go too close to
\( x_{B}=1 \) in order to stay well above the resonance region. \\
\indent In Fig.~\ref{fig:diffkin}, the leading order predictions for the
\( \rho ^{0}_{L} \) and \( \pi ^{+} \) fivefold differential
electroproduction cross sections are also compared in the same kinematics and
one again observes that the virtual photon flux boosts the cross sections as
one goes to higher beam energies. In the valence region (\( x_{B}\approx 0.3 \)),
the \( \rho ^{0}_{L} \) cross section, is about an order of magnitude larger
than the DVCS cross section. For the $\pi^+$ electroproduction cross
section, one remarks the prominent contribution of the charged pion
pole (OFPD $\tilde E$), which gives a flatter $t$-dependence than the
contribution of the OFPD $\tilde H$, because the $\tilde E$
contribution comes with a momentum transfer in the amplitude. 
\newline
\indent
In Fig.~\ref{fig:comparexidep} we compare in COMPASS kinematics, 
the leading order
predictions for $\rho^0_L$ electroproduction using the $\xi$-dependent
ansatz and a $\xi$-independent ansatz which we used in our previous
work \cite{ourprl,vcsrev}. 
Both calculations in Fig.~\ref{fig:comparexidep}
use the MRST98 parton distributions \cite{MRST98} as input 
and thus the OFPD's in both cases reduce to the same quark
distributions in the forward limit and satisfy the first sum rule. 
One sees that the cross sections differ by a factor of two, 
which indicates that the skewedness of the OFPD's a priori cannot be
neglected in the analysis.
\newline
\indent 
Although it is clear from Fig.~\ref{fig:diffkin} that a high energy
beam such as planned at COMPASS is preferable, one can try to undertake a preliminary
study of the hard electroproduction reactions using the existing facilities
such as HERMES or JLab, despite their lower energy. Concerning JLab, the high
luminosity available compensates for the low cross section, and its good energy
and angular resolution permits to identify in a clean way
exclusive reactions. To make a preliminary exploration of the reaction
mechanisms in the few GeV regions and to test the onset of the
scaling, the measurement
of the \( \rho ^{0}_{L} \) leptoproduction through its decay into charged pions
seems the easiest from the experimental point of view as the count rates are
the highest. An experiment to explore the \( \rho ^{0}_{L} \) electroproduction
at JLab at 6 GeV with the CLAS detector has been proposed \cite{propo98107}.
\newline
\indent 
For the \( \gamma  \) leptoproduction at low energies, we suggest
that an exploration of DVCS might be possible if the beam is 
polarized.
The electron single spin asymmetry (SSA) does not vanish out of plane 
due to the interference between the purely real BH process and the imaginary
part of the DVCS amplitude. Therefore, even if the cross section is dominated
by the BH process, the SSA is \textit{linear} in the OFPD's. To illustrate the
point, we show on Fig.~\ref{fig:dvcsasymm} the unpolarized cross section for
a 6 GeV beam and the SSA at an azimuthal angle \( \phi  \) = 120\( ^{0} \).
When the angle between the real and virtual photons is in the \( 0^{0} \) -
\( 5^{0} \) region, a rather large asymmetry is predicted, even though
the cross section is dominated by the BH process. 
We also show on Fig.~\ref{fig:dvcsasymm}
the effect of the \( \pi ^{0} \) pole in the OFPD \( \tilde{E} \) to the DVCS.
One sees that at small angles (small \( t \)), the effect of the \( \pi ^{0} \)
pole is quite modest and increases at larger angles.
\newline
\indent
We furthermore illustrate the effect of the gauge restoring term 
for the DVCS amplitude in Eq.~(\ref{eq:dvcsgaugeinv}).
We have plotted in Fig.~\ref{fig:dvcsasymm} the SSA both for the
gauge invariant and non gauge invariant amplitudes. For the non gauge invariant
amplitude of Eq.~(\ref{eq:dvcsampl}), the SSA is shown both in the radiative
gauge and in the Feynman gauge. As expected, all predictions are identical at
small angle. At larger angles the gauge dependence clearly shows up, especially
in the Feynman gauge. \\
 \indent As the SSA accesses the imaginary part of the DVCS amplitude, the real
part of the BH-VCS interference can be accessed by reversing the charge of the
lepton beam since this changes the relative sign of the BH and DVCS amplitudes.
We have given in Ref.~\cite{vcsrev} an estimate for the e\( ^{+} \)e\( ^{-} \)
asymmetry at 27 GeV, which yields a comfortable asymmetry in the small angle
region. This may offer an interesting opportunity for HERMES, although the
experimental (e\(
^{+} \)e\( ^{-} \)) subtraction
might be delicate to perform. 
\newline
\indent 
Before considering the extraction of the OFPD's from electroproduction
data, it is compulsory to demonstrate that the scaling regime has been reached.
In Fig.~\ref{fig:scaling}, we show the forward longitudinal 
electroproduction cross sections 
as a function of \( Q^{2} \) and compare the L.O. predictions
for different mesons. The leading order amplitude 
for longitudinal electroproduction of mesons was
seen to behave as \( 1/Q \). Therefore, the
leading order longitudinal cross section \( d\sigma _{L}/dt \) 
for meson electroproduction
behaves as \( 1/Q^{6} \). In Fig.~\ref{fig:scaling}, we give all predictions
with a coupling constant frozen at a scale 1 GeV\( ^{2} \)
as explained in section \ref{sec:kperpmeson}. 
When using a running coupling evaluated at the scale \( Q^{2} \), 
one probably underestimates the cross section in the range 
$Q^2 \approx$ 1 - 10 GeV$^2$, as the average gluon
virtuality in the L.O. meson electroproduction amplitudes is considerably less
than \( Q^{2} \). This is similar to what  was  observed for the pion form factor. The
effect of the running of the coupling will be discussed further on when we also
include the intrinsic transverse momentum dependence. 
This will allow us to adapt the renormalization scale entering the
running coupling to the gluon virtuality, as discussed in section \ref{sec:kperpmeson}.
\newline
\indent 
By comparing the different vector meson channels in Fig.~\ref{fig:scaling},
one sees that the \( \rho ^{0}_{L} \) channel yields the largest cross section.
The \( \omega _{L} \) channel in the valence region (\( x_{B}\approx  \) 0.3)
is about a factor of 5 smaller than the \( \rho ^{0}_{L} \) channel, which
is to be compared with the ratio at small \( x_{B} \) (in the diffractive regime
) where \( \rho ^{0} \) : \( \omega  \) = 9 : 1. The \( \rho _{L}^{+} \)
channel, which is sensitive to the isovector combination of the unpolarized
OFPD's, yields a cross section comparable to the \( \omega _{L} \)
channel. The \( \rho _{L}^{+} \) channel is interesting as there is no
competing diffractive contribution, 
and therefore allows to test directly the quark OFPD's.
The three vector meson channels (\( \rho _{L}^{0} \), \( \rho _{L}^{+} \),
\( \omega _{L} \)) are highly complementary in order to perform a
flavor separation of the unpolarized OFPD's \( H^{u} \) and \( H^{d} \). 
\newline
\indent
For the pseudoscalar mesons which involve 
the polarized OFPD's, one again remarks in Fig.~\ref{fig:scaling} 
the prominent contribution 
of the charged pion pole to the \( \pi ^{+} \) cross section. 
For the  contribution proportional to the OFPD \( \tilde{H} \), 
it is also seen that the \( \pi ^{0} \)
channel is about a factor of 5 below the \( \pi ^{+} \) channel due to
isospin factors. In the \( \pi ^{0} \) channel,
the \( u \)- and \( d \)-quark polarized OFPD's enter with the same sign,
whereas in the \( \pi ^{+} \) channel, they enter with opposite signs. As the
polarized OFPD's are constructed here from the corresponding polarized parton
distributions, the difference of our predictions for the \( \pi ^{0} \) and
\( \pi ^{+} \) channels results from the fact that the polarized \( d \)-quark
distribution is opposite in sign to the polarized \( u \)-quark
distribution. For the $\eta$ channel, the ansatz for the OFPD $\tilde
H$ based on the polarized quark distributions yields a prediction comparable
to the $\pi^0$ cross section. 
\newline
\indent 
For the DVCS, the leading order amplitude is constant in \( Q \) and
is predominantly transverse. Therefore, the L.O. DVCS transverse cross section
\( d\sigma _{T}/dt \) shows a \( 1/Q^{4} \) behavior 
in Fig.~\ref{fig:scaling}. To test this scaling
behavior, one needs of course a kinematical situation where the DVCS dominates
over the BH. As the L.O. DVCS amplitude does involve hard gluon
exchange as for meson electroproduction, it is a rather clean observable 
 to study the onset of the scaling in \( Q^{2} \).
In Fig.~\ref{fig:dvcskperp}, we compare the L.O. DVCS transverse cross section
(multiplied with the scaling factor \( Q^{4} \)) with the result including
the transverse momentum dependence in the handbag diagrams as described in section
\ref{sec:dvcskperp}. One observes the onset of the scaling as \( Q^{2} \)
runs through the range 2 - 10 GeV\( ^{2} \), similar to what was also observed
for the \( \pi ^{0}\gamma ^{*}\gamma  \) FF, which is also described at leading
order by handbag type diagrams (see Fig.~(\ref{fig:piogaga})). It will
therefore be interesting to measure electroproduction reactions in the range 
\( Q^{2}\approx  \) 2 - 10 GeV\( ^{2} \) to study the onset of this scaling. 
In addition, the preasymptotic effects (in the valence region) can teach us 
about the quark's intrinsic transverse momentum dependence. 
\newline
\indent 
In Fig.~\ref{fig:mesonkperp}, we show how the power corrections
modify the L.O. prediction for the longitudinal \( \rho _{L}^{0} \) 
electroproduction. One sees that the inclusion of the intrinsic 
transverse momentum dependence leads to
an  appreciable reduction of the cross section at the lower \( Q^{2} \) values,
before the scaling regime is reached. Therefore, the question arises as to how
important are the other competing mechanisms in this lower/intermediate \( Q^{2} \)
region. We show in Fig.~\ref{fig:mesonkperp} the estimate of the soft overlap
mechanism of Fig.~\ref{fig:overlap}, where the meson is produced without invoking
a gluon exchange. As discussed in section \ref{sec:overlap}, in this work we
are only able to estimate the soft overlap contribution from the region \( |x|\leq \xi  \)
. To estimate the contribution
which is neglected (\( |x|\geq \xi  \)), we show predictions
for different values of \( x_{B} \), as for larger values of \( x_{B} \),
the region which is neglected becomes smaller (note on Fig.~\ref{fig:mesonkperp}
that at larger \( x_{B} \), the minimal \( Q^{2} \) value which is kinematically
possible in order to be above threshold, increases). One sees from Fig.~\ref{fig:mesonkperp}
that the overlap contribution to the longitudinal electroproduction amplitude
drops approximately as 1/\( Q^{8} \), which is indeed the expected result at
large \( Q^{2} \) \cite{Collins97}. At \( x_{B} \) = 0.3 and \( Q^{2}\approx  \)
3 GeV\( ^{2} \), our estimate for the soft overlap contribution is already
more than a decade below the hard electroproduction amplitude including the
transverse momentum dependence. At \( x_{B} \) = 0.6, where the approximation
that we make in the calculation of the soft overlap contribution is on  much
safer side, one still observes that the soft overlap contribution does not dominate
over the hard amplitude and is more than a factor of 5 below the hard amplitude
including the \( \vec{k}_{\perp } \)-dependence, as one approaches \( Q^{2}\approx  \)
10 GeV\( ^{2} \). This behavior is quite different compared with the overlap
contribution to the pion electromagnetic FF. From Fig.~\ref{fig:piemff}, we
indeed see that a similar calculation (using also a gaussian ansatz for the
\( \vec{k}_{\perp } \)-dependence, with the same parameter \( \sigma  \))
yields at \( Q^{2}\approx  \) 10~GeV\( ^{2} \) an overlap contribution to
the pion electromagnetic FF which is twice as large as the leading order PQCD prediction
including the \( \vec{k}_{\perp } \)-dependence. Our estimate for the overlap
contribution indicates therefore that one is in a more favorable situation with
longitudinal meson electroproduction compared with the pion electromagnetic
FF case. Numerically, the main reason for this is that  in the FF case, the
gluon in the L.O. diagram is purely spacelike yielding a real quantity whereas
in the L.O. meson electroproduction diagram, the exchanged gluon can be both
spacelike or timelike according to the value of the quark's momentum fraction
\( x \), yielding a complex amplitude. The form factor is therefore comparable
to the real part of the hard electroproduction amplitude. As the imaginary part
of the hard electroproduction amplitude is numerically by far larger than its
real part in the kinematics considered here, the competing (real) soft overlap
mechanism might well dominate the real part of the amplitude at the
lower \( Q^{2} \), without dominating the predictions calculated with 
the total L.O. electroproduction amplitude. 
\newline
\indent 
At present, no experimental data for the \( \rho ^{0}_{L} \) electroproduction
at larger \( Q^{2} \) exist in the valence region (\( x_{B}\approx  \) 0.3).
However, at smaller values of \( x_{B} \), the reaction \( \gamma ^{*}\, p\longrightarrow \rho ^{0}_{L}\, p \)
has been measured at rather large \( Q^{2} \). We will therefore compare our
results to see how these data approach the valence region, where one is sensitive
to the quark OFPD's. For the purpose of this discussion, we call the mechanism
of Fig.~\ref{fig:factmeson} which proceeds through the quark OFPD's, the Quark
Exchange Mechanism (QEM). It is well known that, besides the QEM of Fig.~\ref{fig:factmeson},
\( \rho ^{0} \) electroproduction (and, more generally, neutral vector meson
electroproduction) can also proceed through a two-gluon exchange
mechanism, where it has mostly been studied in the past. 
For this latter, Frankfurt, Koepf and Strikman
\cite{Fra96} calculated longitudinal \( \rho ^{0}_{L} \) electroproduction
at low \( x_{B} \) and large \( Q^{2} \) through a Perturbative Two Gluon
Exchange Mechanism (PTGEM). In this PTGEM, the two quarks connecting to the
lower blob in Fig.~(\ref{fig:factmeson}) are replaced by two gluon lines. We
have implemented this model for the PTGEM, in which the longitudinal forward
differential \( \rho ^{0}_{L} \) electroproduction cross section was found
as \cite{Fra96} : 
\begin{eqnarray}
\label{eq:fks1}
\frac{d\sigma ^{L}_{\gamma ^{*}N\rightarrow VN}}{dt}\mid _{t=0}\; =&&\; 
12\pi ^{3}\Gamma _{V\rightarrow e^{+}e^{-}}M_{V}\alpha ^{2}_{s}(Q)\eta
^{2}_{V} \,/\, \left(\alpha _{em}Q^{6}N^{2}_{c} \right) \nonumber\\
&&\times \mid (1+i\frac{\pi }{2}\frac{d}{d \, \ln x})\; x\,
G_{T}(x,Q^{2})\mid ^{2} \; T(Q^{2})\;,
\end{eqnarray}
 where \( G_{T}(x,Q^{2}) \) is the unpolarized gluon distribution and 
\begin{equation}
\eta _{V}\equiv \frac{1}{2}\; \frac{\int dz\, d^{2}\vec{k}_{\perp }\; \Psi _{V}(z,\vec{k}_{\perp })\, /\, \left( z(1-z)\right) }{\int dz\, d^{2}\vec{k}_{\perp }\; \Psi _{V}(z,\vec{k}_{\perp })}\; ,
\end{equation}
 is equal to 3 if one uses  the normalization of Eq.~(\ref{eq:normpsi}) for
the vector meson wavefuntion \( \Psi _{V}(z,\vec{k}_{\perp }) \) with 
the asymptotic distribution amplitude. The factor \( T(Q^{2}) \) in Eq.~(\ref{eq:fks1})
accounts for preasymptotic effects and was calculated in Ref.~\cite{Fra96}
as : 
\begin{equation}
T(Q^{2})=\left( \frac{\int _{0}^{1}dz\, \int
 _{0}^{Q^{2}}d^{2}\vec{k}_{\perp }\; \Psi _{V}(z,\vec{k}_{\perp
 })(-\frac{1}{4}{\Delta }_{\perp }) \;A(z, \vec k_\perp, Q^2)}
{\int _{0}^{1} d z \, \int _{0}^{Q^{2}}d^{2}\vec{k}_{\perp
 }\; \Psi _{V}(z,\vec{k}_{\perp }) 
\,/\, \left( z (1 - z) \right) }\right) ^{2}\; ,
\label{eq:fermimotion}
\end{equation}
 where \( \Delta _{\perp } \) is the two-dimensional Laplacian operator in
transverse momentum space, and where we introduced the notation
\begin{equation}
A(z, \vec k_\perp, Q^2) \;=\; 
\frac{Q^{4}}{Q^{2}\, +\, \left( \vec{k}_{\perp}^{2}+m^{2}\right) 
\, /\, \left( z(1-z)\right) } \;. 
\end{equation}
The correction of Eq.~(\ref{eq:fermimotion}) is due to 
the Fermi motion of the quarks in the vector meson and is equivalent to the 
intrinsic transverse momentum degree
of freedom that we introduced in section \ref{sec:sec4}. Likewise, it results
in a significant suppression of the cross section as \( Q^{2} \) decreases
(\( T(Q^{2})\rightarrow 1 \) as \( Q^{2}\rightarrow \infty  \)). 
\newline
\indent 
Intuitively, it is clear that the PTGEM should dominate at high c.m. 
energies, that is  small
\( x_{B} \), where the \( gluon \) distribution dominates. The QEM mechanism,
which is proportional to the \( quark \) distributions will dominate in the
valence region. Due to the \( sea \) \( quark \) distribution, it also contributes
and rises in the large W region but quite less than the PTGEM. 
In Fig.~\ref{fig:rhotot2},
the forward differential longitudinal \( \rho ^{0}_{L} \) electroproduction
cross section is shown as a function of the c.m. energy \( W \) for three values
of \( Q^{2} \) (5.6, 9 and 27 GeV\( ^{2} \)). The figure shows the contributions
of both mechanisms. The PTGEM and QEM cross sections are both calculated with
the MRST98 parton distribution parametrizations \cite{MRST98}, including their
evolution. For the QEM, we include the \( \vec{k}_{\perp } \)-dependence effects
as
discussed in section \ref{sec:sec4}. One sees from Fig.~\ref{fig:rhotot2}
that the PTGEM explains well the fast increase at high energy of the cross section
but it is clear that it substantially underestimates the data at lower energies
(around \( W\approx  \) 10 GeV). This is where the QEM is expected to contribute
since \( x_{B} \) is then in the valence region. The QEM describes well the
change of behavior of the data (like a plateau) at lower W. The incoherent sum
of both mechanisms is also indicated. Fig.~\ref{fig:rhotot1} shows this incoherent
sum of the two mechanisms for other \( Q^{2} \) values, up to the largest \( Q^{2} \)
value of 27 GeV\( ^{2} \), measured at ZEUS. It is seen that both \( W \)
and \( Q^{2} \) dependences are rather well reproduced. This provides a good 
indication that the deviation from the PTGEM of the data at lower energies
can be attributed to the onset of the QEM. The present calculation strengthens
our previous conclusion \cite{ourprl} about the onset of the QEM at lower \( W \)
and at large \( Q^{2} \). The calculation uses now a phenomenological \( \xi  \)-dependent
ansatz for the OFPD's based on the most recent MRST98 parametrization for the
parton distributions. Furthermore, the calculation also includes the power corrections
due the intrinsic transverse momentum dependence both in the QEM as discussed
in section \ref{sec:sec4}, and in the PTGEM using the formalism (Eq.~(\ref{eq:fks1}))
of Ref.~\cite{Fra96}.
\newline
\indent
Finally we show in Fig.~\ref{fig:ep_epipn_qdep} how our calculations
for $\gamma^*_L \, p \rightarrow \pi^+ \, n$ compare with the few data at
larger $Q^2$. For the sake of comparison with the calculations of
Ref.~\cite{Mank99b}, we also show the
L.O. predictions for $\tilde H$ and for the pion pole using a running
coupling constant (extrapolated to the lower $Q^2$ values) with
$\Lambda_{QCD}$ = 0.2 GeV. 
Our results are compatible with those of
Ref.~\cite{Mank99b} where it was also found that in the valence
region, the pion pole dominates the $\gamma^*_L \, p \rightarrow \pi^+
\, n$ cross section. We furthermore show in Fig.~\ref{fig:ep_epipn_qdep} our 
predictions including the power corrections. For the power
corrections to the pion pole contribution we use
Eq.~(\ref{eq:lopipole}), where we include in the pion electromagnetic FF 
the transverse momentum dependence and the soft overlap contribution 
as shown in Fig.~\ref{fig:piemff}. It is seen
that in the forward kinematics of Fig.~\ref{fig:ep_epipn_qdep}, these
power corrections enhance the leading order predictions for the 
cross sections substantially, which is also indicated by the
sparse data. It will be most valuable to have more 
accurate data for the longitudinal cross sections for pion
electroproduction to check the consistency of the 
power correction calculations. Of particular interest should be the
ratio of $\pi^+$ versus $\pi^0$, because the $\pi^+$ process, 
which is dominated by the pion pole contribution, has an amplitude
with a large real part in contrast to the $\pi^0$ channel, where the
pion pole contribution is absent.

\section{Conclusion}

\label{sec:sec6}

We have given in this work predictions for leading order observables for DVCS
and various meson electroproduction reactions in the valence region at large
\( Q^{2} \), using a \( \xi  \)-dependent ansatz for the OFPD's. We have indicated
some observables and kinematical conditions where experiments are already planned
or can be performed. As these \textit{exclusive} experiments can currently only
be performed at not too high values of \( Q^{2} \), we have estimated the power
corrections due to the intrinsic transverse momentum dependence
of the partons in the DVCS and meson electroproduction amplitudes. We have taken
the \( \pi ^{0}\gamma ^{*}\gamma  \) transition FF and the pion electromagnetic
FF as our guidance to estimate these power corrections to both DVCS and hard
meson electroproduction reactions respectively. In this way, we have generalized
the skewed parton distribution formalism to include the parton intrinsic transverse
momentum dependence. For the meson electroproduction amplitude, we have estimated,
in addition, the competing soft overlap mechanism, which - in contrast to the
leading order perturbative mechanism - does not proceed through one-gluon exchange.
The soft overlap contribution is found not to be dominant in contrast to the
pion FF case. Our estimates for these different power corrections show that
by measuring exclusive electroproduction reactions in the range \( Q^{2}\approx  \)
2 - 10 GeV\( ^{2} \), one will be able to study the onset of the predicted
scaling. In addition, the preasymptotic effects (in the valence region) can
teach us about the quark's intrinsic transverse momentum dependence in the nucleon.
\\
 \indent In conclusion, we believe that a broad new physics program, i.e. the
study of \textit{exclusive} reactions at large \( Q^{2} \) in the valence region,
where the quark exchange mechanism dominates, opens up. Although these exclusive
experiments at large \( Q^{2} \) are quite demanding, we think that both the
scaling region and the onset of the scaling region promise to be sufficiently
rich to motivate an extensive experimental investigation at different facilities.

\section*{acknowledgments}

We like to acknowledge useful discussions with M. Polyakov, A. Radyushkin, N.
Stefanis, M. Strikman and C. Weiss. This work was supported in part by the Deutsche
Forschungsgemeinschaft (SFB 443), by the French Commissariat \`{a} l'Energie
Atomique (CEA), and by the French CNRS/IN2P3.

\newpage
\begin{table}
\begin{tabular}{|c|c|c|c|c|}
\hline 
\(  \)&
 \( \int _{-1}^{1}dx\, x\, H \)&
 \( \int _{-1}^{1}dx\, x\, (H+E) \)&
 SMC &
 \( L_{q} \)\\
\hline 
\( u_{v} \)&
 .28 &
 .51 &
 \( \Delta u_{v}=.77\pm .10\pm .08 \)&
 -.13\\
\hline 
\( d_{v} \)&
 .11 &
 -.12 &
 \( \Delta d_{v}=-.52\pm .14\pm .09 \)&
 .20 \\
\hline 
\( u_{v}+d_{v} \)&
 .39 &
 .51 &
 \( \Delta u_{v}+\Delta d_{v} \)=.25 &
 .07\\
\hline 
\( u+\bar{u} \)&
 .34 &
 .62 &
 \( \Delta u_{v}+2\Delta \bar{u} \)=.79 &
 -.09\\
\hline 
\( d+\bar{d} \)&
 .18 &
 -.19 &
 \( \Delta d_{v}+2\Delta \bar{d} \)=-.50 &
 .16\\
\hline 
(\( u+\bar{u} \))+(\( d+\bar{d} \)) &
 .52 &
 .43 &
 \( \sum  \)=.29 &
 .07  \\
\hline 
\end{tabular}
\vspace{1cm}

\caption{\label{table1}The evaluation of the spin sum rule using the \protect\protect\( \xi \protect \protect \)-independent
factorized ansatz for the OFPD's as described in the text. The column denoted
by SMC contains the experimentally determined \protect\cite{SMC98} 
contributions of the different quark flavors to the nucleon spin, 
measured through semi-inclusive spin asymmetries. 
Remark that the model calculation implies a fraction of 0.57 for the gluon
contribution to the nucleon spin.}
\end{table}

\newpage

\vspace{5cm}

\vspace{-6cm}

\vspace {5cm}

\vspace {5cm}

\end{document}